\documentclass{aa}
\usepackage{graphicx}
\graphicspath{{figures/}}	% Path to figure files
\usepackage{caption}
\usepackage{subcaption}
\usepackage{amssymb}
\usepackage{amsfonts}

\usepackage[varg]{txfonts} %AA request
\usepackage{amssymb}
\usepackage{hyperref}
\usepackage{wrapfig}

\usepackage{soul}

\def\alphahot{\alpha_\mathrm{hot}}
\def\alphacold{\alpha_\mathrm{cold}}

 \def\Ledd{L_\mathrm{Edd}}

\def\texp{t_\mathrm{exp}}
 \def\nut{\nu_{\mathrm{t}}}

\newcommand{\vertcode}{{\textsc{AlphaDisc}}\xspace}

\newcommand{\dotMEdd}{\dot M_\mathrm{Edd}}
\newcommand{\dotMbefore}{\dot M_\mathrm{before}}
\newcommand{\Mdisc}{M_\mathrm{flare}}
\newcommand{\tpeak}{t_\mathrm{peak}}

\newcommand{\prm}{\mathrm{Pr}_\mathrm{m}}
\newcommand{\Rem}{\mathrm{Re}_\mathrm{m}}
\newcommand{\magdif}{{\nu_\mathrm{m}}}

\newcommand{\Rout}{R_\mathrm{out}}
\newcommand{\Rmax}{R_\mathrm{out,max}}
 \newcommand{\rmax}{r_\mathrm{max}}
 
 \newcommand{\Rs}{R_\mathrm{S}}
 \newcommand{\rg}{R_\mathrm{S}}
 \newcommand{\Rtrunc}{R_\mathrm{trunc}}
 \newcommand{\Rsg}{R_\mathrm{sg}}
 \newcommand{\tvis}{t_\mathrm{vis}}
\newcommand{\tfront}{t_\mathrm{front}}
 
 \newcommand{\Sigmadv}{\Sigma_\mathrm{adv}}
 \newcommand{\SigmaA}{\Sigma_\mathrm{A}}
 \newcommand{\Sigmastat}{\Sigma_\mathrm{stat}}

 \renewcommand{\emph}[1]{\textbf{#1}}
\newcommand{\pn}[1]{}

\newcommand{\dotMed}{\dot M_{\rm Edd}}

\newcommand{\Ugs}{\ensuremath{\mathrm{g}\,\mathrm{s}^{-1}}}
\newcommand{\Usigma}{\ensuremath{\mathrm{g}\,\mathrm{cm}^{-2}}}
\newcommand{\Uergs}{\ensuremath{\mathrm{erg}\,\mathrm{s}^{-1}}}

\graphicspath{{figures/}}
\usepackage{xcolor}

\newcommand{\revtri}[1]{\textbf{#1}}

\renewcommand{\revtri}[1]{{#1}}

\begin{document}

    \title{Fast giant flares in discs around supermassive black holes}
   %[Giant flares in discs around SMBH]
   % \subtitle{}

   \author{G. V. Lipunova\inst{1}
          \and
          A. S. Tavleev\inst{2}%\fnmsep\thanks{Just to show the usage of the elements in the author field}
          \and
          K. L. Malanchev\inst{3,4}
          }

   \institute{Max-Planck-Institut f\"ur Radioastronomie, Auf dem H\"ugel 69, 53121 Bonn, Germany\\
              \email{gvlipunova@gmail.com}
          \and
          Institut f\"ur Astronomie und Astrophysik, Kepler Center for Astro and Particle Physics, Universit\"at T\"ubingen, Sand 1, 72076 T\"ubingen, Germany
          \and
          McWilliams Center for Cosmology \& Astrophysics, Department of Physics, Carnegie Mellon University, Pittsburgh, PA 15213, USA
          \and
          Department of Astronomy, University of Illinois at Urbana-Champaign, 1002 W. Green St., IL 61801, USA
             }

  % \date{Received \dots , 2024 / Accepted \dots}

  \abstract
   {}
   {
We studied the thermal stability of non-self-gravitating turbulent $\alpha$-discs around supermassive black holes (SMBHs) to test a new type of high-amplitude galactic nucleus flares.
    }
{
By calculating the disc structures, we computed the critical points of equilibrium curves for discs around SMBHs, which cover a wide range of accretion rates and resemble the shape of a $\xi$ curve.  
}
   {
We find that a transition of a disc ring from a recombined cold state to a hot, fully ionised, advection dominated, geometrically thick state is possible. 
Such a transition can trigger a giant flare for SMBHs with masses $\sim 10^6-10^8\, M_\odot$ if the prior geometrically thin and optically thick disc with convective energy transport surrounded a central radiatively inefficient accretion flow.
{An increase in the viscosity parameter $\alpha$ is a necessary condition for this scenario. This increase may be related to the fact that the magnetic Prandtl number increases and exceeds 1 during ionisation.}
When self-gravity effects in the disc are negligible, the duration and power of the flare 
exhibit a positive correlation with the prior truncation radius of the geometrically thin disc. 
According to our rough estimates, the mass of about $\sim 4-3000\, M_\odot$ can be involved in the giant flare lasting 1 to 400 years if the flare is triggered somewhere between $60$ and $600$ gravitational radii from the SMBH of $10^7\, M_\odot$.
The accretion rate on the SMBH peaks about ten times faster at the potentially super-Eddington level. 
An optically thick outflow with the comparable mass loss rate leads to anisotropy of the emission.
{At the beginning of the giant flare, the region near the truncation radius is heated to $\sim 10^5\,$K, and its UV/optical luminosity is at least $\sim 0.3-4 \,L_\mathrm{Edd}$ depending on the SMBH mass.  
}
}
   { 
The sudden heating of a cold disc around a SMBH can trigger a massive outburst, similar \revtri{in appearance} to what is proposed to occur after a tidal disruption event.
   }

   \maketitle
%
%-------------------------------------------------------------------

%%%%%%%%%%%%%%%%% BODY OF PAPER %%%%%%%%%%%%%%%%%%

\section{Introduction}\label{s.intro}

Disc instability is believed to be one of the underlying causes of transient events in binary systems with accretion discs.
For dwarf and X-ray novae, a well-studied scenario of outbursts, the Disc Instability Model~\citep[DIM; see e.g.][and references therein]{Hameury2020_review} has been developed, according to which the non-monotonic opacity behaviour related to the partial ionisation of hydrogen drives the thermal-viscous instability. 
Admittedly, discs in active galactic nuclei (AGNs) can also undergo viscous-thermal instabilities~\citep[e.g.][]{Mineshige1990, Siemiginowska_etal1996, Menou-Quataert2001TDE,Janiuk+2004}.

 In  low-mass X-ray binaries, matter from a normal star may accumulate in a slowly expanding torus around a compact object. As the mass accumulates, the temperature gradually rises and ionisation of the hydrogen eventually becomes possible.
Then a notorious opacity dependence on temperature allows a thermal instability~\citep[see e.g.][]{Faulkner_etal1983_1}. 
The outburst mechanism can be suitably illustrated as a limit cycle on the relationship between the effective temperature~$T_{\rm eff}$ or the accretion rate $\dot M$ and the surface density $\Sigma$ (called the S-curve; see Fig.~\ref{fig:S-curve} or figure 4 in the review by \citet{Lasota2016}, also \citet{Meyer-MeyerH1981,Smak1984PASP}). 
 If the parameters of a particular ring of a disc fall on the negative branch of the S-curve, the ring is unstable. The ionisation instability  occurs at temperatures $\sim 10^4 $~K.
 
Another instability is associated with the radiation-pressure-dominated optically thick regime  and occurs at higher temperatures.
A corresponding limit-cycle behaviour for black-hole discs with accretion rates close to the critical Eddington accretion rate has been previously proposed~\citep{Taam-Lin1984, Lasota-Pelat1991, Cannizzo1996, Belloni+1997,Szuszkiewicz-Miller1998,Janiuk+2000,Janiuk-Czerny2011,Czerny+2019}.

For the present work, we studied the equilibrium curves for discs around SMBHs in a wide range of accretion rates. These equilibrium curves resemble the letter $\xi$ and have two unstable branches, one connected to the partial ionisation instability and the other to the radiation-pressure instability.
We find a new scenario of disc heating, triggered by the ionisation instability, which can only occur in a disc around a SMBH and requires that the turbulent $\alpha$-parameter becomes higher in the ionised state.
As the radiation pressure plays a much larger role in discs around SMBH than around stellar-mass black holes (BHs), a situation arises when the two instabilities can occur at the same surface density. This can cause a cold disc ring heated by the ionisation instability to bypass the hot thin state and continue to heat until it becomes geometrically thick, leading to a huge increase in the accretion rate, by a factor of $10^5-10^6$.
Such a transition is likely to induce the super-Eddington state, when strong outflows are produced.
The high aspect ratio of the optically thick advection-dominated disc explains the relatively short flare duration.

We expect such flares to be observationally similar to tidal disruption events (TDEs), except for the asymmetric manifestations. According to our toy-model estimates, the giant flares can be more powerful and longer than the typical average TDE.
The emission from the initial heating is likely to be in the extreme ultraviolet and optical diapason, followed by soft X-rays, which may be strongly attenuated, as is usually expected from the outflowing accretion discs in the unfortunate geometrical situation.

In Sect.~\ref{s.Ecurves_flares} we describe an analysis that can be carried out using equilibrium curves, which shows why giant flares are possible. A more detailed description of the conditions for the development of `normal' and giant flares in discs around SMBH is given in Sect.~\ref{s.SMBH_flares}. The basic properties of giant flares are considered in Sect.~\ref{s.properties}. 
We further discuss key assumptions for the mechanism and observational implications in Sect.~\ref{s.discussion}, and we summarise in Sect.~\ref{s.summary}.

In the paper, we use the following  definition for the Eddington accretion rate:
\begin{equation}
 0.1\, \dotMEdd \, c^2  \equiv L_{\rm Edd}\, ,\quad  \dotMEdd  \simeq 1.39 \cdot 10^{18}\, \frac{M}{M_{\sun}}\, \Ugs\, .
 \label{eq.dotMEd}
\end{equation}
Here $M$ is the BH mass and
 the Eddington luminosity is
\begin{equation}
L_{\rm Edd} = \frac{4\pi GMcm_\mathrm{p}}{\sigma_{\rm T}} \simeq 1.25\cdot 10^{38} \, \frac{M}{M_{\sun}}\, \Uergs\, .
\end{equation}
The Thomson cross-section is denoted by $\sigma_{\rm T}$  and other symbols have their standard meaning.

\section{Equilibrium curves and flares}\label{s.Ecurves_flares}
\subsection{S-curves} \label{s.scurves}
\begin{figure}
\center{\includegraphics[width=0.75\columnwidth]{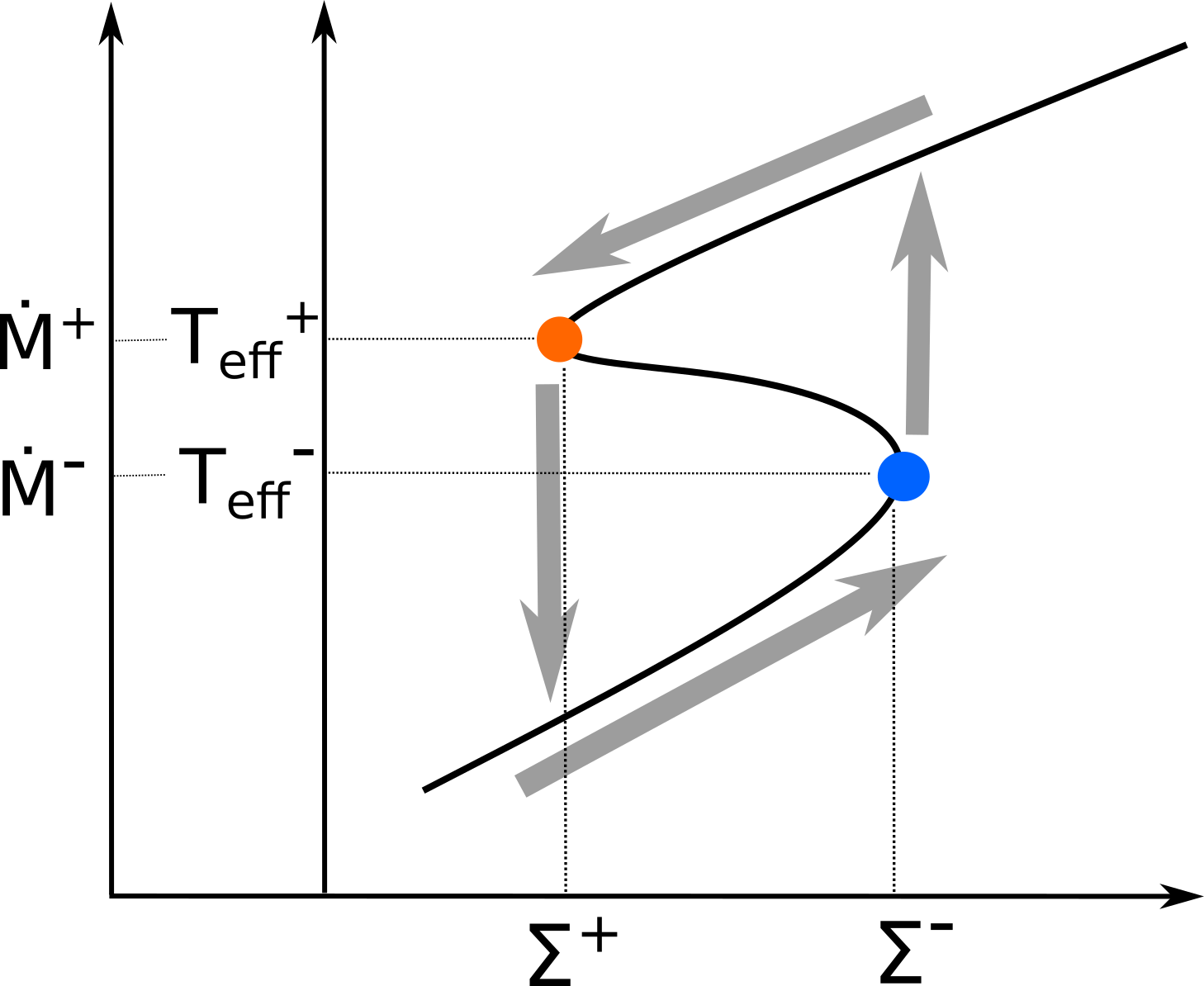}}
	\caption{General S-curve at some radius in accretion disc. The limit cycle instability is shown schematically: on the lower branch the mass accumulates, on the upper branch the mass flows out of a ring.}
\label{fig:S-curve}
\end{figure}
\begin{figure}
	\center{\includegraphics[width=0.8\linewidth]{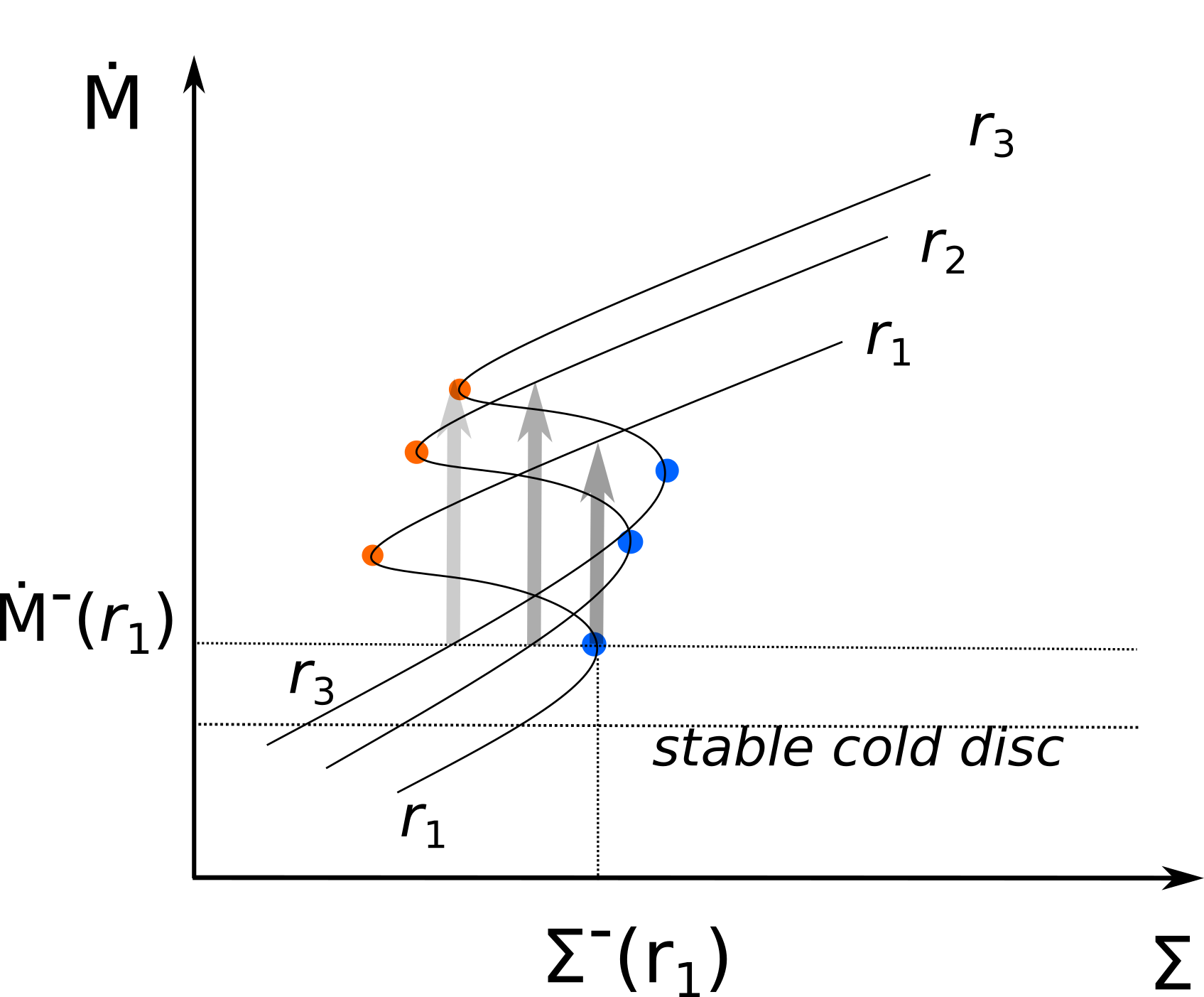}}
	\caption{S-curves at different radii $r_1<r_2<r_3$, shown schematically for ionisation instability. When the accretion rate is very low, the whole disc is stable and cold (the bottom line on the plot). If the surface density exceeds the critical value  at the innermost radius $r_1$, a heating wave, indicated by  arrows, starts outwards, potentially reaching $r_3$. %This picture presumes frozen $\Sigma$ at all radii.
 }
	\label{fig:S-set}
\end{figure}

S-curve is a graphically depicted locus of the disc accretion rate~$\dot{M}$ (or effective temperature $T_{\rm eff}$) and the surface density~$\Sigma$ at a single disc radius, see Fig.~\ref{fig:S-curve}. The intervals of an S-curve with a positive slope represent the thermally and viscously stable state of the disc, whereas the negative slope intervals are unstable. To obtain an S-curve one needs to solve the system of equations for the vertical structure of the accretion disc to obtain the relationship between the required values.
%(approximate S-curves can be constructed by averaging the equations over the $z$ coordinate, which allows the relationship between the required values to be obtained). 

\begin{figure*}
\center{\includegraphics[width=0.95\linewidth]{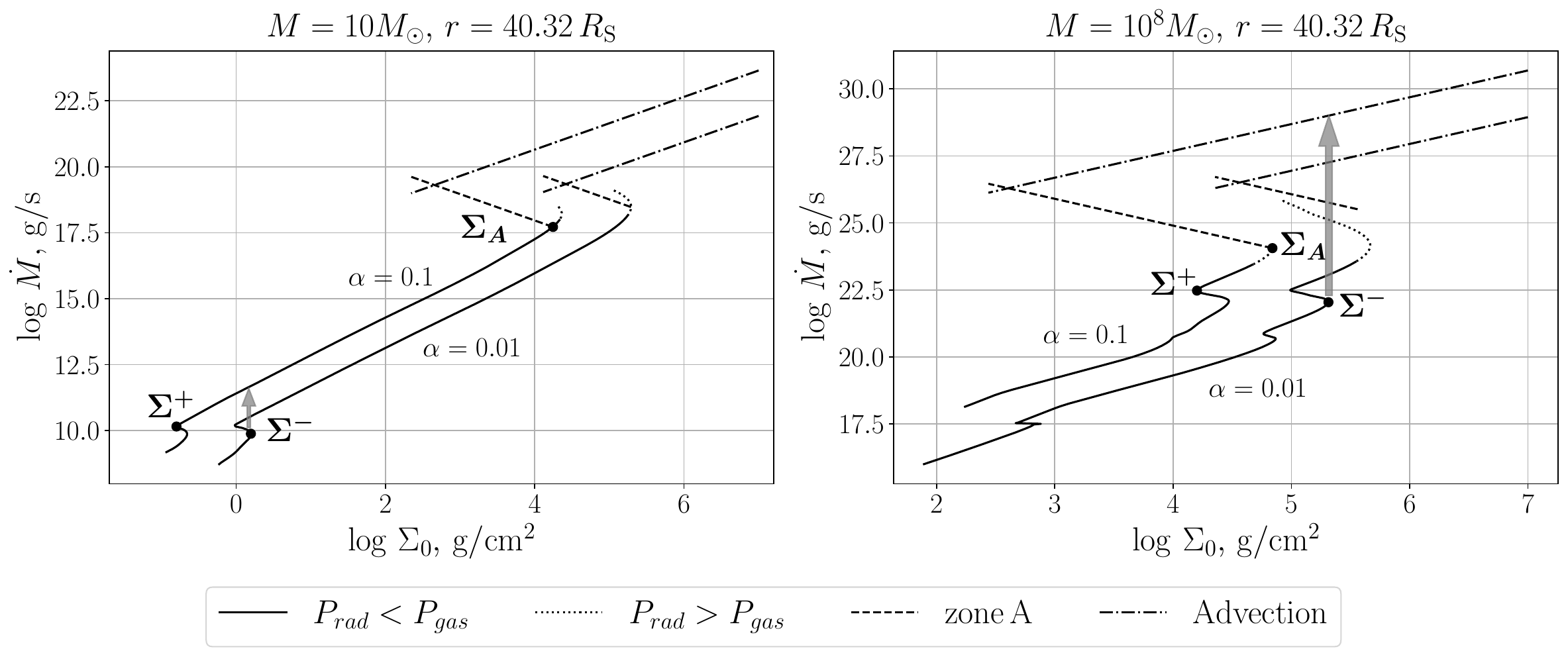}}
	\caption{Equilibrium  curves for discs around stellar-mass BH (left panel) and SMBH (right panel). In each panel, the  $\xi$-curves are constructed for two values of the turbulent parameter $\alpha$: 0.1 and 0.01. 
    \revtri{The solid lines are calculated by \vertcode{} and their upper dotted tails  correspond to $P_{\rm rad}>P_{\rm gas}$.}
    Analytic relations in the radiation-pressure regime (A-zone, dashed, $\propto \Sigma^{-1}$) and the advective regime (dash-dotted, $\propto \Sigma$) are calculated using Eqs.~\eqref{eq.A_zone} and \eqref{eq.Adv_zone}, respectively. 
    The dots mark critical points for the  geometrically thin states:  the minimum and the maximum surface density, $\Sigma^+$ and $\Sigma_A$, in the ionised state and the maximum surface density $\Sigma^-$ in the recombined state.
    The arrows schematically show the direction of ring heating due to ionisation instability. At the same time, $\alpha$ increases.   }
\label{fig:S-curve_SMBH}
\end{figure*}
In Fig.~\ref{fig:S-set} we depict S-curves in a simplified form  to demonstrate how the development of instability leads to an outburst. 
We neglect for a while a possible change of the turbulent parameter $\alpha$. 
\footnote{The change of $\alpha$ is not needed to produce an instability cycle when only one type of two instabilities is involved.}
For each radius in a quasi-stationary disc, a proper S-curve can be constructed.
The orange and blue points mark the critical surface densities $\Sigma^-$ and $\Sigma^+$ -- the turning points. If the actual $\Sigma$ of a ring lies within this interval, the ring can exist in either of the two stable states (for a viscous time).
We assume that the whole disc is in the stable cold state and gaining material. If the critical density has been accumulated at the radius $r_1$, the ring can `move' from the blue turning point to the upper branch of the S-curve (on a thermal timescale). 
The ring will heat the neighbouring rings ($r_2$ and $r_3$ in the picture) by the heat diffusion, advection, turbulent diffusion, and work of pressure forces~\citep{Menou+1999}, and such an avalanche-like process leads to the heating of the disc rings in a significant range of radii. Simplifying the picture, if the heating is triggered at $r_1$  in Fig.~\ref{fig:S-set}, the heating wave could reach $r_3$ but no further, 
since the density there is the minimum possible to allow the ionisation of matter.
The velocity of the radial motion of matter in the heated portion of the disc increases and the accretion rate grows, that is, a flare occurs.

In Fig.~\ref{fig:S-curve_SMBH} we plot
S-curves for geometrically thin and thick $\alpha$-discs. The open code \vertcode by \citet{Tavleev+2023, DiscCode_ascl} solves the equations and constructs the \hbox{S-curves} for the geometrically thin $\alpha$-disc (solid lines, see Appendix~\ref{s.thin_disc_code}).
At higher temperatures, the radiation-dominated regime comprises another negative branch of the $\dot M(\Sigma)$ dependence.
This regime is also thermally and viscous unstable~\citep{Lightman_etal1974,Shakura-Sunyaev1976, Piran1978}. 
A hotter positive-slope branch emerges at even higher accretion rates when photon trapping with the radially moving matter (advection) acts as an energy sink at each radius~\citep{Abramowicz+1988, Abramowicz+1995,Bjoernsson+1996}.

To construct S-curves for the accretion rates around~$\dotMEdd$ and higher, one needs to include the non-local energy balance in the equations of the radial structure, while the vertical structure can be taken as vertically integrated~\citep[see][for a vertical solution]{Sadowski+2011}. The relativistic super-Eddington disc radial solution was obtained by~\citet{Sadowski2009}. 
However, to show the possibility of giant flares, it is not necessary to consider the non-Keplerian character of the flow. The simple analytical relations obtained for the branches of the S-curve of \citet{Abramowicz+1988} are sufficient for this purpose.

When the radiation pressure dominates the gas pressure in the geometrically thin disc~\citep[the A-zone,][]{sha-sun1973}, the height-integrated viscous stress (or $\dot M$) can be expressed algebraically from $\Sigma$ considering the vertical structure equations~\citep{Lightman_etal1974}.
First, we consider a well-known relation between the accretion rate in a stationary disc and the surface density that follows from the definition of the viscous-stress tensor~\citep{lyn-pri1974}:\begin{equation}
    \dot M = 3\pi\, \nut\, \Sigma_0\, .
\end{equation}
Here $\nut$ is the kinematic coefficient of viscosity. In the scope of $\alpha$-ansatz for turbulent viscosity, see Eq.~\eqref{eq.alpha_nut_relation}, this can be re-written as:
\begin{equation}
    \dot M \propto 2\pi\, \alpha \, \sqrt{G\,M\, r}\, \Sigma_0 \,\left(\frac{z_0}{r}\right)^2\, .
    \label{eq.Mdot_Sigma_ralation}
\end{equation}
Here we have taken into account that  the sound velocity \hbox{$ \propto z_0 \,\omega_{\rm K}$}, which follows from the vertical hydrostatic balance, where $\omega_{\rm K}$ is the Keplerian angular velocity.

For the purposes of the qualitative analysis, we use the well-known facts about the thickness of the disc. In the zone where radiation pressure dominates, the disc thickness is proportional to $\dot M$~\citep{sha-sun1973}. Accordingly, we get from  Eq.~\eqref{eq.Mdot_Sigma_ralation} that $\dot M \propto 1/\Sigma_0$ and the disc is unstable~\citep[see, e.g.][]{Lightman_etal1974,KFM2008}.  In the advection-dominated zone, $z_0/r$ saturates at a constant value~\citep[][]{Abramowicz+1988, Lipunova1999, Sadowski+2011, Lasota+2016, Chashkina+2019}, which results in $\dot M \propto \Sigma_0$.  These dependencies are essentially shown in Fig.~\ref{fig:S-curve_SMBH} by the dashed and dot-dashed lines, respectively %These relations are shown in Fig.~\ref{fig:S-curve_SMBH}. 
(Eqs.~\eqref{eq.A_zone} and \eqref{eq.Adv_zone} are used to plot them, see Appendix~\ref{s.selfsim}). 
For the disc aspect ratio in the advective regime we adopt $z_0/r=0.5$.

\subsection{Possibility of giant flares in accretion discs around
SMBH}
\label{s.discs_SMBH}
It has been proposed that in the discs around stars and \revtri{compact stellar remnants}, the turbulent parameter $\alpha$ is higher in the ionised material than in recombined material~\citep[e.g.][]{Smak1984_part4, Meyer-MeyerH2015}. 
\revtri{Similarly,} we assume different $\alpha$ for a cold recombined disc and a hot fully ionised disc around SMBH, although the decreasing $\alpha$ in the recombined material in AGN discs has been criticised~\citep{Menou-Quataert2001}. \citet{Janiuk+2004} studied the magnetic Reynolds number and ambipolar diffusion number and found that they exceed the critical values required for sustained turbulence unless the accretion rate is very low.
Nevertheless, $\alpha$ might still depend on the degree of ionisation, as shown for protoplanetary discs by \citet{Landry+2013}; see also  discussion in \S2.1 of \citet{Hameury+2009}.
Additionally, we show in Sect.~\ref{s.limits} that, within the partial ionisation regime, the magnetic Prandtl number~-- which is expected to influence the intensity of saturated turbulence and the value of $\alpha$~--  varies strongly and crosses 1, which favours the change of $\alpha$.

{Therefore, assuming that $\alpha$ can change its value, we compare the equilibrium curves for a stellar-mass BH and a SMBH in Fig.~\ref{fig:S-curve_SMBH}, constructed as described in Sect.~\ref{s.scurves}.
In our model, the value $\alphacold$ is considered viable below the $\Sigma^-$ points, while the value $\alphahot$ is deemed possible above the $\Sigma^+$ points of every curve.}

The distinction between the S-curves of the stellar-mass BH and the SMBH is strikingly evident. For SMBH disc, the length of the gas-pressure-dominated interval is much shorter; this essentially enables the proposed mechanism, as illustrated for particular parameters.
Specifically, for the SMBH disc on the right, the maximum critical surface density $\Sigma^-$ in the cold disc with $\alpha=0.01$ is higher than the maximum critical surface density $\SigmaA$ of the geometrically thin hot ionised disc with $\alpha=0.1$.

Thus, if the surface density in the cold state of a SMBH disc exceeds the critical maximum value $\Sigma^-$, the heating instability can lead to a direct transition of the ring to the  advective state, skipping the thin-ionised-disc state. 
A local stability analysis is insufficient for calculating the course of such a process. 
However, the central point of the proposed mechanism is the assumption that, if the ring in the diagram ($\Sigma$, $T$) turns out to have such a high surface density that a stable solution only exists at a higher temperature, then the ring should heat up if it does not lose its density faster. This is similar to the classic mechanism in the DIM. The increase in $\alpha$ can occur over thermal time, at least we assume that in the ionised state  $\alpha$ is higher.
 
 {Some time is required for a transition between the states. In the case of a geometrically thin disc, this time is of the order of the thermal time. In the case of a slim advective disc, whose energy balance is not local (the heating and cooling depends on the radial structure), this time is longer: apparently, it is of the order of the viscous time.
 } 
 Because the slim disc is quite thick, the viscous time-scale $(r/z_0)^2\,(\alpha\,\omega_{\rm K})^{-1}$ is not much longer than the thermal time $(\alpha\,\omega_{\rm K})^{-1}$.
Thus, we expect that over a characteristic viscous time, the heated rings of the disc travel upwards in temperature with some change in surface density. Whether this change is a decrease or an increase depends on the radius of the ring. 
 At the same time, the accretion rate on the central BH increases.

The evolution of the accretion rate on the SMBH depends on how the size of the advective zone decreases with time and on the outflows from the disc surface. 
\revtri{As with outbursts in the DIM,} the duration of the flare can be estimated using the viscous time at the maximum radius $r_{\rm max}$ of the advective zone:
% see general_disc/disc_times.mw & FGF/FGF_times
\begin{equation}
    t_{\rm vis} = \frac{\rmax^2}{\nut} 
    = \frac{3}{2}\, \frac{1}{\alpha\omega_{\rm K}}\, \left(\frac{z_0}r\right)^{-2} \Pi_1\, .
    %\sim \frac{1}{\alpha\omega_{\rm K}}  \, \left(\frac{z_0}r\right)^{-2}\, . 
    \label{eq.tvis}
\end{equation}
This time can be an order or two longer than the orbital time $1/\omega_\mathrm{K}$ because $\alpha\sim 0.1$ and $\Pi_1 \sim 4-$8 (see Appendix~\ref{s.selfsim}).
This time is much shorter than the viscous evolution time in the geometrically thin disc around the SMBH due to the fact that $z_0/r \sim 1$ in the advective regime and $z_0/r\sim 10^{-3}$ for a thin disc. \revtri{Substituting the characteristic values, one obtains}
\begin{equation} \label{eq.tvis_typical}
\tvis 
%\approx 0.4 \, \times \, & \frac{3}{2}\, \frac{1}{\alpha\omega_{\rm K}}\, \left(\frac{z_0}r\right)^{-2} \Pi_1
 \approx  150   \,\left(\frac{\alpha}{0.1}\right)^{-1} \left(\frac{z_0}{r}\right)^{-2} \left(\frac{\rmax}{100\, \Rs}\right)^{3/2} \left(\frac{M}{10^7 \, M_{\odot}}\right)\, \mathrm{days} ,
\end{equation}
 where
\begin{equation}
 \rg = 2 G M / c^2 \, .  
\end{equation}

In the following section, we abandon the analytical approximations for the upper branches of the $\xi$-curves and calculate them numerically. 
 We continue to call them S-curves.

 \section{Flares in discs around SMBH}\label{s.SMBH_flares}

\subsection{Turning points of equilibrium curves}\label{s.points}
The critical values of the equilibrium curves affect the possibility of a giant flare and its amplitude. To find these values, 
\begin{equation}
\begin{aligned}
&\Sigma^- = \Sigma^-(r, \alpha_{\rm cold}), \quad
\Sigma^{+} = \Sigma^{+}(r, \alpha_{\rm hot}), \\
&\Sigma_{\rm A} = \Sigma_{\rm A}(r, \alpha_{\rm hot}), \quad
\Sigma_{\rm adv} = \Sigma_{\rm adv}(r, \alpha_{\rm hot}), \\
&\alpha_{\rm cold} = 0.01\, , \qquad \alpha_{\rm hot} = 0.1\, ,
\label{eq.Sigma_R}
\end{aligned}
\end{equation}
we constructed a set of S-curves (see example in Fig.~\ref{fig:S-4032}) for different SMBH masses.

The lower parts of the S-curves, where the gas pressure dominates, are calculated using the open code \vertcode (see Appendix~\ref{s.thin_disc_code}). 
To calculate upper parts of S-curves, for accretion rates around~$\dotMEdd$ and higher, \revtri{advective models (for many $\dot M$) must be calculated. An advective disc model requires the calculation of the radial structure, since the local energy balance is violated in a thick disc. Such calculations were performed using
the code from \citet{Lipunova1999} (hereafter, L99) for the super-Eddington  Keplerian disc. 
The upper and lower parts of the S-curve are stitched at a point in the hot-disc interval between $\Sigma^+$ and $\SigmaA$, where the disc is geometrically thin. For further details, see Appendix~\ref{s.thick_disc_code}.}

Code L99 can calculate the disc structure with or without a super-Eddington outflow. 
Although the super-Eddington accretion disc is expected to generate strong winds, the critical values of the turning points are not affected by this. Therefore, to plot the S-curves in the present work we used the option without outflow in L99.

%%%%%%%%%% REFEREE version
%\def\widthfigs{0.65}
%\def\widthfigbigger{0.75}
%\def\xshift{2cm}
%%%%%%%%%%%%%% 2 COLUMN VERSION
\def\widthfigs{0.85}
\def\widthfigbigger{0.95}
\def\xshift{0.5cm}

\begin{figure}
\begin{subfigure}[t]{\linewidth}
\hskip \xshift
\includegraphics[width=\widthfigbigger\columnwidth]{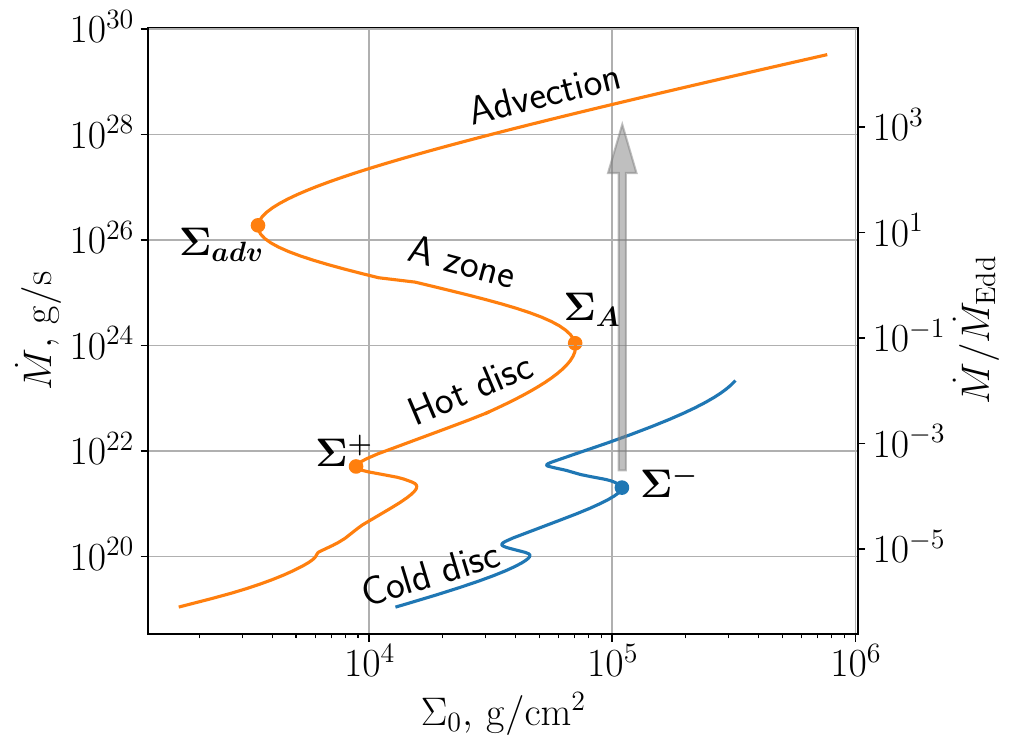}
	\caption{Accretion rate in absolute units shown on the left axis and in the Eddington rates \eqref{eq.dotMEd}, on the right.}
 \label{fig:S-4032}
  \end{subfigure}
    \begin{subfigure}[t]{\linewidth}
    \centering
\includegraphics[width=\widthfigs\columnwidth]{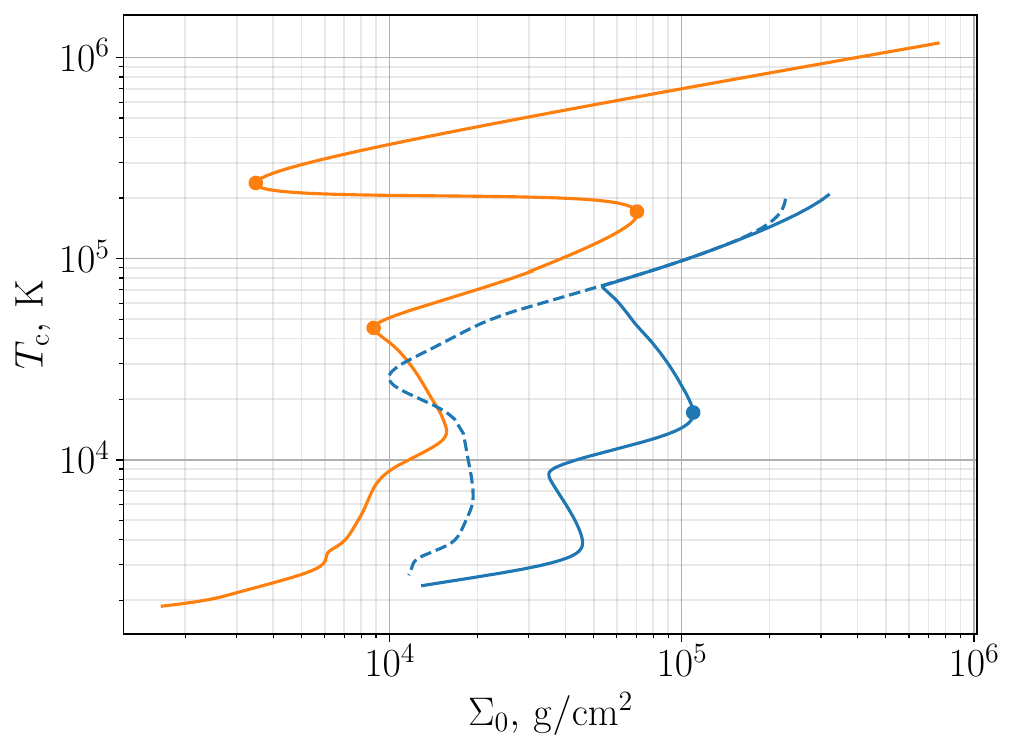}
	\caption{Temperature $T_\mathrm{c}$ in the symmetry plane. The dashed line shows the S-curve for $\alphacold$ without convection.}
	\label{fig:S-4032_Tc}    
    \end{subfigure}
\begin{subfigure}[t]{\linewidth}
\centering
\includegraphics[width=\widthfigs\linewidth]{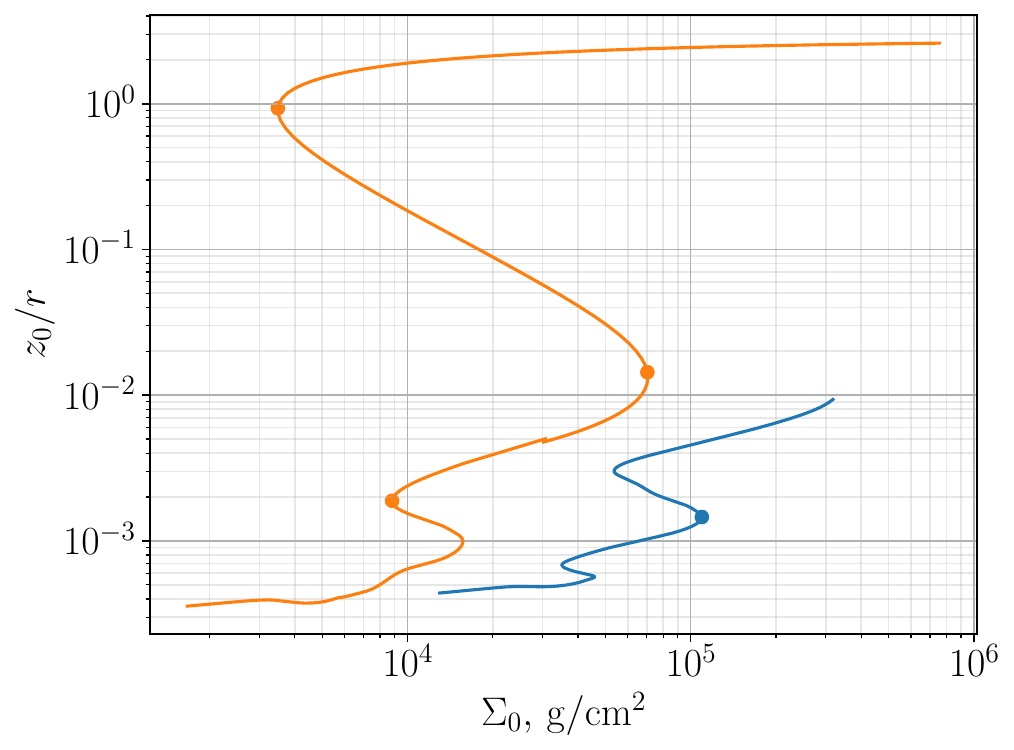}
    \caption{Relative semi-thickness  $z/r$.}
    \label{fig:z_r_curve}
    \end{subfigure}
    \caption{S-curves for $10^7 \, M_{\sun}$ at $r=100\,\rg$ for two values of turbulent parameter $\alphacold=0.01$ (blue) and $\alphahot=0.1$ (orange). The branches  `A-zone' and `Advection' were calculated using the code from~\citet{Lipunova1999}.}
    \label{fig:S-curvesM1e7}
\end{figure}
\begin{figure}
\center{\includegraphics[width=0.96\columnwidth]{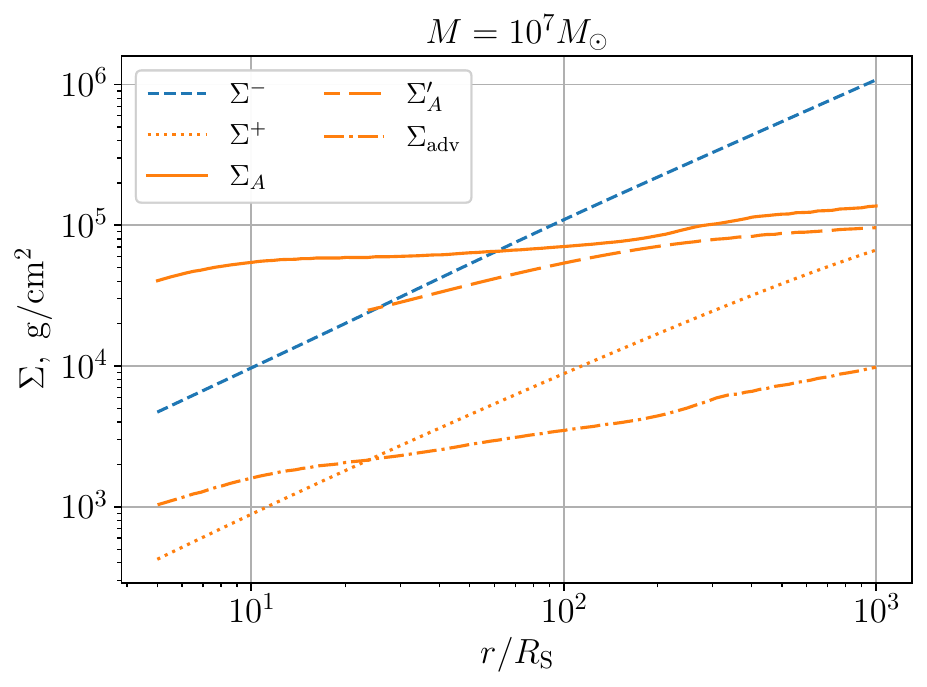}}
	\caption{Critical values of surface density \eqref{eq.Sigma_R}  for $M=10^7\, M_{\sun}$. Viscous parameter is $\alphacold=0.01$ for $\Sigma^-$ (the blue line) and  $\alphahot=0.1$ in all other dependencies (orange lines). Relations $\SigmaA$ and $\SigmaA'$ show results of different codes, see Sect.~\ref{s.points}.}
\label{fig:Sigmas_turns}
\end{figure}

The resulting dependencies of the critical values~(\ref{eq.Sigma_R}) are shown in Fig.~\ref{fig:Sigmas_turns} for SMBH mass $10^7 M_\odot$. 
As described above, $\Sigma^-$ and $\Sigma^+$ are calculated using the \vertcode code, $\SigmaA$ and $\Sigmadv$ are calculated using the L99 code.
Only one curve $\Sigma^-$  is calculated with $\alpha=\alphacold$, and others, with $\alpha=\alphahot$.  A giant flare can happen if \hbox{$\Sigma^-(\alphacold) > \SigmaA(\alphahot)$}. 
Critical $\Sigma^- = \SigmaA \sim 7\times 10^{4}$~\Usigma for $10^6-10^8~M_\odot$ SMBH for specific $\alphahot$, $\alphacold$, and chemical abundance (see Fig.~\ref{fig:SigmaAllDots}).

%The Ledoux Criterion for Convection in a Star
In Fig.~\ref{fig:S-4032_Tc} we also show the equilibrium curve for a disc calculated by \vertcode without convection by the dashed line. This shows that ignoring convection reduces the critical surface density, as already discussed, for example, by \citet{Pojmanski1986,Cannizzo-Reiff1992}. Our values of $\Sigma^-$ agree well with the results of \citet{Hameury+2009}, who included convection.

The aspect ratio of the advective disc is greater than 1 in Fig.~\ref{fig:z_r_curve}. This can be attributed to the approximate character of the vertical structure approach in the code L99  (in the original paper $\Pi_1 = 1$, which decreases the thickness). Here we keep $\Pi_1 $  constant (equal to 6) along the upper S-curve as we are most interested in calculating $\SigmaA$.  
In the following analysis, see Sect.~\ref{s.properties}, we assume a value of $z_0/r=0.5$ for the thickness of the advective slim disc.

In general, the \vertcode code encounters numerical difficulties when the radiation pressure is significant. However, in certain instances, we have been able to identify the turning points, designated as $\Sigma_A'(r)$. The resulting dependence is plotted in Fig.~\ref{fig:Sigmas_turns} by the long dashes. It can be seen that at the radii, where the turning points $\Sigma_A'$ have been successfully identified by \vertcode, the value is less than that obtained via the L99 code.
For further analysis, we will use the $\SigmaA(r)$ dependence.

\begin{figure}[h!]
\centering
\begin{subfigure}[t]{\linewidth}
\centering
\subcaption{}
\includegraphics[width=\widthfigs\textwidth]{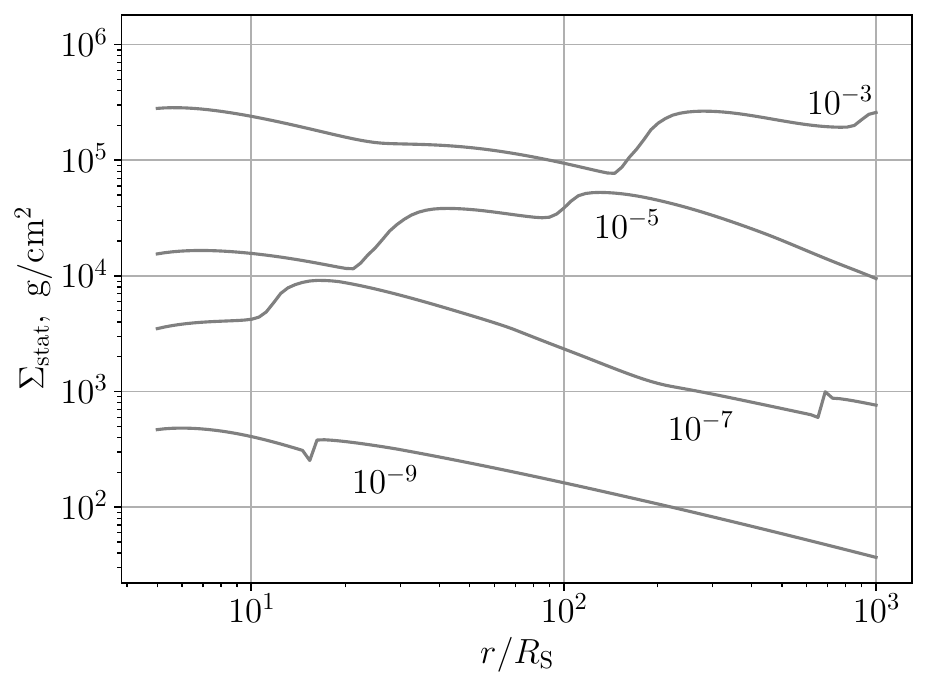} 
\end{subfigure}
\vskip -0.5cm
\begin{subfigure}[t]{\linewidth}
\centering
\subcaption{}
\includegraphics[width=\widthfigs\linewidth]{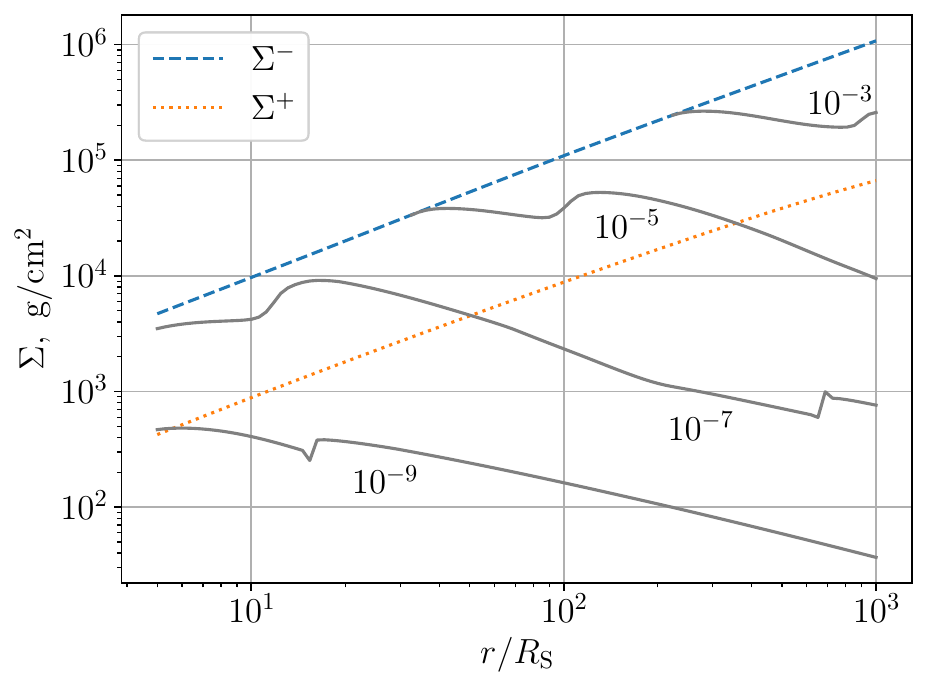}
\end{subfigure}
\vskip -0.5cm
\begin{subfigure}[t]{\linewidth}
\centering
\subcaption{}
\includegraphics[width=\widthfigs\linewidth]{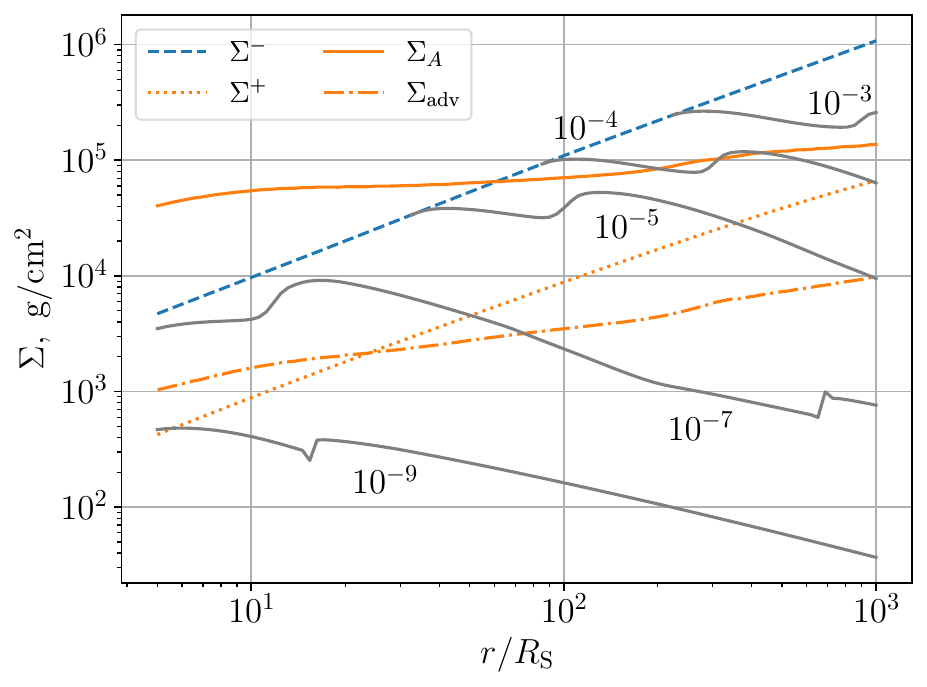}
\end{subfigure}
	\caption{ (a) Quasi-stationary surface density $\Sigma_{\rm stat}(r)$  for $M = 10^7 M_{\sun}$ and  $\alpha=\alpha_{\rm cold}$ at several values of $\dot{M}$ (indicated in the graphs in units of $\dotMEdd$); (b) Radial distributions of the critical points $\Sigma^+$ for $\alphahot$ and $\Sigma^-$ for $\alphacold$ together with the stationary distributions. The latter are shown only for $\Sigma_{\rm stat}<\Sigma^-$; (c)  Superimposed are the radial dependencies of the  critical points $\SigmaA$ and $\Sigmadv$  for $\alphahot$. In all panels, $\alphacold=0.01$ (blue) and $\alphahot=0.1$ (orange). \revtri{Curves are plotted starting from $5\rg$.}  }
 \label{fig:Sigmas_oneM}
\end{figure}

\subsection{Flares in  geometrically thin discs reaching  innermost orbit }\label{s.SMBH_DIM_scenario}

In this subsection, we shall examine the implications of the found  critical densities (Fig.~\ref{fig:Sigmas_turns}), if a disc is geometrically thin down to the innermost orbit. 
We shall consider discs around SMBH with mass $10^7 M_\odot$.

 In Fig.~\ref{fig:Sigmas_oneM}a we show the quasi-stationary-disc surface density distributions $\Sigmastat(r)$ computed by \vertcode for several values of the accretion rate.  In fact, in the real disc, such a density distribution cannot be found to the left of the point where the density exceeds the critical value $\Sigma^-(r)$ and the disc becomes unstable. Accordingly, in panels (b) and (c) we cut the surface density distributions at the intersections with $\Sigma^-(r)$.
Similar plots for other SMBH masses can be found in Fig.~\ref{fig:SigmaAllDots}.

In the quiescent epoch, as the mass accumulates in the disc due to income of the galactic matter, in its innermost parts the accretion rate $\dot{M}$ slowly increases. We imagine that the grey curves change successively in time, replacing each other in order from bottom to top.\footnote{\label{foot.Sigma}
\revtri{In the DIM, the accretion rate increases with radius in a quiescent disc.  Consequently, at the large radii, quiescent $\Sigma(r)$ could be somewhat higher than depicted $\Sigma_\mathrm{stat}(r)$ in Fig.~\ref{fig:Sigmas_oneM}a.
However, this fact does not change our estimates because the condition at the trigger radius does not depend on the radial distribution at larger radii.} }
%{In reality,  the accretion rate probably increases with radius in a quiescent disc. Consequently, realistic  $\Sigma(r)$ would have less negative slope (compare, for example with figure 5b in \citealt{Hameury1998}).  This observation does not lead to qualitative changes in the picture. It can be kept in mind, though, that the accretion rates, indicated in  Fig.~\ref{fig:Sigmas_oneM}, correspond to the accretion rate at the inner disc radius.} 

As long as $\dot{M}\lesssim  10^{-7}\dot{M}_{\rm Edd}$, the whole disc  is in the stable cold state (this accretion rate roughly corresponds to the intersection of $\Sigma^-(r)$ with a grey curve  $\Sigmastat(r)$ at the inner disc edge). 
Nowhere in the disc do the surface density and temperature reach values where hydrogen ionisation triggers the instability. 

When the quasi-stationary accretion rate exceeds \hbox{$\sim 10^{-7}\dot{M}_{\rm Edd}$}, the surface density $\Sigma_{\rm stat}$ becomes greater than $\Sigma^-$ near the inner edge of the disc (see Fig.~\ref{fig:Sigmas_oneM}b) . At this radius $\Sigma^-<\Sigma_{\rm A}$, see Fig.~\ref{fig:Sigmas_oneM}c. This means that the outburst can occur in the classical framework of the DIM. There will be a transition from the `Cold disc' branch to the `Hot disc' branch near the inner edge of the disc ($\Sigma$ is between $\Sigma^{+}$ and  $\SigmaA$), and the hot front will start to propagate outwards. A rebrightening of the disc also occurs.
The accretion rate on the SMBH  increases, resulting in a slow outburst with the characteristic time $t_{\rm vis} \sim 5\cdot 10^4$~yr, found from \eqref{eq.tvis} 
%or \eqref{eq.visc_time_scale} 
by substituting  $z_0/r \sim 0.002 $ (cf. Fig.~\ref{fig:z_r_curve}) and $\rmax \sim 60\, \rg$. (Value $\rmax$ can be roughly estimated  as  the intersection of the grey curve for $10^{-7}\dot{M}_{\rm Edd}$ and the curve $\Sigma^+(r)$, see  Fig.~\ref{fig:Sigmas_oneM}b). 

At the end of the outburst, as the cooling front moves towards the centre and the disc surface density falls below $\Sigma^+$, the disc moves to the lower cold branch of the equilibrium curves at each radius.
The transition to the lower stable branch is actually possible  for any $\Sigma$ in the interval from $\Sigma^+$ to $\Sigma^-$. In that respect the left arrow in Fig.~\ref{fig:S-curve} is drawn very tentatively.
%(Just after the outburst the radial distribution in the disc is not quasi-stationary, but this is not important here).
Due to the external influx of matter, the process can start anew. The time between these flares corresponds to a long viscous time in a geometrically thin cold disc with $\alpha = \alpha_{\rm cold}$.

\subsection{Conditions for a  giant flare}
\label{s.giant_flares_cond}
There are strong indications that to explain the spectral features of some low-accretion-rate accretion discs around BHs (both supermassive and stellar), it is necessary to assume that a geometrically thin disc is truncated from the inside and its inner part is replaced by a high-temperature and low-density accretion flow~\citep[see e.g.][]{Yuan-Narayan2014,Nemmen+2014,Lopez+2024}.

The absence of a standard disc at small radii means that, compared to the situation outlined in Sect.~\ref{s.SMBH_DIM_scenario}, the quiescent accretion rate should increase to higher values until an instability is triggered. (Because the critical $\Sigma^-$ grows with the radius.)
The same considerations underlie the DIM model of \citet{Dubus+2001} for the evolution of X-ray transients in binary systems to explain their quiescent times.

For example, the disc around a $10^7 M_{\sun}$ black hole can be truncated at the radius $R_{\rm trunc}{\sim}\,40\,\Rs$.   
In this case, the quiescent accretion rate must exceed $10^{-5} \dotMEdd$ to trigger an outburst similar to the one described in Sect.~\ref{s.SMBH_DIM_scenario}, when the ring goes into the geometrically thin hot-disc state ($\Sigma^- < \SigmaA$ at $40\,\rg$, see Fig.~\ref{fig:Sigmas_oneM}c).
In particular, if $\Sigma$ is close enough to $\Sigma_{\rm A}$, the increased accretion rate from this ring inwards will be such that the inner part of the disc, at $r< 40\,\Rs$, will be in the radiation-pressure instability zone (the A zone). The radiation-pressure instability was suggested to drive fluctuations, explaining a particular class of Changing-Look AGN~\citep[e.g.][]{Sniegowska+2023}.

In contrast, when the truncation radius $R_{\rm trunc}$ is even larger, the critical density is so high ($\Sigma^- > \SigmaA$ at $R_{\rm trunc}$) that a giant flare occurs.
For $M=10^7 M_{\sun}$ this happens when $R_{\rm trunc}\,{\gtrsim}\,60\,\Rs$. This lower limit for the truncation radius is determined by the intersection of the curves $\Sigma^-(r)$ and $\SigmaA(r)$ (see the bottom panel in Fig.~\ref{fig:Sigmas_oneM}) and depends on the assumed parameters: $\alphahot$, $\alphacold$, and chemical composition.

\begin{figure}
\includegraphics[width=1.0\columnwidth]{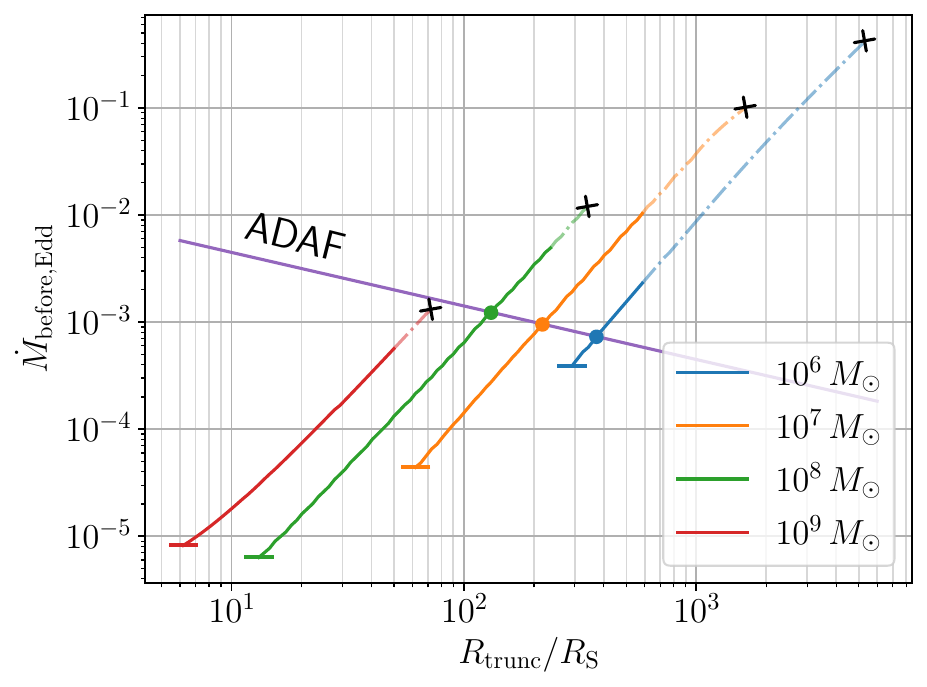}
	\caption{Accretion rate versus truncation radius of thin disc before  giant flare. The predicted dependence between the accretion rate and the ADAF outer boundary (\ref{R_ADAF}) is also shown. The crosses mark the radius constraint due to self-gravity following \eqref{eq.Rself-grav}. The transparent dashed style indicates the $\Rtrunc$ region excluded from further analysis.}
	\label{fig:R_trunc-ADAF}
 %distribution of mass accretion rates https://ui.adsabs.harvard.edu/abs/2015MNRAS.452..575S/abstract
\end{figure}

The truncation radius $\Rtrunc$  itself cannot be determined within our model but it is crucial for the flare properties. 
It directly determines the accretion rate before a giant flare,  which is the critical value of the accretion rate at the turning point (point $\Sigma^-$ in Fig.~\ref{fig:S-4032}).
Figure~\ref{fig:R_trunc-ADAF} shows the accretion rate before the flare $\dot{M}_{\rm before}$ versus the truncation radius $R_{\rm trunc}$ for different SMBH masses. 
We plotted the dependencies only for those truncation radii that are larger than the minimum that allows a giant flare to occur. 
When the truncation radius is smaller than shown in Fig.~\ref{fig:R_trunc-ADAF}, the thermal instability results in a `normal' outburst and cannot heat the disc to the super-Eddington regime.  The maximum value of  $\Rtrunc$ is determined by the self-gravity limit (see  Eq.~(\ref{eq.Rself-grav}) below). 

We can analyse the truncation radii proposed \revtri{in the model of an advection-dominated accretion flow \citep[ADAF;][]{naraya1994} replacing the central part of the thin disc}~\citep{Narayan1996,Esin+1997,Narayan+1997,Poutanen+1997,Dubus+2001,Menou-Quataert2001TDE,Janiuk+2004,Hameury+2007,Narayan-McClintock2008}.
To find a transition radius, one needs to consider the physical conditions leading to the formation and thermal structure of the transition 
%\S4.2.2.
region~\citep[see e.g.][]{Yuan-Narayan2014}.  \citet{Czerny+2004} examined the data from the Broad Line Regions using optical/UV spectra from the Large Bright Quasar Survey and found evidence for a `strong ADAF principle': the geometrically thin solution is chosen by the disc only if it is the only solution available for certain $\Sigma$.
%Poutanen J, Krolik JH, Ryde F. 1997. MNRAS 292:L21–25
Accordingly, an estimate of the radius of the boundary between the ADAF and the standard disc %(\citealt{Abramowicz+1995},\citealt{Czerny+2004}) 
is as follows:
\begin{equation}
    R_{\rm ADAF} / \Rs \sim 200 \left(\frac{\alpha}{0.01}\right)^4 \left(\frac{\dot{M}}{10^{-3}\,\dot{M}_{\rm Edd}}\right)^{-2}. 
    \label{R_ADAF}
\end{equation}
Inverted relation~(\ref{R_ADAF}) is  shown in Fig.~\ref{fig:R_trunc-ADAF}.    \revtri{Taking into account condition  $\dot M_{\rm ADAF}/\dotMEdd  \lesssim 10^{-2}$~\citep{Narayan-Yi1995}}, we infer that there is potentially a wide  parameter space allowing giant flares to occur. Obviously, the required accretion rates $\dotMbefore$ and $\Rtrunc$ depend on the ratio $\alphahot/\alphacold$; they grow as the ratio decreases.

\revtri{
%It should be noted that radius \eqref{R_ADAF} is not obtained for the transition between ADAF and a geometrically thin cold recombined disc. 
It should be noted that the processes underlying the transition between an ADAF and a cold disc are far from being  fully understood. Furthermore, proposals similar to \eqref{R_ADAF} usually presume that $\alpha$ is the value in the ADAF. 
%see https://ui.adsabs.harvard.edu/abs/1997ApJ...482..448N about connecting ADAF and cold disc
However, various estimates for the transition radius fall in a rather broad range qualitatively agreeing with the interval $\lg (\Rtrunc/\rg) \approx 1-3$~\citep[see refs. in][their figure~7]{Czerny+2004,Taam+2012,Yuan-Narayan2014}.}
{The values predicted by the purple `ADAF' line in Fig.~\ref{fig:R_trunc-ADAF}  all give  $\dotMbefore/\dotMEdd \sim  10^{-3}$, and thus can serve as reference values.}

\begin{figure*}
\includegraphics[width=0.49\linewidth]{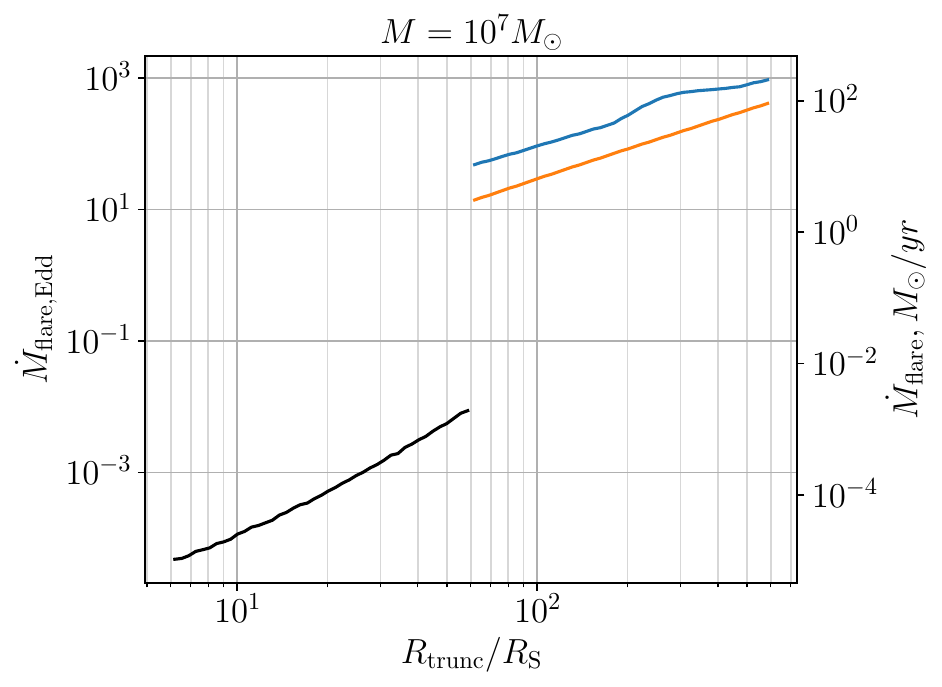}\quad
\includegraphics[width=0.49\linewidth]{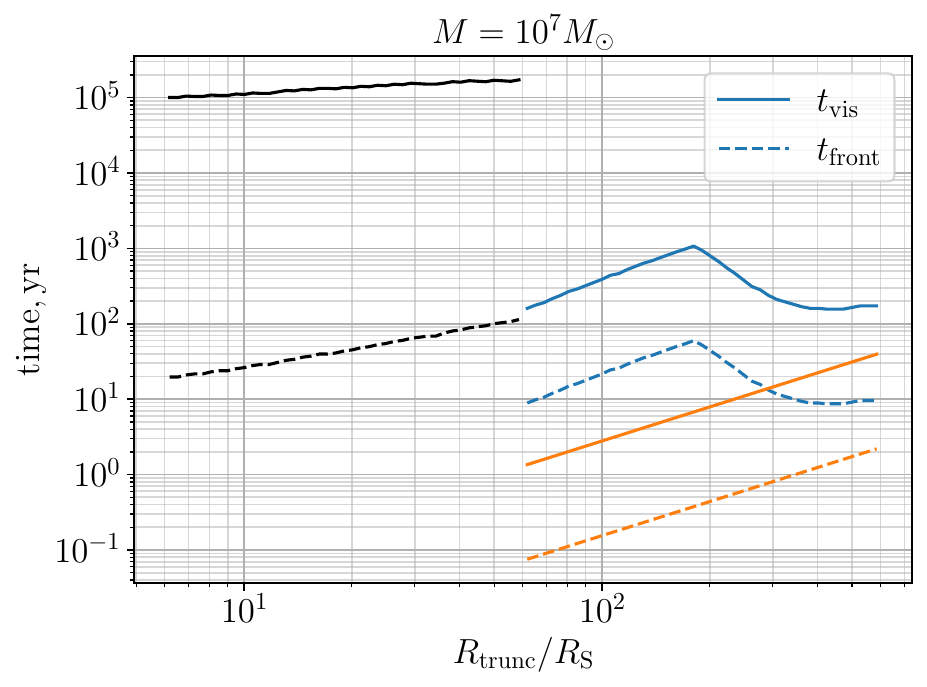}
\includegraphics[width=0.49\linewidth]{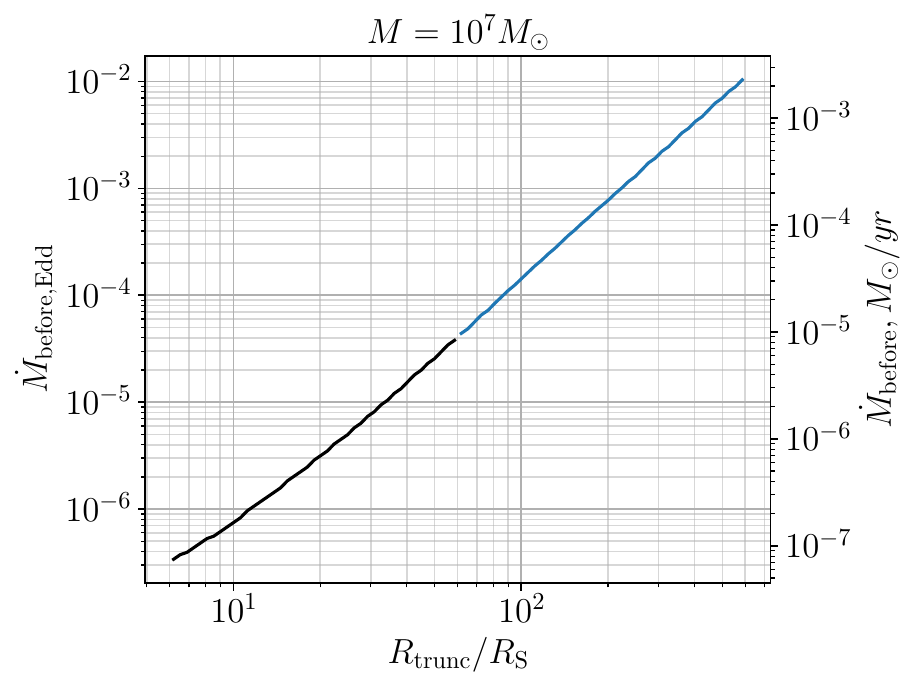}\quad
\includegraphics[width=0.49\linewidth]{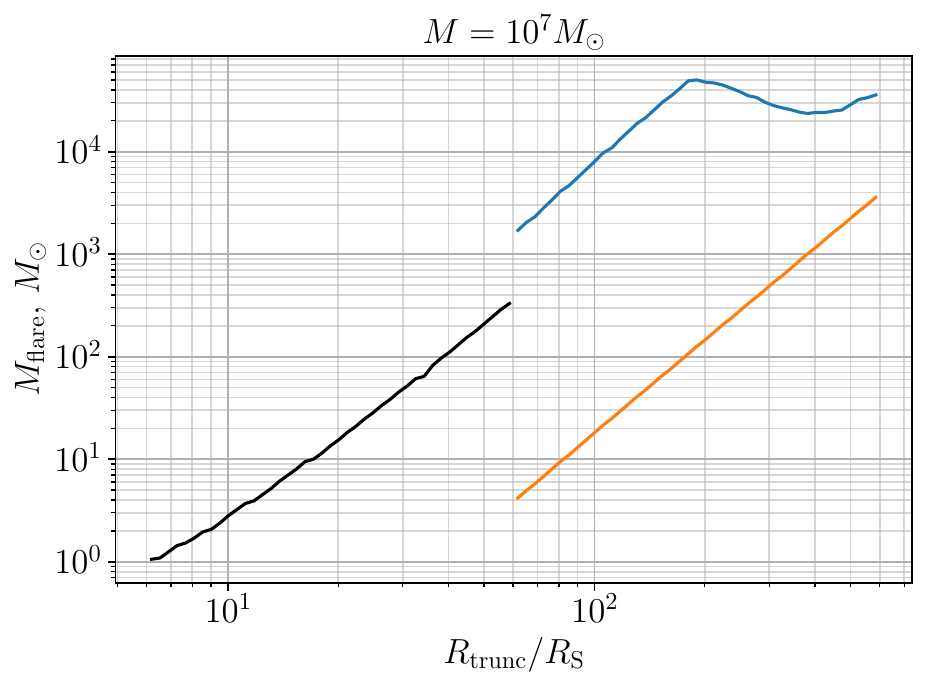}
	% \end{minipage}
	\caption{
    Outburst properties for normal (black) and giant flares (blue and orange) vs  inner truncation radius of  quiescent geometrically thin disc. Peak accretion rate \eqref{eq.dotMpeak} (upper left);  characteristic times, calculated using \eqref{eq.t_front} for the flare evolution before the peak and \eqref{eq.tvis} for the characteristic flare duration (upper right);  inner disc accretion rate before the flare (bottom left);   mass involved in a flare (bottom right).  
 The blue curves show the upper limit, obtained if the disc is heated to the radius $\Rmax$. The orange curves represent the lower limit following assumption $\Rout = \Rtrunc+z_0$ (see Sect.~\ref{s.properties}). 
    }
	\label{fig:Flare_1e7}
\end{figure*}

\section{Energetics and duration  of  giant flares}
\label{s.properties}
Similar to the DIM scenario, 
the ring heated by the instability can trigger a transition wave (a heating wave) that moves outwards.  How far out can such a wave travel? The maximum possible distance $\Rmax$ is determined by the condition that $\Sigma_{\rm stat}$ there (before the flare) should not be less than $\Sigmadv$~(see Fig.~\ref{fig:S-4032}).
This can be difficult to achieve in a `rapidly changing landscape' because, for example, viscous evolution occurs on almost the thermal timescale. 
We suggest that the minimum extent of the heating wave $\Rout-\Rtrunc$ is of the order of $z_0(\Rtrunc)$, assuming isotropic heat diffusion.

At a certain radius in the disc, $\Rsg$, the effects of self-gravity begin to take effect. The distance beyond which the razor-thin disc becomes gravitationally unstable is defined by the following condition~\citep{Safronov1960,Toomre1964}:
\begin{equation}
\frac{\pi G\Sigma}{\omega_{\rm K}^2 z_0}  \geq 1\, . 
\label{eq.Rself-grav}  
\end{equation}
The behaviour of the disc and $\Sigma(R)$ beyond $\Rsg$ depend on the cooling and heating mechanisms there.
% more hotter discs are thicker and more stable
% Numerical simulations show that heating mechanisms, such as  spiral structure (compressional heating and shock dissipation) can self-regulate the disc  with $Q ≈ 1$ over a wide radial range. (Cossins et al) Self-fragmentation is possible, when cooling is fast and Mdisc<<Mbh.
Given the uncertainties in the disc structure, which are beyond the scope of the present work, we chose to present the results only for
radii less than $R_\mathrm{sg}$.
Also, due to numerical difficulties, we have not calculated flare parameters for $\Rtrunc$ very close to $\Rsg$, nor have we considered radii larger than $10^3\,\rg$, as our primary focus is on lower estimates. The radii $\Rtrunc$ that we have included in calculations are indicated by the solid intervals of the curves in Fig.~\ref{fig:R_trunc-ADAF}.

The time before the flare peaks can be estimated as the \hbox{heating} front propagation time~\citep{Meyer1984}:
\begin{equation}
\tfront = t_{\rm th} \left(\frac{z_0}{R_{\rm out}}\right)^{-1} = \frac{1}{\alpha\,\omega_{\rm K}(R_{\rm out})} \left(\frac{z_0}{R_{\rm out}}\right)^{-1}.
\label{eq.t_front}
\end{equation}
\revtri{Alternatively, the time before the peak for a ring viscously spreading from radius $\Rout$ in the classical solution of \citet{lyn-pri1974} is $\tpeak = \tvis/9$, see Appendix~\ref{s.analyt}. Both times are of the same order for the slim disc: $\tpeak \sim \tfront\,\Rout/z_0$ if $\Pi_1 =6$, see Eq.~\eqref{eq.visc_time_scale}.
}

The mass of the disc involved in the flare is as follows:
\begin{equation}
\Mdisc = \int\limits_{R_{\rm trunc}}^{R_{\rm out}} 2\pi\,\Sigma_{\rm stat}\,r\, \mathrm{d}r \, .
\label{eq.Mdisc}
\end{equation}
We neglect here the evolution of the surface density that occur as the  heating front propagates \revtri{and that the quiescent  $\Sigma(r)$ could be somewhat higher than  $\Sigma_\mathrm{stat}$, see footnote~\ref{foot.Sigma}.}

The  characteristic flare decay time is the viscous time at the outer boundary of the disc involved in the flare, $\Rout^2/\nut(\Rout)$, see Eq.~\eqref{eq.tvis}.
Finally, the peak accretion rate can be approximated as the mass divided by the characteristic time, see Eq.~\eqref{eq.Mdotpeak}:
%and \eqref{eq.visc_time_scale}:
\begin{equation}
    \dot{M}_{\rm flare} = \frac{\Mdisc(\Rout)}{\Rout^2/\nut(\Rout)}\, .
    \label{eq.dotMpeak}
\end{equation}
It should be remembered that a substantial fraction of the `heated' mass can be blown away with the outflow, up to \hbox{$(1-\dot{M}_{\rm Edd}/\dot{M}_{\rm flare})$} if one applies the estimate for a super-Eddington disc by~\citep{sha-sun1973}.

 It is evident that the truncation radius $R_{\rm trunc}$ determines $\Sigma_{\rm stat}$ and $\dotMbefore$ before the flare, as well as  the power and duration of the flare itself.

Figure~\ref{fig:Flare_1e7} shows the calculated parameters of the flares for $10^7 M_\odot$: peak accretion rate estimate \eqref{eq.dotMpeak}, front time \eqref{eq.t_front} and viscous time ~\eqref{eq.tvis}, accretion rate in a thin disc before the flare,  and the mass involved in a flare.
\revtri{The left sections of the curves (shown in black)} represent the values calculated for the `normal' sub-Eddington flares (see Sect.~\ref{s.SMBH_DIM_scenario}). \revtri{Figure~\ref{fig:Flares_all_masses} summarises the flare characteristics for different masses of SMBH.}
%Due to the assumed fixed aspect ratio, the {factor} between $\tvis$ and $\tfront$ is $3\Pi_1$ for the slim disc.

For the giant flares, which are faster and more luminous, we plotted two sequences: assuming $\Rout = \Rmax\equiv$ \hbox{$\min(R\,(\Sigmastat=\Sigmadv), \,\Rsg)$}, in blue, and $\Rout = \Rtrunc + z_o(\Rtrunc) = 1.5\, \Rtrunc$, in orange. For `normal' flares, we use $\Rout = \min(R\,(\Sigmastat=\Sigma^+), \,\Rsg)$, cf. Fig.~\ref{fig:S-set}.
The non-monotonic behaviour of the blue curves for $\Rtrunc\gtrsim200\rg$ is explained by the fact that, according to our assumption, the maximum radius that the heating wave can reach is limited by the radius of the gravitational instability $\Rsg$.

Table~\ref{tab:props} is a compilation of the flare characteristics for $\Rout=1.5\,\Rtrunc$:
 \begin{equation}
 \label{eq.flare_estimates}
 \begin{split}
 \dot{M}_{\rm flare}/\dotMEdd &\sim 26\, (\Rout/100\rg)^{1/2}\,  \, (z_0/r/0.5)^2\, \alpha_{0.1}\, \Sigma_5\,,\\
 \tvis &\sim 1.6 \,\mathrm{yr}  \,(\Rout/100\rg)^{3/2}\,(z_0/r/0.5)^{-2}\,M_7 \,\alpha_{0.1}^{-1}\, ,\\
 \Mdisc/M_\odot  &\sim 17 \, (\Rout/100\rg)^2\,  M_7^2 \, \Sigma_5 \,.
 \end{split}    
 \end{equation}
 Here the surface density is normalised to $10^5$~g\,cm$^{-2}$, $\alpha$ to 0.1,  and mass to $10^7 M_\odot$.
These estimates are the most conservative and do not virtually depend on the radial distribution of the surface density in the quiescent disc.

\revtri{The maximum effective temperature of the heated ring at radius $r$ can be roughly estimated as the effective temperature at the uppermost branch of the S-curve, assuming that all heat is radiated. 
 For a stationary disc without advection} the local viscous flux $\sigma T_\mathrm{eff}^4 = 3/(8\pi)\, \dot M G M / r^3$, where $\dot M$ is substituted from Eq.~\eqref{eq.Adv_zone}.  Consequently, 
$T_\mathrm{eff} \approx 1.6 \times 10^5\,\mathrm{K}\,(r/100\rg)^{-5/8}\,  \sqrt{ z_0/r}\,M_7^{-1/4}\,\alpha_{0.1}^{1/4}\,\Sigma_5^{1/4}$ and
 the maximum $T_\mathrm{eff} \sim 10^5$~K, provided the heating front does not advance significantly beyond $\Rtrunc$.
%Due to the advection effect, we find that the effective temperature can be twice as low; for $10^6$ and $10^7 M_\odot$, for $\dot M_\mathrm{flare}=30-40 $ and $10-80\, \dotMed$, respectively, $T_\mathrm{eff} \sim 10^5$ K.
The luminosity of the heated ring, approximated as $\sigma T_\mathrm{eff}^4 \, r^2$, is as follows:
\begin{equation}
  {L_{\rm UV/opt}}   \sim 0.6\,  L_{\rm Edd} \, (r/{100\,\rg})^{-1/2} \, (z_0/r/0.5)^2\, \alpha_{0.1}\,   {\Sigma_5}\,  .
  \label{eq.LUVopt1}   
 \end{equation}
 {This translates into  the range  $\sim 0.3-4\, L_\mathrm{Edd}$ for the cases listed in Table~\ref{tab:props}.
 The corresponding \revtri{central} accretion rate is super-Eddington (see the 1st line in \eqref{eq.flare_estimates}); the following relation can be constructed:
 $ \dot M_{\rm flare}/\dotMEdd \sim 40 \,(\Rtrunc/100\rg)\,\, {L_{\rm UV/opt}}/L_{\rm Edd} $ for $\Rout = 1.5\, \Rtrunc$.}

 In the super-Eddington regime, matter is expected to be blown away from the disc, \revtri{mainly from the radii of order of} the spherisation radius~\citep{sha-sun1973}. The  luminosity $L_{\rm UV/opt}$  would not show a plateau at values \eqref{eq.LUVopt1}, because there is no continuous input of matter from the outer radii beyond $\Rout$ at a sufficient rate \revtri{(an optical plateau is possible, however, at a lower luminosity later, see Sect.~\ref{s.gf-tde})}.  Emission in the harder spectral range, produced by the central part of the disc, can be levelled either because of the outflow regulation mechanism, or because of the photon trapping (advection), or a combination of both.  Similarly to the situation of super-Eddington sources in binary systems~\citep{poutanen_et2007}, an anisotropy of the observed emission from the giant flares is expected during the outflow phase. 
 %\revtwo{\st{X-rays can also be reprocessed and}}
 % Roth+2016  https://ui.adsabs.harvard.edu/abs/2016ApJ...827....3R/abstract

It is important to note that while the annuli up to $\Rout$ become as hot as required by a slim disc solution, further out the cold disc may experience a less intense heating and reach the stable geometrically thin hot-disc branch (see Fig.~\ref{fig:S-4032}). This part of the disc becomes optically bright, $T_{\rm eff} \sim 10^4 $~K, for characteristic $z_0/r\sim 0.005$, and evolves quite slowly (see also Sect.~\ref{s.gf-tde}).
\begin{table}
    \centering
     \caption{Peak accretion rate, characteristic duration, and mass involved
     %Characteristics of 
     in giant flares.}
     \begin{tabular}{ccccc}
    \hline
         $M,\,M_\odot$&  $10^6$&  $10^7$&  $10^8$& $10^9$\\
         \hline\\[-0.9em]
         $\dot M_\mathrm{flare} , \dotMed$&  $>30$&  $>$\revtri{14}&  $>6$&$ >4$\\
         &  45&  90&  200& --\\
         $\tvis$, yr&  $>1$&  $>1$&  $>1$& $>$\revtri{1}\\
         &  2&  \revtri{9}&  40& --\\
         $\Mdisc, M_\odot$&  $>1$&  $>4$& >20& $>120$\\
         &  2&  190&  $2\cdot 10^4$& --\\
         %$L_\mathrm{loc},\, \Ledd$&  $0.18-0.3$&  $0.5-2$&  $1.2-5$& $4-10$\\
         %&  0.2&  1&  4& --\\
         \hline
    \end{tabular}
    \label{tab:props}
    \tablefoot{
 It is assumed that $\Rout = \Rtrunc + z_o(\Rtrunc)$, see Sect.~\ref{s.properties}. For each parameter listed in the first column, we give in the first line the values corresponding to the possible minimum $\Rtrunc$ and in the second line, a value corresponding to the ADAF intersections in Fig.~\ref{fig:R_trunc-ADAF}: $\Rtrunc\approx370$, 220, and 130 $\rg$, for $10^6$, $10^7$, and $10^8\, M_\odot$, respectively. See also Fig.~\ref{fig:Flares_all_masses}.
}
\end{table}

\section{Discussion}\label{s.discussion}

\subsection{Critical assumptions of the model}
\label{s.limits}
The proposed model is currently based on local S-curve analysis.
Numerical simulations show that the trajectories of the parameters of the disc ring during the development of a flare can be quite complicated, even in the case of `normal' flares, see for example, \citet{Lasota2001}, figures 11 and 12, \citet{Sniegowska+2023}, figure 5. 
In order to confirm the viability of giant flares, it is important to conduct numerical modelling.

Another possible difficulty is a difference between $\alphacold$ and $\alphahot$ required by the mechanism. Overall, a change of $\alpha$ fits well within the modern paradigm~\citep[e.g.][]{Martin+2019}. 
However, doubts were expressed that $\alpha$ changes between ionised and cold state in AGN discs~\citep{Menou-Quataert2001} on the grounds that the magnetic Reynolds number does not decrease below the critical value in the cold state, and the MHD turbulence is thus still viable.
%https://www.overleaf.com/project/64ad69ed1d72b004d97b0afa#cite.Taam+2012

\begin{figure}
\includegraphics[width=0.95\columnwidth]{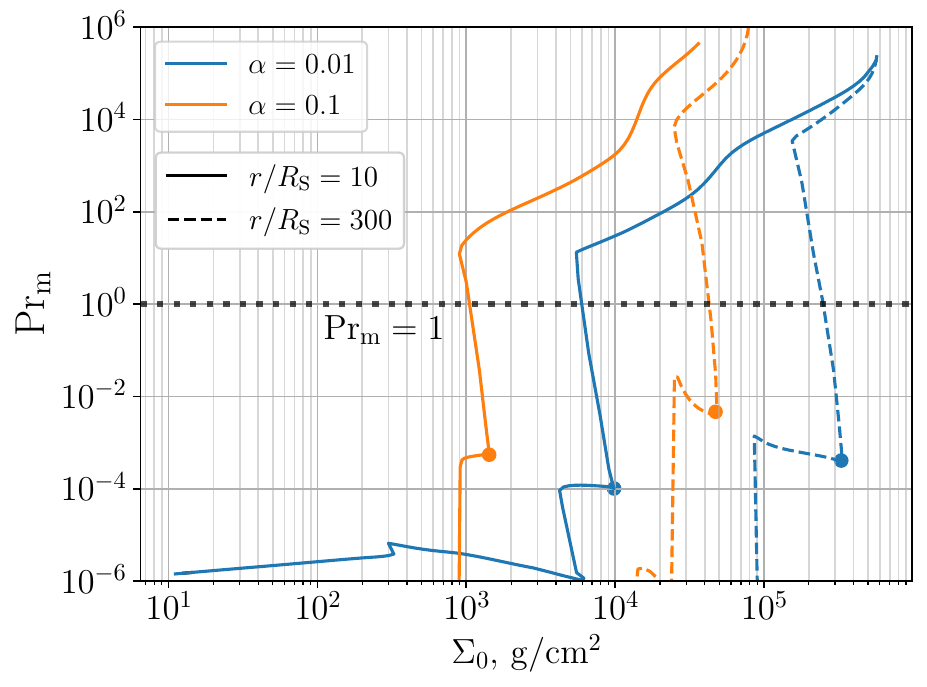}
	\caption{Magnetic Prandtl number calculated for disc around $M=10^7~M_\odot$  for $\alpha=0.1$  and 0.01   and radii 10 and 300 $\rg$. Bullets mark locations of the critical  surface densities $\Sigma^-$.} 
	\label{fig:Prandtl}
 %distribution of mass accretion rates https://ui.adsabs.harvard.edu/abs/2015MNRAS.452..575S/abstract
\end{figure}

We present here the argument in favour of $\alpha$ changing with the temperature  in AGN discs. It was suggested by \citet{balbus-hawley1998} that the magnetic Prandtl number $\prm$ influences the saturated value of the stress tensor for MHD turbulence and affects the value of $\alpha$~\citep[see also][and references therein]{Potter-Balbus2014,Potter-Balbus2017}.  The value $\prm \sim 1$ may be a critical value in this regard.
At the same time, \citet{Potter-Balbus2014} underlie the dominance of radiative viscosity in the central zones of the accretion discs around stellar-mass objects.
%(where the the electron scattering opacity is larger than  bound-free and free-free absorption)
Taking into account the radiative viscosity, we have calculated the magnetic Prandtl number for the AGN disc along the S-curves and  found that $\prm$  passes through the value 1 in the range of parameters corresponding to the lower unstable branch (related to the hydrogen partial ionisation), see Fig.~\ref{fig:Prandtl} and Appendix~\ref{s.Prandtl}.

%{It might be that  $\alpha$ parameter should grows on the same thermal, though this is not certain~\cite[see, e.g.][]{Held-Latter2022} }
%have made unstratified 3D MHD shearing box simulations with effective $\alpha \sim 0.01$  that revealed that the stress oscillations lag behind the pressure by 5 orbits, when the pressure oscillates with a period of several tens of orbits or more. They further suggested that this could imped but not stop the thermal instability in the radiation-dominated regime for large $\alpha$ > 0.1.
%}

\revtri{In non-ideal MHD simulations $\alpha$ was found to change by a factor of \hbox{$\sim$few$\times (1-10)$}  with  the magnetic Prandtl number varying from  $\sim 1$ to $10^2$~\citep{Guilet+2022,Held+2024}.}
%{In simulations $\alpha$ was found to change by $\sim 4-$10 times~\citep{Hirose+2009,Held+2024}. 
\revtri{In this paper, we assume that $\alpha$ varies by a factor of 10.
Reducing the value of the assumed $\alpha$ shift would lead to an increase in the minimum truncation radius, which, in turn, would increase the predicted magnitude of a flare.}
%Less $\alpha$ shift would make the minimum truncation radius larger and will affect their power.
%However, one cannot expect the factor to be a universal value. 

Observationally, assuming a thermal timescale for the underlying variability mechanism,  estimates for $\alphahot$ in the case of AGN have been obtained with a broad variety of results.
%Observational constraints for $\alphahot$  in the case of AGN featuring great variety of results were obtained assuming thermal time-scale of the underlying variations' mechanism. 
By studying the UV/optical spectral variability of 26 AGN~\citet{Siemiginowska-Czerny1989} obtained $\alpha = 0.001-0.1 $ and for 49 AGN~\citet{Starling+2004} obtained $0.01-0.03$,  with $L/\Ledd=0.01-1$.  \citet{Xie+2009} compared observed variability timescales of 31 gamma-ray loud blazars (with $L/\Ledd=0.02-0.2$) with the thermal timescale of $\alpha$-disc models
%temporal distance between optical light curve maxima of 
 and got $\alpha=0.1-0.3$. An analysis by \citet{Kelly+2009} of $R$-band light curves as of a continuous-time stochastic process for 100 quasars provided a lower $\alpha\sim 0.001$. Another study of stochastic variability in the optical and UV domain yielded $\alpha\sim 0.05$
in 67 AGN~\citep{Burke+2021}. 

%https://iopscience.iop.org/article/10.3847/1538-4357/ab1844   van Velzen+2019 alpha>0.1 from TDE models....

\subsection{Giant flares and Tidal Disruption Events (TDE)}\label{s.gf-tde}
TDEs can explain bright X-ray, UV and/or optical events in galactic centres lasting months to years~\citep{Hills1975, Rees1988}.  If a star passes close enough to a SMBH, and if the tidal forces exerted by the massive object are strong enough, the star can be ripped apart.
After returning to the circularisation radius, which is estimated to be twice the tidal radius $R_\mathrm{tidal}=R_\star \, (M/M_\star)^{1/3}$, 
\begin{equation}
    R_{\rm circ} / \Rs = 2 R_{\rm tidal} / \Rs \sim 47 \left(\frac{M}{10^6 M_{\odot}}\right)^{-2/3}\left(\frac{M_\star}{M_{\odot}}\right)^{-1/3}\left(\frac{R_\star}{R_{\odot}}\right)\, , 
    \label{eq.Rcirc}
\end{equation}
some debris is thought to form a viscously evolving disc~\citep{Cannizzo_etal1990,Ulmer1999,Shen2014}. 
A disc of stellar debris, if promptly circularised, heats up and emits in a wide range of wavelengths, from optical to X-ray.  Alternatively, radiation is produced directly by stream collisions~\citep{Piran+2015,Jiang+2016,Ryu+2020}. The sudden increase in brightness is one of the key features used to identify a TDE.

The giant flares involve accretion in the super-Eddington regime, which makes their manifestation similar to that of TDEs~\citep[e.g.][]{Shen2014,Komossa2015,Metzger-Stone2016,Dai+2018}. 
   %{The bolometric luminosity of a giant flare is determined by the disc luminosity, which logarithmically exceeds $\Ledd$, plus the emission from the pressure-supported envelope. Similarly, a cooling envelope is suggested to explain TDE emission~\citep[see][and references therein]{Sarin-Metzger2024}.}  
   %The expected  wind velocity is also of order of the escape velocity $\sim \sqrt{2GM/\Rtrunc} \approx 
   %0.1\,c\,%3\times 10^4 \, ~{km/s}
%(\Rtrunc/100\,\rg)^{-1/2}$. 
As TDEs, giant flares from disc instability can be sources of ultra-high-energy cosmic rays~\citep{Farrar-Gruzinov2009}.

While the origin of the early emission of TDEs is debated, it is believed that at later times radiation has to originate in the accretion disc.  
\citet{VanVelzen+2019} have found that optical light curves of TDE with BH masses $<10^{6.5} M_\odot$ manifest a flattening at late times that can be explained by an evolving disc ~\citep{Mummery-Balbus2020,Mummery+2024}. In giant flares, \revtri{a similar evolution is possible if some disc annuli transfer to the middle stable branch of the S-curve, where the hot `Shakura-Sunyaev model' applies, and $\dot M \sim 10^{-3} - 10^{-1} \dotMed$, see Fig.~\ref{fig:S-curvesM1e7}, the upper panel}. The viscous time of the standard hot disc with $z_0/r<0.01$  is very long:
\begin{equation}
 {\tvis}_\mathrm{,hot} \sim 1600\, {\rm yr}~ (r/100 \rg)^{3/2}\, (z_0/r/0.005)^{-2}\, M_6\,  \alpha_{0.1}^{-1}\,    
\end{equation}
with $\Pi_1=6$ substituted. On this timescale, the effective temperature of such a remnant disc evolves quite slowly until the disc returns to the cold-disc state.

In general, the viscous time of the advection-dominated flow, as shown in Table~\ref{tab:props}, is not less than one year, which contrasts with the typical duration of most TDEs. However, care should be taken when comparing observed times and the viscous time-scales: the $e$-folding time of a flare is about 0.4 times less than $\tvis$, see Eq.~\eqref{eq.texp_an}, and the rise time is about $0.1 \,\tvis$, see Eqs.~\eqref{eq.t_front} and \eqref{eq.tpeak}. It should be noted that longer TDEs sometimes occur~\citep{Kankare+2017,Lin+2017,He+2021,Yan-Xie2018,Zhang2023,Subrayan+2023,Bandopadhyay+2024,Kumar+2024}. At this stage, we do not consider our flare parameters to be restrictive, as they are derived from the simplest estimates of the viscous disc evolution model. The parameters given in Table~\ref{tab:props} are physically the upper limits for actual giant flares, taking into account that (i) accretion discs with a shrinking outer boundary evolve faster than those with a fixed outer boundary, (ii) outflows from radii comparable to the outer disc radius accelerate the evolution\footnote{For $10^7 M_\odot$, at $100 \, \rg$ the slim advective disc cannot exist at $\lesssim 10 \dotMed$ (see Fig.~\ref{fig:S-4032}). This shows that the zone of $z/r\sim 1$, which is the `short-viscous-time zone', is shrinking. In this sense, the solution for an expanding disc from \citet{Cannizzo_etal1990} is not applicable.}.

Estimates of the accretion rates in the TDE disc can be obtained from the spectral fitting of X-rays~\citep{Wen+2020,Wen+2021}. Using a spectral model for the relativistic slim disc, the SMBH mass and spin can be restricted, as well as the accretion rate entering the hole, which is sometimes super-Eddington. In three cases, it was inferred that a subsolar mass has passed through a disc.  A way to estimate the mass rate in a possible super-Eddington outflow from X-ray spectral observations was also suggested ~\citep{Xiang+2024}.

\begin{figure}
	\center{\includegraphics[width=1.0\linewidth]{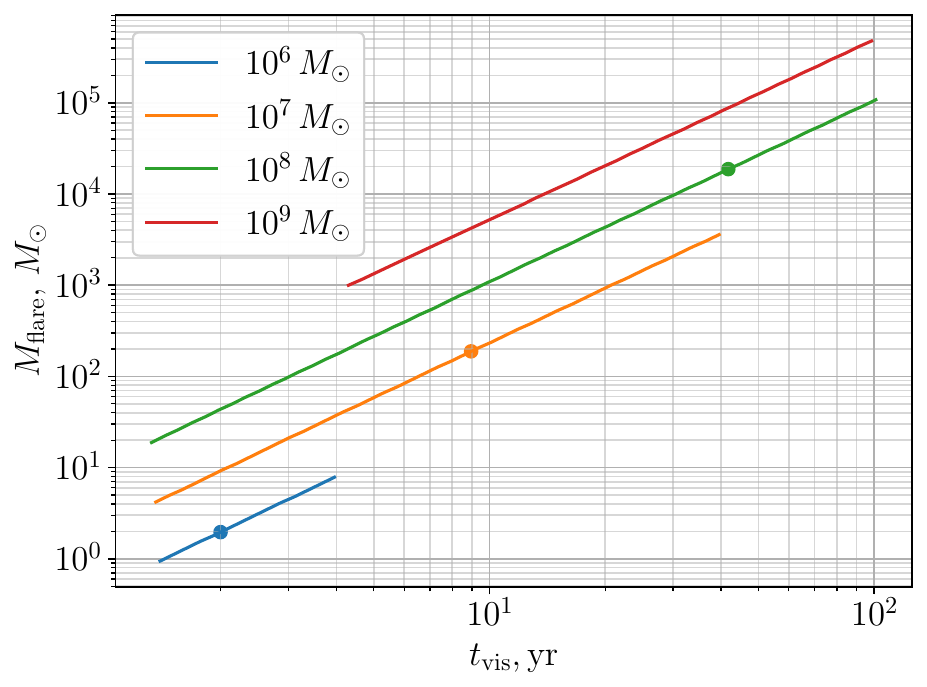}}
	\caption{Disc mass \eqref{eq.Mdisc} involved in  giant flare vs viscosity time \eqref{eq.tvis}.  The dots mark the values corresponding to the ADAF truncation radius~\eqref{R_ADAF}, shown in Fig.~\ref{fig:R_trunc-ADAF}. 
 }
 \label{fig:Mdisc_tvis}
\end{figure}

Spectral line shapes of TDEs are used to extract the geometry and kinematics of the photoionised gas. In some works, the interpretation based on the expanding spherical outflow is preferred~\citep{Hung+2019,Charalampopoulos+2024}. \citet{Kara+2018} found evidence for an ionised outflow with a velocity of $0.2 c$ in ASASSN-14li at early times. Using the ionisation parameter $\xi  = L_{\rm x} /(n\, R^2) \sim 10^3$, they obtained the upper limit on the outflow rate $\dot M = 4\pi\, m_\mathrm{H} \,  v_\mathrm{out}\, L_{\rm x} /\xi \sim 10^3~M_\odot$yr$^{-1}$. 
 \citet{Matsumoto-Piran2021} applied the ‘reprocessing-outflow’ model of the optically thick quasi-spherical wind and, for a sample of observed TDEs, they concluded that the outflows were too massive, about an order of magnitude higher than the mass available in the TDE scenario ($1 M_\odot$).
The giant flare scenario could help to resolve this issue. The masses involved in giant flares are not less than one solar, typically at least $\sim 10 \,M_\odot$, see Fig.~\ref{fig:Mdisc_tvis}, which shows the dependence between the mass and viscous time-scale, \revtri{obtained within the conservative estimate for the size of the heated zone of the disc, $\Rout = \Rtrunc + z_o(\Rtrunc)$. }

The emission from a giant flare is anisotropic due to outflows.  However, this fact cannot be used to distinguish them, since anisotropy is an essential part of the `unification scheme' for TDEs as well~\citep{Dai+2018, Thomsen+2022, Parkinson+2025}.

\begin{figure}
\includegraphics[width=1.0\columnwidth]{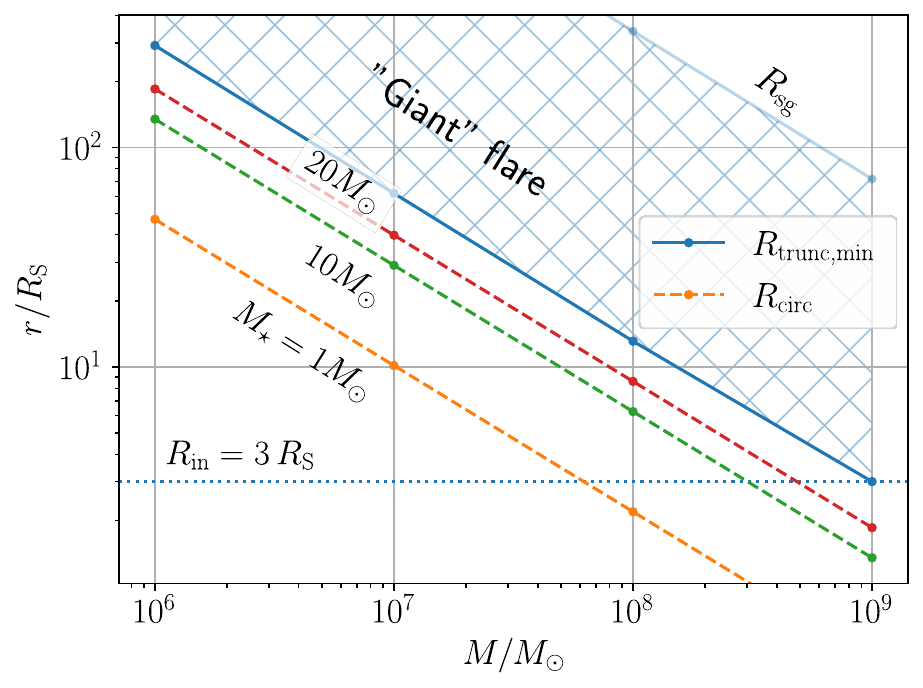}
	\caption{{Truncation radii of thin discs and SMBH masses allowing giant flares (hatched area).}  The upper boundary corresponds to self-gravitation instability radius $R_\mathrm{sg}$.
Also shown are the circularisation radii $R_{\rm circ}$ for stars of $1$, $10$, and $20\,M_\odot$. }
	\label{fig:R_circ-R_trunc}
\end{figure}
Sometimes, observations strongly suggest that a TDE event  preceded the outburst. 
For example, spectroscopic observations of  AT\,2020zso indicate the elliptical geometry of the emitter~\citep{Wevers+2022}.
Further,  event AT 2019azh shows signatures of both the stream-stream collision and delayed accretion~\citep{Liu+2022}.  Linearly polarised optical emission at the level of $\sim 25$ per cent observed from AT\,2020mot possibly indicate accretion disc formation, when `the stream of stellar material collides with itself'~\citep{Liodakis+2023}. \revtri{The alignment  of the jet or outflow direction with the perpendicular to the galactic dust plane ~\citep{Uno+2025} favours the giant flare scenario and argues for the TDE otherwise~\citep{Mattila+2018}.}

Figure~\ref{fig:R_circ-R_trunc} depicts the minimum truncation radius necessary for the ionisation instability to trigger a giant flare versus the SMBH mass. \revtri{The required truncation radius decreases with mass; in particular,} for SMBH with $M\gtrsim 10^9\,M_\odot$, any outburst in a disc with $\alphahot=0.1$ and $\alphacold=0.01$ will be a giant flare, provided that the geometrically thin accretion disc exists at the accretion rates $\dot M$ lower than $10^{-5} \dotMEdd$. 
We also plotted the circularisation radius $R_{\rm circ}$ for stars of different masses, where dependence $R_\star(M_\star)$ for the main sequence stars was used~\citep{Kippenhahn2012}. 
It can be seen that the tidal disruption of a sufficiently massive star can trigger a giant flare \revtri{by adding material to the cold disc.}
For example, in the simulations by \citet{Andalman+2022}, the fallback following the complete destruction of a $1\,M_\odot$ star near a $10^6\,M_\odot$ SMBH results in a disc up to $200\,\rg$.

\subsection{Giant flares rates}\label{s.dew}
The process of supplying mass to the central galactic discs during quiescent periods remains unclear~\citep{Lodato2012,Alexander-Hickox2012}. At distances of $\sim 0.01$~pc in the disc, the self-gravity effects such as spiral waves, fragmentation, and subsequent star formation are possible. 
The influence of these effects on the mass accumulation in the central cold disc is quite complex, as they can both nourish and hinder the feeding of SMBHs~\citep{Nayakshin+2007,Hopkins-Quataert2010}. Other possible ways of feeding the central SMBH have also been proposed, for example, by the very low angular momentum gas~\citep{King-Pringle2006}.

Candidates for galaxies capable of producing giant flares include low-luminosity AGN (LLAGN)~\citep{Yuan-Narayan2014} and other quiescent SMBHs~\citep{Soria+2006,Volonteri+2011} if a geometrically thin accretion disc manages to form at $R\gtrsim 10-10^2~\rg$.
%Yuan+2002 https://ui.adsabs.harvard.edu/abs/2002A%26A...391..139Y/abstract
For very massive SMBHs, giant flares may be impossible due to the absence of a cold disc, because the radius of the gravitational instability $R_\mathrm{sg}/\rg$ decreases with increasing mass. 
%For $10^9\,M_\odot$, the radius $\Rsg$ of the self-gravity  instability is less than the assumed ADAF boundary  (see Fig.~\ref{fig:R_trunc-ADAF}).

For low-mass stellar X-ray binaries, it is common to assume that a binary is transient, that is, it experiences outbursts, if the transfer rate from a neighbouring star through the Lagrangian point is between $\dot M^-$ and $\dot M^+$. The ratio between these values is only $\sim 2.5$ in Fig.~\ref{fig:S-set}. Does this mean that only galaxies with such a rate of mass supply can have giant flares? Obviously not, since galaxies in the process of increasing their central accretion rate from very low levels could also be susceptible to giant flares.
Assuming that there is a viscous mass supply from a distant reservoir (e.g. a \revtri{torus} of matter) to the cold disc, the upper limit on the time between giant flares is the Lynden-Bell-Pringle peak time $t_\mathrm{peak} \approx 0.1 r^2/\nut$~(see Appendix~\ref{s.analyt}) which gives $\sim 4\times 10^5 \,\mathrm{yr} \, (r/100\,\rg)^{3/2}\, (M/10^7 M_\odot)$, where we have substituted $ z_0/r = 10^{-3} $, $\alphacold = 10^{-2}$, and $\Pi_1=6$. 
We note on the one hand the strong dependence on the reservoir distance from the centre. On the other hand, this time is an upper limit since the critical accretion rate need not be the maximum value attained at the peak, and the time required may be an order of magnitude less. Therefore, we cannot consider the above estimate to be very informative. 

%The radius $R$ corresponds to the extent of the mass reservoir, which appears to be larger than $\Rtrunc$, but possibly only by a factor of a few to 10. If a giant flare does not reach a certain radius of the disc, this means that the mass reservoir there is intact. Consequently, the lower the estimates of the flare amplitude, the shorter the replenishment times.

The situation becomes further complicated when one considers the multiplicity of potential scenarios for fuelling the central discs. For example, events of tidal stripping of stars \citep[`partial TDE', see][]{Chen_etal2021} can significantly reduce the time between giant flares.  

The frequency of `complete' TDE events is currently estimated to be on the order of one per $10^4-10^5$ yr~\citep{Velzen+2020,Saxton+2021,Sazonov_etal2021,Masterson+2024}. Whether giant flares may contribute to events currently attributed solely to TDEs requires further investigation.

%Given that the rate of `complete' TDEs is estimated to be 1~event per $10^4-10^5$ yr, see ~\citet{Velzen+2020,Saxton+2021} and references therein, \cite{Sazonov_etal2021,Masterson+2024}, we suggest that giant flares, if the mechanism is confirmed, could contribute to the observed TDE rate.
%Based on the different TDE samples, the TDE rates are typically around 10−5 –10−4 yr−1 per galaxy (e.g. Magorrian & Tremaine 1999; Wang & Merritt 2004; Arcavi et al. 2014; French et al. 2016; van Velzen et al. 2018; Stone et al. 2020; Yao et al. 2023

%The narrow-line and lineless AGN  are possible candidates for galaxies that can produce giant flares~\citep{Trump+2011}.

% EROSITA detection rates https://ui.adsabs.harvard.edu/abs/2020ApJ...889..166J/abstract

\section{Summary}
\label{s.summary}

We propose a novel mechanism for outbursts associated with the ionisation-viscous instability in the accretion disc. According to our proposal, a massive, cold, recombined, geometrically thin disc with vertical convective energy transfer can transition to a slim, advection-dominated regime within a few thermal timescales, provided that the critical surface density is accumulated. The mechanism is only viable in discs around SMBHs and if the turbulent parameter $\alpha$ {is higher in the ionised state than in the recombined state.}

For example, if $\alphahot=0.1$ and $\alphacold=0.01$, and the SMBH mass is $\gtrsim 10^9\, M_\odot$, the giant flare can develop if the geometrically thin accretion disc exists at accretion rates $\dot M/\dotMEdd$ less than $10^{-5}$. Also, any flare due to the ionisation instability in a geometrically thin convective recombined disc around a SMBH with $M\gtrsim 10^9\, M_\odot$ will be a giant flare.

For a lighter SMBH, a giant outburst can develop if there is a geometrically thin accretion disc and its inner part is truncated by an optically thin, tenuous flow. An ADAF is a compelling candidate for such a scenario. The region near the truncation radius, when heated after the instability has developed, produces a luminosity of \revtri{order of} $\Ledd$.

Estimated peak accretion rates depend on the size of the disc region transitioning to the ionised state and are $>10\,\dotMEdd$, but may be somewhat reduced by strong outflows.  The characteristic outburst duration is estimated by the viscous time of a slim disc, which also depends on the size of the region involved in a flare and can range from several months to several years. \revtri{After a giant flare, the relic is likely to be a slowly evolving geometrically thin hot disc.}

\revtri{The analysis presented here is not intended to provide precise quantitative estimates.} The possibility of giant flare development needs to be further investigated using numerical models that can account for non-equilibrium processes in accretion discs.

If confirmed, giant flares could contribute to a variety of bright extragalactic phenomena, such as TDE.

\begin{acknowledgements} 

We thank Arman Tursunov, Loren Held, and Steven Shore for the comments. We are grateful to the referee for many helpful suggestions, and to the language editor for corrections. Support was provided by Schmidt Sciences, LLC. for KM. We acknowledge the usage of computing resources of the M.~V.~Lomonosov Moscow State University Program of De\-ve\-lop\-ment and the Sternberg Astronomical Institute where this work has been started. AT was supported by the \textit{Deut\-sche For\-schungs\-ge\-mein\-schaft\/} under grant WE1312/56--1.

\end{acknowledgements}
\bibliographystyle{aa}
\bibliography{references} % if your bibtex file is called example.bib

\begin{thebibliography}{143}
\expandafter\ifx\csname natexlab\endcsname\relax\def\natexlab#1{#1}\fi

\bibitem[{{Abramowicz} {et~al.}(1995){Abramowicz}, {Chen}, {Kato}, {Lasota}, \&
  {Regev}}]{Abramowicz+1995}
{Abramowicz}, M.~A., {Chen}, X., {Kato}, S., {Lasota}, J.-P., \& {Regev}, O.
  1995, \apjl, 438, L37

\bibitem[{{Abramowicz} {et~al.}(1988){Abramowicz}, {Czerny}, {Lasota}, \&
  {Szuszkiewicz}}]{Abramowicz+1988}
{Abramowicz}, M.~A., {Czerny}, B., {Lasota}, J.~P., \& {Szuszkiewicz}, E. 1988,
  \apj, 332, 646

\bibitem[{{Agol} \& {Krolik}(1998)}]{Agol-Krolik1998}
{Agol}, E. \& {Krolik}, J. 1998, \apj, 507, 304

\bibitem[{{Alexander} \& {Hickox}(2012)}]{Alexander-Hickox2012}
{Alexander}, D.~M. \& {Hickox}, R.~C. 2012, \nar, 56, 93

\bibitem[{{Andalman} {et~al.}(2022){Andalman}, {Liska}, {Tchekhovskoy},
  {Coughlin}, \& {Stone}}]{Andalman+2022}
{Andalman}, Z.~L., {Liska}, M. T.~P., {Tchekhovskoy}, A., {Coughlin}, E.~R., \&
  {Stone}, N. 2022, \mnras, 510, 1627

\bibitem[{{Balbus} \& {Hawley}(1998)}]{balbus-hawley1998}
{Balbus}, S.~A. \& {Hawley}, J.~F. 1998, Reviews of Modern Physics, 70, 1

\bibitem[{{Balbus} \& {Henri}(2008)}]{Balbus-Henri2008}
{Balbus}, S.~A. \& {Henri}, P. 2008, \apj, 674, 408

\bibitem[{{Bandopadhyay} {et~al.}(2024){Bandopadhyay}, {Fancher}, {Athian},
  {Indelicato}, {Kapalanga}, {Kumah}, {Paradiso}, {Todd}, {Coughlin}, \&
  {Nixon}}]{Bandopadhyay+2024}
{Bandopadhyay}, A., {Fancher}, J., {Athian}, A., {et~al.} 2024, \apjl, 961, L2

\bibitem[{{Belloni} {et~al.}(1997){Belloni}, {M{\'e}ndez}, {King}, {van der
  Klis}, \& {van Paradijs}}]{Belloni+1997}
{Belloni}, T., {M{\'e}ndez}, M., {King}, A.~R., {van der Klis}, M., \& {van
  Paradijs}, J. 1997, \apjl, 479, L145

\bibitem[{{Bjoernsson} {et~al.}(1996){Bjoernsson}, {Abramowicz}, {Chen}, \&
  {Lasota}}]{Bjoernsson+1996}
{Bjoernsson}, G., {Abramowicz}, M.~A., {Chen}, X., \& {Lasota}, J.-P. 1996,
  \apj, 467, 99

\bibitem[{{Blaes}(2014)}]{Blaes2014}
{Blaes}, O. 2014, \ssr, 183, 21

\bibitem[{{Burke} {et~al.}(2021){Burke}, {Shen}, {Blaes}, {Gammie}, {Horne},
  {Jiang}, {Liu}, {McHardy}, {Morgan}, {Scaringi}, \& {Yang}}]{Burke+2021}
{Burke}, C.~J., {Shen}, Y., {Blaes}, O., {et~al.} 2021, Science, 373, 789

\bibitem[{{Cannizzo}(1996)}]{Cannizzo1996}
{Cannizzo}, J.~K. 1996, \apjl, 466, L31

\bibitem[{{Cannizzo} {et~al.}(1990){Cannizzo}, {Lee}, \&
  {Goodman}}]{Cannizzo_etal1990}
{Cannizzo}, J.~K., {Lee}, H.~M., \& {Goodman}, J. 1990, \apj, 351, 38

\bibitem[{{Cannizzo} \& {Reiff}(1992)}]{Cannizzo-Reiff1992}
{Cannizzo}, J.~K. \& {Reiff}, C.~M. 1992, \apj, 385, 87

\bibitem[{{Charalampopoulos} {et~al.}(2024){Charalampopoulos}, {Kotak},
  {Wevers}, {Leloudas}, {Kravtsov}, {Pursiainen}, {Ramsden}, {Reynolds},
  {Aamer}, {Anderson}, {Arcavi}, {Cai}, {Chen}, {Dennefeld}, {Galbany},
  {Gromadzki}, {Guti{\'e}rrez}, {Ihanec}, {Kangas}, {Kankare}, {Kool},
  {Lawrence}, {Lundqvist}, {Makrygianni}, {Mattila}, {M{\"u}ller-Bravo},
  {Nicholl}, {Onori}, {Sahu}, {Smartt}, {Sollerman}, {Wang}, \&
  {Young}}]{Charalampopoulos+2024}
{Charalampopoulos}, P., {Kotak}, R., {Wevers}, T., {et~al.} 2024, \aap, 689,
  A350

\bibitem[{{Chashkina} {et~al.}(2019){Chashkina}, {Lipunova}, {Abolmasov}, \&
  {Poutanen}}]{Chashkina+2019}
{Chashkina}, A., {Lipunova}, G., {Abolmasov}, P., \& {Poutanen}, J. 2019, \aap,
  626, A18

\bibitem[{{Chen} \& {Shen}(2021)}]{Chen_etal2021}
{Chen}, J.-H. \& {Shen}, R.-F. 2021, \apj, 914, 69

\bibitem[{{Clou{\"e}t} \& {Soulard}(2021)}]{Clouet-Soulard2021}
{Clou{\"e}t}, J.-F. \& {Soulard}, O. 2021, \apj, 919, 78

\bibitem[{{Czerny}(2019)}]{Czerny+2019}
{Czerny}, B. 2019, Universe, 5, 131

\bibitem[{{Czerny} {et~al.}(2004){Czerny}, {R{\'o}za{\'n}ska}, \&
  {Kuraszkiewicz}}]{Czerny+2004}
{Czerny}, B., {R{\'o}za{\'n}ska}, A., \& {Kuraszkiewicz}, J. 2004, \aap, 428,
  39

\bibitem[{{Dai} {et~al.}(2018){Dai}, {McKinney}, {Roth}, {Ramirez-Ruiz}, \&
  {Miller}}]{Dai+2018}
{Dai}, L., {McKinney}, J.~C., {Roth}, N., {Ramirez-Ruiz}, E., \& {Miller},
  M.~C. 2018, \apjl, 859, L20

\bibitem[{{Dubus} {et~al.}(2001){Dubus}, {Hameury}, \& {Lasota}}]{Dubus+2001}
{Dubus}, G., {Hameury}, J.~M., \& {Lasota}, J.~P. 2001, \aap, 373, 251

\bibitem[{{Esin} {et~al.}(1997){Esin}, {McClintock}, \& {Narayan}}]{Esin+1997}
{Esin}, A.~A., {McClintock}, J.~E., \& {Narayan}, R. 1997, \apj, 489, 865

\bibitem[{{Farrar} \& {Gruzinov}(2009)}]{Farrar-Gruzinov2009}
{Farrar}, G.~R. \& {Gruzinov}, A. 2009, \apj, 693, 329

\bibitem[{{Faulkner} {et~al.}(1983){Faulkner}, {Lin}, \&
  {Papaloizou}}]{Faulkner_etal1983_1}
{Faulkner}, J., {Lin}, D.~N.~C., \& {Papaloizou}, J. 1983, \mnras, 205, 359

\bibitem[{{Guilet} {et~al.}(2022){Guilet}, {Reboul-Salze}, {Raynaud}, {Bugli},
  \& {Gallet}}]{Guilet+2022}
{Guilet}, J., {Reboul-Salze}, A., {Raynaud}, R., {Bugli}, M., \& {Gallet}, B.
  2022, \mnras, 516, 4346

\bibitem[{{Hameury}(2020)}]{Hameury2020_review}
{Hameury}, J.~M. 2020, Advances in Space Research, 66, 1004

\bibitem[{{Hameury} {et~al.}(2007){Hameury}, {Lasota}, \&
  {Viallet}}]{Hameury+2007}
{Hameury}, J.-M., {Lasota}, J.-P., \& {Viallet}, M. 2007, in Black Holes from
  Stars to Galaxies -- Across the Range of Masses, ed. V.~{Karas} \& G.~{Matt},
  Vol. 238, 297--300

\bibitem[{{Hameury} {et~al.}(1998){Hameury}, {Menou}, {Dubus},
  {et~al.}}]{Hameury1998}
{Hameury}, J.-M., {Menou}, K., {Dubus}, G., {et~al.} 1998, \mnras, 298, 1048

\bibitem[{{Hameury} {et~al.}(2009){Hameury}, {Viallet}, \&
  {Lasota}}]{Hameury+2009}
{Hameury}, J.~M., {Viallet}, M., \& {Lasota}, J.~P. 2009, \aap, 496, 413

\bibitem[{{He} {et~al.}(2021){He}, {Dou}, {Ai}, {Shu}, {Jiang}, {Wang},
  {Zhang}, \& {Shen}}]{He+2021}
{He}, J.~S., {Dou}, L.~M., {Ai}, Y.~L., {et~al.} 2021, \aap, 652, A15

\bibitem[{{Held} {et~al.}(2024){Held}, {Mamatsashvili}, \&
  {Pessah}}]{Held+2024}
{Held}, L.~E., {Mamatsashvili}, G., \& {Pessah}, M.~E. 2024, \mnras, 530, 2232

\bibitem[{{Hills}(1975)}]{Hills1975}
{Hills}, J.~G. 1975, \nat, 254, 295

\bibitem[{{Hopkins} \& {Quataert}(2010)}]{Hopkins-Quataert2010}
{Hopkins}, P.~F. \& {Quataert}, E. 2010, \mnras, 407, 1529

\bibitem[{Huba {et~al.}(2013)Huba, of~Naval~Research, \& (U.S.)}]{NRLPF2013}
Huba, J., of~Naval~Research, U. S.~O., \& (U.S.), N. R.~L. 2013, 2013 NRL
  Plasma Formulary (Naval Research Laboratory)

\bibitem[{{Hung} {et~al.}(2019){Hung}, {Cenko}, {Roth}, {Gezari}, {Veilleux},
  {van Velzen}, {Gaskell}, {Foley}, {Blagorodnova}, {Yan}, {Graham}, {Brown},
  {Siebert}, {Frederick}, {Ward}, {Gatkine}, {Gal-Yam}, {Yang}, {Schulze},
  {Dimitriadis}, {Kupfer}, {Shupe}, {Rusholme}, {Masci}, {Riddle}, {Soumagnac},
  {van Roestel}, \& {Dekany}}]{Hung+2019}
{Hung}, T., {Cenko}, S.~B., {Roth}, N., {et~al.} 2019, \apj, 879, 119

\bibitem[{{Janiuk} \& {Czerny}(2011)}]{Janiuk-Czerny2011}
{Janiuk}, A. \& {Czerny}, B. 2011, \mnras, 414, 2186

\bibitem[{{Janiuk} {et~al.}(2000){Janiuk}, {Czerny}, \&
  {Siemiginowska}}]{Janiuk+2000}
{Janiuk}, A., {Czerny}, B., \& {Siemiginowska}, A. 2000, \apjl, 542, L33

\bibitem[{{Janiuk} {et~al.}(2004){Janiuk}, {Czerny}, {Siemiginowska}, \&
  {Szczerba}}]{Janiuk+2004}
{Janiuk}, A., {Czerny}, B., {Siemiginowska}, A., \& {Szczerba}, R. 2004, \apj,
  602, 595

\bibitem[{{Jiang} {et~al.}(2016){Jiang}, {Guillochon}, \& {Loeb}}]{Jiang+2016}
{Jiang}, Y.-F., {Guillochon}, J., \& {Loeb}, A. 2016, \apj, 830, 125

\bibitem[{{Jiang} {et~al.}(2013){Jiang}, {Stone}, \& {Davis}}]{Jiang+2013}
{Jiang}, Y.-F., {Stone}, J.~M., \& {Davis}, S.~W. 2013, \apj, 767, 148

\bibitem[{{Kankare} {et~al.}(2017){Kankare}, {Kotak}, {Mattila}, {Lundqvist},
  {Ward}, {Fraser}, {Lawrence}, {Smartt}, {Meikle}, {Bruce}, {Harmanen},
  {Hutton}, {Inserra}, {Kangas}, {Pastorello}, {Reynolds},
  {Romero-Ca{\~n}izales}, {Smith}, {Valenti}, {Chambers}, {Hodapp}, {Huber},
  {Kaiser}, {Kudritzki}, {Magnier}, {Tonry}, {Wainscoat}, \&
  {Waters}}]{Kankare+2017}
{Kankare}, E., {Kotak}, R., {Mattila}, S., {et~al.} 2017, Nature Astronomy, 1,
  865

\bibitem[{{Kara} {et~al.}(2018){Kara}, {Dai}, {Reynolds}, \&
  {Kallman}}]{Kara+2018}
{Kara}, E., {Dai}, L., {Reynolds}, C.~S., \& {Kallman}, T. 2018, \mnras, 474,
  3593

\bibitem[{{Kato} {et~al.}(2008){Kato}, {Fukue}, \& {Mineshige}}]{KFM2008}
{Kato}, S., {Fukue}, J., \& {Mineshige}, S. 2008, {Black-Hole Accretion Disks
  --- Towards a New Paradigm ---}

\bibitem[{{Kelly} {et~al.}(2009){Kelly}, {Bechtold}, \&
  {Siemiginowska}}]{Kelly+2009}
{Kelly}, B.~C., {Bechtold}, J., \& {Siemiginowska}, A. 2009, \apj, 698, 895

\bibitem[{{Ketsaris} \& {Shakura}(1998)}]{KetsarisShakura1998}
{Ketsaris}, N.~A. \& {Shakura}, N.~I. 1998, Astronomical and Astrophysical
  Transactions, 15, 193

\bibitem[{{King} \& {Pringle}(2006)}]{King-Pringle2006}
{King}, A.~R. \& {Pringle}, J.~E. 2006, \mnras, 373, L90

\bibitem[{{Kippenhahn} {et~al.}(2012){Kippenhahn}, {Weigert}, \&
  {Weiss}}]{Kippenhahn2012}
{Kippenhahn}, R., {Weigert}, A., \& {Weiss}, A. 2012, {Stellar Structure and
  Evolution}

\bibitem[{{Komossa}(2015)}]{Komossa2015}
{Komossa}, S. 2015, Journal of High Energy Astrophysics, 7, 148

\bibitem[{{Kumar} {et~al.}(2024){Kumar}, {Berger}, {Hiramatsu}, {Gomez},
  {Blanchard}, {Cendes}, {Bostroem}, {Farah}, {Padilla Gonzalez}, {Howell},
  {McCully}, {Newsome}, \& {Terreran}}]{Kumar+2024}
{Kumar}, H., {Berger}, E., {Hiramatsu}, D., {et~al.} 2024, \apjl, 974, L36

\bibitem[{{Landry} {et~al.}(2013){Landry}, {Dodson-Robinson}, {Turner}, \&
  {Abram}}]{Landry+2013}
{Landry}, R., {Dodson-Robinson}, S.~E., {Turner}, N.~J., \& {Abram}, G. 2013,
  \apj, 771, 80

\bibitem[{{Lasota}(2001)}]{Lasota2001}
{Lasota}, J.-P. 2001, \nar, 45, 449

\bibitem[{{Lasota}(2016)}]{Lasota2016}
{Lasota}, J.-P. 2016, in Astrophysics and Space Science Library, Vol. 440,
  Astrophysics of Black Holes: From Fundamental Aspects to Latest Developments,
  ed. C.~{Bambi}, 1

\bibitem[{{Lasota} \& {Pelat}(1991)}]{Lasota-Pelat1991}
{Lasota}, J.~P. \& {Pelat}, D. 1991, \aap, 249, 574

\bibitem[{{Lasota} {et~al.}(2016){Lasota}, {Vieira}, {Sadowski}, {Narayan}, \&
  {Abramowicz}}]{Lasota+2016}
{Lasota}, J.~P., {Vieira}, R.~S.~S., {Sadowski}, A., {Narayan}, R., \&
  {Abramowicz}, M.~A. 2016, \aap, 587, A13

\bibitem[{{Lightman} \& {Eardley}(1974)}]{Lightman_etal1974}
{Lightman}, A.~P. \& {Eardley}, D.~M. 1974, \apjl, 187, L1

\bibitem[{{Lin} {et~al.}(2017){Lin}, {Guillochon}, {Komossa}, {Ramirez-Ruiz},
  {Irwin}, {Maksym}, {Grupe}, {Godet}, {Webb}, {Barret}, {Zauderer}, {Duc},
  {Carrasco}, \& {Gwyn}}]{Lin+2017}
{Lin}, D., {Guillochon}, J., {Komossa}, S., {et~al.} 2017, Nature Astronomy, 1,
  0033

\bibitem[{{Liodakis} {et~al.}(2023){Liodakis}, {Koljonen}, {Blinov},
  {Lindfors}, {Alexander}, {Hovatta}, {Berton}, {Hajela}, {Jormanainen},
  {Kouroumpatzakis}, {Mandarakas}, \& {Nilsson}}]{Liodakis+2023}
{Liodakis}, I., {Koljonen}, K.~I.~I., {Blinov}, D., {et~al.} 2023, Science,
  380, 656

\bibitem[{{Lipunova}(1999)}]{Lipunova1999}
{Lipunova}, G.~V. 1999, Astronomy Letters, 25, 508

\bibitem[{{Lipunova}(2015)}]{lipunova2015}
{Lipunova}, G.~V. 2015, \apj, 804, 87

\bibitem[{{Liu} {et~al.}(2022){Liu}, {Dou}, {Chen}, \& {Shen}}]{Liu+2022}
{Liu}, X.-L., {Dou}, L.-M., {Chen}, J.-H., \& {Shen}, R.-F. 2022, \apj, 925, 67

\bibitem[{{Lodato}(2012)}]{Lodato2012}
{Lodato}, G. 2012, Advances in Astronomy, 2012, 846875

\bibitem[{{L{\'o}pez} {et~al.}(2024){L{\'o}pez}, {Yang}, {Mountrichas},
  {Brusa}, {Alexander}, {Baldi}, {Bertola}, {Bonoli}, {Comastri}, {Shankar},
  {Acharya}, {Alonso Tetilla}, {Lapi}, {Laloux}, {L{\'o}pez L{\'o}pez},
  {Mu{\~n}oz Rodr{\'\i}guez}, {Musiimenta}, {Osorio Clavijo}, {Sala}, \&
  {Sengupta}}]{Lopez+2024}
{L{\'o}pez}, I.~E., {Yang}, G., {Mountrichas}, G., {et~al.} 2024, \aap, 692,
  A209

\bibitem[{{Lynden-Bell} \& {Pringle}(1974)}]{lyn-pri1974}
{Lynden-Bell}, D. \& {Pringle}, J.~E. 1974, \mnras, 168, 603

\bibitem[{{Martin} {et~al.}(2019){Martin}, {Nixon}, {Pringle}, \&
  {Livio}}]{Martin+2019}
{Martin}, R.~G., {Nixon}, C.~J., {Pringle}, J.~E., \& {Livio}, M. 2019, \na,
  70, 7

\bibitem[{{Masterson} {et~al.}(2024){Masterson}, {De}, {Panagiotou}, {Kara},
  {Arcavi}, {Eilers}, {Frostig}, {Gezari}, {Grotova}, {Liu}, {Malyali},
  {Meisner}, {Merloni}, {Newsome}, {Rau}, {Simcoe}, \& {van
  Velzen}}]{Masterson+2024}
{Masterson}, M., {De}, K., {Panagiotou}, C., {et~al.} 2024, \apj, 961, 211

\bibitem[{{Matsumoto} \& {Piran}(2021)}]{Matsumoto-Piran2021}
{Matsumoto}, T. \& {Piran}, T. 2021, \mnras, 502, 3385

\bibitem[{{Mattila} {et~al.}(2018){Mattila}, {P{\'e}rez-Torres}, {Efstathiou},
  {Mimica}, {Fraser}, {Kankare}, {Alberdi}, {Aloy}, {Heikkil{\"a}}, {Jonker},
  {Lundqvist}, {Mart{\'\i}-Vidal}, {Meikle}, {Romero-Ca{\~n}izales}, {Smartt},
  {Tsygankov}, {Varenius}, {Alonso-Herrero}, {Bondi}, {Fransson},
  {Herrero-Illana}, {Kangas}, {Kotak}, {Ram{\'\i}rez-Olivencia},
  {V{\"a}is{\"a}nen}, {Beswick}, {Clements}, {Greimel}, {Harmanen},
  {Kotilainen}, {Nandra}, {Reynolds}, {Ryder}, {Walton}, {Wiik}, \&
  {{\"O}stlin}}]{Mattila+2018}
{Mattila}, S., {P{\'e}rez-Torres}, M., {Efstathiou}, A., {et~al.} 2018,
  Science, 361, 482

\bibitem[{{Menou} {et~al.}(1999){Menou}, {Hameury}, \& {Stehle}}]{Menou+1999}
{Menou}, K., {Hameury}, J.-M., \& {Stehle}, R. 1999, \mnras, 305, 79

\bibitem[{{Menou} \& {Quataert}(2001{\natexlab{a}})}]{Menou-Quataert2001TDE}
{Menou}, K. \& {Quataert}, E. 2001{\natexlab{a}}, \apjl, 562, L137

\bibitem[{{Menou} \& {Quataert}(2001{\natexlab{b}})}]{Menou-Quataert2001}
{Menou}, K. \& {Quataert}, E. 2001{\natexlab{b}}, \apj, 552, 204

\bibitem[{{Metzger} \& {Stone}(2016)}]{Metzger-Stone2016}
{Metzger}, B.~D. \& {Stone}, N.~C. 2016, \mnras, 461, 948

\bibitem[{{Meyer}(1984)}]{Meyer1984}
{Meyer}, F. 1984, \aap, 131, 303

\bibitem[{{Meyer} \& {Meyer-Hofmeister}(1981)}]{Meyer-MeyerH1981}
{Meyer}, F. \& {Meyer-Hofmeister}, E. 1981, \aap, 104, L10

\bibitem[{{Meyer} \& {Meyer-Hofmeister}(2015)}]{Meyer-MeyerH2015}
{Meyer}, F. \& {Meyer-Hofmeister}, E. 2015, \pasj, 67, 52

\bibitem[{{Mihalas} \& {Mihalas}(1984)}]{mihalas-mihalas1984}
{Mihalas}, D. \& {Mihalas}, B.~W. 1984, {Foundations of radiation
  hydrodynamics}, ed. {Mihalas, D.~\& Mihalas, B.~W.}

\bibitem[{{Mineshige} \& {Shields}(1990)}]{Mineshige1990}
{Mineshige}, S. \& {Shields}, G.~A. 1990, \apj, 351, 47

\bibitem[{{Mummery} \& {Balbus}(2020)}]{Mummery-Balbus2020}
{Mummery}, A. \& {Balbus}, S.~A. 2020, \mnras, 492, 5655

\bibitem[{{Mummery} {et~al.}(2024){Mummery}, {van Velzen}, {Nathan}, {Ingram},
  {Hammerstein}, {Fraser-Taliente}, \& {Balbus}}]{Mummery+2024}
{Mummery}, A., {van Velzen}, S., {Nathan}, E., {et~al.} 2024, \mnras, 527, 2452

\bibitem[{{Narayan}(1996)}]{Narayan1996}
{Narayan}, R. 1996, \apj, 462, 136

\bibitem[{{Narayan} {et~al.}(1997){Narayan}, {Barret}, \&
  {McClintock}}]{Narayan+1997}
{Narayan}, R., {Barret}, D., \& {McClintock}, J.~E. 1997, \apj, 482, 448

\bibitem[{{Narayan} \& {McClintock}(2008)}]{Narayan-McClintock2008}
{Narayan}, R. \& {McClintock}, J.~E. 2008, \nar, 51, 733

\bibitem[{{Narayan} \& {Yi}(1994)}]{naraya1994}
{Narayan}, R. \& {Yi}, I. 1994, \apjl, 428, L13

\bibitem[{{Narayan} \& {Yi}(1995)}]{Narayan-Yi1995}
{Narayan}, R. \& {Yi}, I. 1995, \apj, 452, 710

\bibitem[{{Nayakshin} {et~al.}(2007){Nayakshin}, {Cuadra}, \&
  {Springel}}]{Nayakshin+2007}
{Nayakshin}, S., {Cuadra}, J., \& {Springel}, V. 2007, \mnras, 379, 21

\bibitem[{{Nemmen} {et~al.}(2014){Nemmen}, {Storchi-Bergmann}, \&
  {Eracleous}}]{Nemmen+2014}
{Nemmen}, R.~S., {Storchi-Bergmann}, T., \& {Eracleous}, M. 2014, \mnras, 438,
  2804

\bibitem[{{Paczy{\'n}ski}(1969)}]{Paczynski1969}
{Paczy{\'n}ski}, B. 1969, \actaa, 19, 1

\bibitem[{{Paczynski} \& {Jaroszynski}(1978)}]{Paczynski-Jaroszynski1978}
{Paczynski}, B. \& {Jaroszynski}, M. 1978, \actaa, 28, 111

\bibitem[{{Parkinson} {et~al.}(2025){Parkinson}, {Knigge}, {Dai}, {Thomsen},
  {Matthews}, \& {Long}}]{Parkinson+2025}
{Parkinson}, E.~J., {Knigge}, C., {Dai}, L., {et~al.} 2025, \mnras
  [\eprint[arXiv]{2408.16371}]

\bibitem[{{Paxton} {et~al.}(2011){Paxton}, {Bildsten}, {Dotter}, {Herwig},
  {Lesaffre}, \& {Timmes}}]{Paxton+2011}
{Paxton}, B., {Bildsten}, L., {Dotter}, A., {et~al.} 2011, \apjs, 192, 3

\bibitem[{{Piran}(1978)}]{Piran1978}
{Piran}, T. 1978, \apj, 221, 652

\bibitem[{{Piran} {et~al.}(2015){Piran}, {Svirski}, {Krolik}, {Cheng}, \&
  {Shiokawa}}]{Piran+2015}
{Piran}, T., {Svirski}, G., {Krolik}, J., {Cheng}, R.~M., \& {Shiokawa}, H.
  2015, \apj, 806, 164

\bibitem[{{Pojmanski}(1986)}]{Pojmanski1986}
{Pojmanski}, G. 1986, \actaa, 36, 69

\bibitem[{{Potter} \& {Balbus}(2014)}]{Potter-Balbus2014}
{Potter}, W.~J. \& {Balbus}, S.~A. 2014, \mnras, 441, 681

\bibitem[{{Potter} \& {Balbus}(2017)}]{Potter-Balbus2017}
{Potter}, W.~J. \& {Balbus}, S.~A. 2017, \mnras, 472, 3021

\bibitem[{{Poutanen} {et~al.}(1997){Poutanen}, {Krolik}, \&
  {Ryde}}]{Poutanen+1997}
{Poutanen}, J., {Krolik}, J.~H., \& {Ryde}, F. 1997, \mnras, 292, L21

\bibitem[{{Poutanen} {et~al.}(2007){Poutanen}, {Lipunova}, {Fabrika},
  {Butkevich}, \& {Abolmasov}}]{poutanen_et2007}
{Poutanen}, J., {Lipunova}, G., {Fabrika}, S., {Butkevich}, A.~G., \&
  {Abolmasov}, P. 2007, \mnras, 377, 1187

\bibitem[{{Rees}(1988)}]{Rees1988}
{Rees}, M.~J. 1988, \nat, 333, 523

\bibitem[{{Ryu} {et~al.}(2020){Ryu}, {Krolik}, \& {Piran}}]{Ryu+2020}
{Ryu}, T., {Krolik}, J., \& {Piran}, T. 2020, \apj, 904, 73

\bibitem[{{Safronov}(1960)}]{Safronov1960}
{Safronov}, V.~S. 1960, Annales d'Astrophysique, 23, 979

\bibitem[{{Saxton} {et~al.}(2021){Saxton}, {Komossa}, {Auchettl}, \&
  {Jonker}}]{Saxton+2021}
{Saxton}, R., {Komossa}, S., {Auchettl}, K., \& {Jonker}, P.~G. 2021, \ssr,
  217, 18

\bibitem[{{Sazonov} {et~al.}(2021){Sazonov}, {Gilfanov}, {Medvedev}, {Yao},
  {Khorunzhev}, {Semena}, {Sunyaev}, {Burenin}, {Lyapin}, {Meshcheryakov},
  {Uskov}, {Zaznobin}, {Postnov}, {Dodin}, {Belinski}, {Cherepashchuk},
  {Eselevich}, {Dodonov}, {Grokhovskaya}, {Kotov}, {Bikmaev}, {Zhuchkov},
  {Gumerov}, {van Velzen}, \& {Kulkarni}}]{Sazonov_etal2021}
{Sazonov}, S., {Gilfanov}, M., {Medvedev}, P., {et~al.} 2021, \mnras, 508, 3820

\bibitem[{{Schwarzschild}(1958)}]{Schwarzschild1958}
{Schwarzschild}, M. 1958, {Structure and evolution of the stars.}

\bibitem[{Shakura {et~al.}(2018)Shakura, Lipunova, Malanchev,
  {et~al.}}]{shakura_etal2018}
Shakura, N.~I., Lipunova, G.~V., Malanchev, K.~L., {et~al.} 2018, Accretion
  flows in astrophysics (New York: New York)

\bibitem[{{Shakura} \& {Sunyaev}(1973)}]{sha-sun1973}
{Shakura}, N.~I. \& {Sunyaev}, R.~A. 1973, \aap, 24, 337

\bibitem[{{Shakura} \& {Sunyaev}(1976)}]{Shakura-Sunyaev1976}
{Shakura}, N.~I. \& {Sunyaev}, R.~A. 1976, \mnras, 175, 613

\bibitem[{{Shen} \& {Matzner}(2014)}]{Shen2014}
{Shen}, R.-F. \& {Matzner}, C.~D. 2014, \apj, 784, 87

\bibitem[{{Siemiginowska} \& {Czerny}(1989)}]{Siemiginowska-Czerny1989}
{Siemiginowska}, A. \& {Czerny}, B. 1989, \mnras, 239, 289

\bibitem[{{Siemiginowska} {et~al.}(1996){Siemiginowska}, {Czerny}, \&
  {Kostyunin}}]{Siemiginowska_etal1996}
{Siemiginowska}, A., {Czerny}, B., \& {Kostyunin}, V. 1996, \apj, 458, 491

\bibitem[{{S{\k{a}}dowski}(2009)}]{Sadowski2009}
{S{\k{a}}dowski}, A. 2009, \apjs, 183, 171

\bibitem[{{S{\k{a}}dowski} {et~al.}(2011){S{\k{a}}dowski}, {Abramowicz},
  {Bursa}, {Klu{\'z}niak}, {Lasota}, \& {R{\'o}{\.z}a{\'n}ska}}]{Sadowski+2011}
{S{\k{a}}dowski}, A., {Abramowicz}, M., {Bursa}, M., {et~al.} 2011, \aap, 527,
  A17

\bibitem[{{Smak}(1984{\natexlab{a}})}]{Smak1984_part4}
{Smak}, J. 1984{\natexlab{a}}, \actaa, 34, 161

\bibitem[{{Smak}(1984{\natexlab{b}})}]{Smak1984PASP}
{Smak}, J. 1984{\natexlab{b}}, \pasp, 96, 5

\bibitem[{{{\'S}niegowska} {et~al.}(2023){{\'S}niegowska},
  {Grz{\c{e}}dzielski}, {Czerny}, \& {Janiuk}}]{Sniegowska+2023}
{{\'S}niegowska}, M., {Grz{\c{e}}dzielski}, M., {Czerny}, B., \& {Janiuk}, A.
  2023, \aap, 672, A19

\bibitem[{{Soria} {et~al.}(2006){Soria}, {Graham}, {Fabbiano}, {Baldi},
  {Elvis}, {Jerjen}, {Pellegrini}, \& {Siemiginowska}}]{Soria+2006}
{Soria}, R., {Graham}, A.~W., {Fabbiano}, G., {et~al.} 2006, \apj, 640, 143

\bibitem[{{Spitzer}(1962)}]{Spitzer1962book}
{Spitzer}, L. 1962, {Physics of Fully Ionized Gases}

\bibitem[{{Starling} {et~al.}(2004){Starling}, {Siemiginowska}, {Uttley}, \&
  {Soria}}]{Starling+2004}
{Starling}, R. L.~C., {Siemiginowska}, A., {Uttley}, P., \& {Soria}, R. 2004,
  \mnras, 347, 67

\bibitem[{{Subrayan} {et~al.}(2023){Subrayan}, {Milisavljevic}, {Chornock},
  {Margutti}, {Alexander}, {Ramakrishnan}, {Duffell}, {Dickinson}, {Lee},
  {Giannios}, {Lentner}, {Linvill}, {Garretson}, {Graham}, {Stern},
  {Brethauer}, {Duong}, {Jacobson-Gal{\'a}n}, {LeBaron}, {Matthews}, {Sears},
  \& {Venkatraman}}]{Subrayan+2023}
{Subrayan}, B.~M., {Milisavljevic}, D., {Chornock}, R., {et~al.} 2023, \apjl,
  948, L19

\bibitem[{{Suleimanov} {et~al.}(2007){Suleimanov}, {Lipunova}, \&
  {Shakura}}]{Suleimanov+2007}
{Suleimanov}, V.~F., {Lipunova}, G.~V., \& {Shakura}, N.~I. 2007, Astronomy
  Reports, 51, 549

\bibitem[{{Szuszkiewicz} \& {Miller}(1998)}]{Szuszkiewicz-Miller1998}
{Szuszkiewicz}, E. \& {Miller}, J.~C. 1998, \mnras, 298, 888

\bibitem[{{Taam} \& {Lin}(1984)}]{Taam-Lin1984}
{Taam}, R.~E. \& {Lin}, D.~N.~C. 1984, \apj, 287, 761

\bibitem[{{Taam} {et~al.}(2012){Taam}, {Liu}, {Yuan}, \& {Qiao}}]{Taam+2012}
{Taam}, R.~E., {Liu}, B.~F., {Yuan}, W., \& {Qiao}, E. 2012, \apj, 759, 65

\bibitem[{{Tavleev} {et~al.}(2023{\natexlab{a}}){Tavleev}, {Lipunova}, \&
  {Malanchev}}]{Tavleev+2023}
{Tavleev}, A.~S., {Lipunova}, G.~V., \& {Malanchev}, K.~L. 2023{\natexlab{a}},
  \mnras, 524, 3647

\bibitem[{{Tavleev} {et~al.}(2023{\natexlab{b}}){Tavleev}, {Lipunova}, \&
  {Malanchev}}]{DiscCode_ascl}
{Tavleev}, A.~S., {Lipunova}, G.~V., \& {Malanchev}, K.~L. 2023{\natexlab{b}},
  {DiscVerSt: Vertical structure calculator for accretion discs around neutron
  stars and black holes}, Astrophysics Source Code Library, record
  ascl:2307.011

\bibitem[{{Thomsen} {et~al.}(2022){Thomsen}, {Kwan}, {Dai}, {Wu}, {Roth}, \&
  {Ramirez-Ruiz}}]{Thomsen+2022}
{Thomsen}, L.~L., {Kwan}, T.~M., {Dai}, L., {et~al.} 2022, \apjl, 937, L28

\bibitem[{{Toomre}(1964)}]{Toomre1964}
{Toomre}, A. 1964, \apj, 139, 1217

\bibitem[{{Ulmer}(1999)}]{Ulmer1999}
{Ulmer}, A. 1999, \apj, 514, 180

\bibitem[{{Uno} {et~al.}(2025){Uno}, {Maeda}, {Nagao}, {Leloudas},
  {Charalampopoulos}, {Mattila}, {Aoki}, {Taguchi}, {Kawabata}, {Moldon},
  {P\textbackslash'erez-Torres}, {Pursiainen}, \& {Reynolds}}]{Uno+2025}
{Uno}, K., {Maeda}, K., {Nagao}, T., {et~al.} 2025, arXiv e-prints,
  arXiv:2503.19024

\bibitem[{{van Velzen} {et~al.}(2020){van Velzen}, {Holoien}, {Onori}, {Hung},
  \& {Arcavi}}]{Velzen+2020}
{van Velzen}, S., {Holoien}, T. W.~S., {Onori}, F., {Hung}, T., \& {Arcavi}, I.
  2020, \ssr, 216, 124

\bibitem[{{van Velzen} {et~al.}(2019){van Velzen}, {Stone}, {Metzger},
  {Gezari}, {Brown}, \& {Fruchter}}]{VanVelzen+2019}
{van Velzen}, S., {Stone}, N.~C., {Metzger}, B.~D., {et~al.} 2019, \apj, 878,
  82

\bibitem[{{Volonteri} {et~al.}(2011){Volonteri}, {Dotti}, {Campbell}, \&
  {Mateo}}]{Volonteri+2011}
{Volonteri}, M., {Dotti}, M., {Campbell}, D., \& {Mateo}, M. 2011, \apj, 730,
  145

\bibitem[{{Vranjes}(2014)}]{Vranjes2014}
{Vranjes}, J. 2014, \mnras, 445, 1614

\bibitem[{{Vranjes} \& {Krstic}(2013)}]{Vranjes-Krstic2013}
{Vranjes}, J. \& {Krstic}, P.~S. 2013, \aap, 554, A22

\bibitem[{{Wen} {et~al.}(2021){Wen}, {Jonker}, {Stone}, \&
  {Zabludoff}}]{Wen+2021}
{Wen}, S., {Jonker}, P.~G., {Stone}, N.~C., \& {Zabludoff}, A.~I. 2021, \apj,
  918, 46

\bibitem[{{Wen} {et~al.}(2020){Wen}, {Jonker}, {Stone}, {Zabludoff}, \&
  {Psaltis}}]{Wen+2020}
{Wen}, S., {Jonker}, P.~G., {Stone}, N.~C., {Zabludoff}, A.~I., \& {Psaltis},
  D. 2020, \apj, 897, 80

\bibitem[{{Wevers} {et~al.}(2022){Wevers}, {Nicholl}, {Guolo},
  {Charalampopoulos}, {Gromadzki}, {Reynolds}, {Kankare}, {Leloudas},
  {Anderson}, {Arcavi}, {Cannizzaro}, {Chen}, {Ihanec}, {Inserra},
  {Guti{\'e}rrez}, {Jonker}, {Lawrence}, {Magee}, {M{\"u}ller-Bravo}, {Onori},
  {Ridley}, {Schulze}, {Short}, {Hiramatsu}, {Newsome}, {Terwel}, {Yang}, \&
  {Young}}]{Wevers+2022}
{Wevers}, T., {Nicholl}, M., {Guolo}, M., {et~al.} 2022, \aap, 666, A6

\bibitem[{{Xiang} {et~al.}(2024){Xiang}, {Miller}, {Zoghbi}, {Reynolds},
  {Bogensberger}, {Dai}, {Draghis}, {Drake}, {Godet}, {Irwin}, {Miller},
  {Mockler}, {Saxton}, \& {Webb}}]{Xiang+2024}
{Xiang}, X., {Miller}, J.~M., {Zoghbi}, A., {et~al.} 2024, \apj, 972, 106

\bibitem[{{Xie} {et~al.}(2009){Xie}, {Ma}, {Zhang}, {Du}, {Hao}, {Yi}, \&
  {Qiao}}]{Xie+2009}
{Xie}, Z.~H., {Ma}, L., {Zhang}, X., {et~al.} 2009, \apj, 707, 866

\bibitem[{{Yan} \& {Xie}(2018)}]{Yan-Xie2018}
{Yan}, Z. \& {Xie}, F.-G. 2018, \mnras, 475, 1190

\bibitem[{{Yuan} \& {Narayan}(2014)}]{Yuan-Narayan2014}
{Yuan}, F. \& {Narayan}, R. 2014, \araa, 52, 529

\bibitem[{{Zhang}(2023)}]{Zhang2023}
{Zhang}, X. 2023, \apj, 948, 68

\bibitem[{Zhdanov(2002)}]{zhdanov2002transport}
Zhdanov, V. 2002, Transport Processes in Multicomponent Plasma (CRC Press)

\end{thebibliography}

%%%%%%%%%%%%%%%%% APPENDICES 

\begin{appendix}

\section{Geometrically thin disc vertical structure}\label{s.thin_disc_code}

The vertical structure of the accretion disc is described by the system of four ordinary differential equations \citep[see e.g.][]{shakura_etal2018}, which follows from the mass, energy, and momentum conservation laws:
\begin{align}
    &\frac{{\rm d}P}{{\rm d}z} = -\rho\,\omega^2_{\rm K} z, \label{eq:P} \\ 
    &\frac{{\rm d}Q}{{\rm d}z} = \frac32 \,w_{r\varphi}\,\omega_{\rm K} = \frac32 \,\alpha P\, \omega_{\rm K}\label{eq:Q}\, , \\
    &\frac{{\rm d}\ln T}{{\rm d}\ln P} \equiv \nabla = 
        \begin{cases}
        \nabla_{\rm rad}, \, \nabla_{\rm rad}\leq\nabla_{\rm ad}, \\
        \nabla_{\rm conv}, \, \nabla_{\rm rad}\geq\nabla_{\rm ad}, \\
        \end{cases} \label{eq:T} \\
    &\frac{{\rm d}\Sigma}{{\rm d}z} = -2\rho, \label{eq:Sigma} \\ 
    &z \in [0, z_0]. \nonumber
\end{align}
Here $P=a_{\rm rad}\, T^4/3 + P_{\rm gas}$ is the total pressure, $Q$ is the heating flux, which for an un-irradiated disc equals the viscous flux $Q_{\rm vis}$, and $T$ is the temperature. The mass coordinate $\Sigma(z)$ equals to zero at $z=z_0$ and the surface density of the disc $\Sigma_0$ at $z=0$. The last part of \eqref{eq:Q} includes the $\alpha$-prescription~\citep{sha-sun1973}, where the absolute value of the $r\varphi$-component of the viscous tensor $w_{r\varphi} = \alpha P$, and $\alpha$ is the turbulent parameter ($0<\alpha<1$).

If the energy is transported solely by radiation diffusion, equation~\eqref{eq:T} implies that
\begin{equation}
    \nabla = \nabla_{\rm rad} \equiv \frac{3\varkappa_{\rm R}}{4a_{\rm rad}c\omega^2_{\rm K} z} \frac{P}{T^4} Q,
    \label{eq:nabla_rad}
\end{equation}
where $\varkappa_{\rm R}$ is the Rosseland opacity coefficient, $a_{\rm rad} = 4\sigma_{\rm SB}/c$ is the radiation constant, $c$ is the speed of light. Otherwise, if $\nabla_{\rm rad}\geq\nabla_{\rm ad}$, the convective motions start to transfer energy according to the Schwarzschild~(\citeyear{Schwarzschild1958}) criterion. The corresponding temperature gradient $\nabla_{\rm conv}$ is calculated according to the mixing length theory~\citep[see][]{Paczynski1969, Kippenhahn2012,Hameury1998}.

Flux radiated from one surface equals half the viscous heat flux at the same radius when the local energy balance holds:
\begin{equation}
Q_{\rm vis} = \frac{3}{8\pi} \frac{F\omega_{\rm K}}{r^2}\, .
\label{eq.Qvis_stationary}
\end{equation}
For a disc around a BH, the viscous torque $F = \dot{M}\,h \,\left(1 - \sqrt{r_{\rm in}/r}\right)$, $h=\sqrt{GMr}$ is the specific angular momentum, and $r_{\rm in}$ is the inner radius of the disc.

The open code \vertcode was developed by~\citet{Tavleev+2023, DiscCode_ascl} to solve the system~(\ref{eq:P}--\ref{eq:Sigma}).
%%boundary
% The semi-thickness of the disc~$z_0$ is a free parameter of the system. The open code \vertcode by~\citet{Tavleev+2023} integrates the system and finds the value of~$z_0$.
% by the shooting method: system~(\ref{eq:P}--\ref{eq:Sigma}) is integrated with different values of the free parameter, starting from initial estimation, in order to satisfy the additional condition~(\ref{eq:Q_bound_another}). Afterwards, the surface density is obtained $\Sigma_0 = \Sigma(0)$.
The main input parameters of the code are: the mass of the central object~$M$, radius~$r$, viscous torque~$F$ (or effective temperature $T_{\rm eff}$, or the accretion rate $\dot{M}$), and turbulence parameter~$\alpha$. The output is $P(z)$, $T(z)$, $Q(z)$, $\rho(z)$, $z_0$, $\Sigma_0$, and other parameters. 
The vertical structure equations are complemented by an equation of state~(EoS) and an opacity law, for which their tabulated values from the \textsc{MESA} code\footnote{\cite{Paxton+2011}} can be used.

By calculating vertical structures for a set of accretion rates~$\dot{M}$,  \revtri{the code provides a set of corresponding surface densities~$\Sigma_0$, that is, the S-curves are obtained.}
 The solar chemical composition is assumed throughout this paper. 

\section{Self-similarity of the vertical structure}\label{s.selfsim} 

The vertical structure of the disc is almost self-similar when the disc is hot and ionised. 
%This fact allows usage of the analytic formulas for the radial structure.  
One can introduce dimensionless $\Pi$-parameters:
%, which are useful for approximate estimations:
\begin{equation}
\begin{aligned}
    \Pi_1 &\equiv \frac{\omega_{\rm K}^2\,z_0^2\,\rho_c}{P_c}, \qquad\quad
    \Pi_2 \equiv \frac{\Sigma_0}{2z_0\,\rho_c}, \\
    \Pi_3 &\equiv \frac{3}{4}\frac{\alpha\,\omega_{\rm K}\,P_c\,\Sigma_0}{\rho_c Q_0}, \quad
    \Pi_4 \equiv \frac{3}{32}\left(\frac{T_{\rm eff}}{T_c}\right)^4\Sigma_0\,\varkappa_c.
\end{aligned} \label{eq:Pi}
\end{equation}
Here $P_c, T_c, \rho_c, \varkappa_c$ are the total pressure, temperature, bulk density and opacity in the symmetry plane. 
Given the  values of $\Pi$ (calculated by \vertcode, for example), and knowing $\omega_{\rm K}$, $\alpha$, $Q_0=\sigma_{\rm SB}\,T_{\rm eff} ^{4}$ (or $\dot M$), one can easily calculate $z_0$, $\rho_c$, $\Sigma_0$, and $T_c$ as functions of radius. 
The pressure $P_c$ is obtained from an equation of state, for example that of an ideal gas. 
If the vertical structure equations are not solved, all parameters of $\Pi$ can be set to 1; this will be the equivalent of averaging  equations (\ref{eq:P}--\ref{eq:Sigma}).
%Omitting the solution of the vertical structure equations, leads to all \hbox{$\Pi$ = 1} and radial distributions  2.16 and 2.19 in  \citet{sha-sun1973}. 

%For this standard regime, a comparison of the analytic radial approximations with $\Pi$ and the  results of numerical integration of the vertical-structure equations is provided by \citep{Suleimanov+2007}.
The self-similarity of the vertical structure is reflected in the fact that the $\Pi$ values found after solving system \eqref{eq:P}-\eqref{eq:Sigma}  vary insignificantly in the stable hot-disc region~\citep{KetsarisShakura1998,Suleimanov+2007,Tavleev+2023}. 
If the vertical structure is approximated by a polytropic solution with the polytropic index $n=1..3$, corresponding $\Pi_1=2\,(n+1)$, $\Pi_2\sim 0.5$, $\Pi_3 \approx 1.1$, and $\Pi_4\approx 0.4$.

Using $\Pi_1$, which can be viewed as $(\omega_{\rm K} \, z_0/v_s)^2$, where  the sound velocity $v_s = (P/\rho)^{1/2}$, one can write the viscous time of a disc:
 \begin{equation}
 \tvis \equiv  \frac{\Rout^2}{\nut (\Rout) } = 
 \frac{3}{2}\, \frac{1}{\alpha\omega_{\rm K}}\, \left(\frac{z_0}r\right)^{-2} \Pi_1\, ,
 \label{eq.visc_time_scale}
\end{equation}
where $\nut$ if determined from the relation between the kinematic coefficient of turbulent viscosity $\nut$ and $\alpha$ is obtained from  the stress-tensor for Keplerian $\alpha$-discs:
\begin{equation}
    w_{r\varphi} = \alpha P = \frac32\, \omega_{\rm K} \, \rho \, \nut\,  .
    \label{eq.alpha_nut_relation}
\end{equation}

We can infer the dependences between $\dot M$ and $\Sigma_0$ for Fig.~\ref{fig:S-curve_SMBH} using the definitions of $\Pi$. 
Combining the first and third equation of  \eqref{eq:Pi}), we obtain:
\begin{equation}
    \dot{M} = \frac{2\,\pi}{\Pi_1\,\Pi_3}\,\alpha\, \sqrt{GMr} \,\Sigma_0 \left(\frac{z_0}{r}\right)^2\, . 
    \label{eq.Adv_zone}
\end{equation}
For the zone A, where $P=P_{\rm rad}=a_{\rm rad}T^4/3$
and opacity is determined by the Thomson scattering, with $Q_0=Q_{\rm vis}$ at $r\gg r_{\rm in}$ in \eqref{eq.Qvis_stationary}, we obtain:
\begin{equation}
    \dot{M} = \frac{\Pi_3\,\Pi_4^2}{\Pi_1\,\Pi_2^2}\, \frac{32}{9\pi\alpha\omega_\mathrm{K}}\left(\frac{L_{\rm Edd}}{GM}\right)^2\frac1{\Sigma_0}\, . 
    \label{eq.A_zone}
\end{equation}

\section{Slim disc structure}\label{s.thick_disc_code}
In the solution for the radial structure of a slim disc, the integration scheme of \citet{Lipunova1999} and equations for $\Pi_1$, $\Pi_2$, and $\Pi_4$ are utilised, see \eqref{eq:Pi}, with fixed values of $\Pi$.  
 The local energy balance equation is substituted by
\begin{equation}
    Q_{\rm vis}= Q_{\rm adv} + Q_0 \, , 
\end{equation}
  where $Q_0$ and $Q_{\rm vis}$ are the radiation and viscous flux, per  one side of the disc,  see Appendix~\ref{s.thin_disc_code}. The disc is assumed to be Keplerian.
The advected flux can be expressed as follows~\citep{Chashkina+2019}
\begin{equation}
     Q_{\rm adv}  = \frac12\, \int\limits_{-z_0}^{+z_0} \rho(z) \, v_r \, \frac {kT(z)} {\tilde{m}}\, \frac {{\rm d} s}{{\rm d} r}\, {\rm d} z\, ,
    \end{equation}
    where the dimensionless entropy per particle is
    \begin{equation}
        s = \frac52 + \ln \left[ \frac{\tilde{m}}{\rho}\, \left(\frac{2\, \pi\,\tilde{m}\,k\, T}{h^2}\right)^{3/2}\right]
        %+ \frac 43\, \frac{\tilde{m} \,a_{\rm rad} T^3}{k\, \rho} 
        + \left( \frac{32 \pi^5}{45\, c^3\, h^3}\right) \frac{\tilde{m} \,(kT)^3}{\rho}
        \label{eq.entropy}
    \end{equation}
    % Sackur–Tetrode equation 
    and $\tilde{m}$ is the mean mass per particle. The radial velocity is assumed to be $z$-independent:
    $
      v_r = \dot M / (2 \pi\, r\, \Sigma_0)\, .   
    $
    Taking into account that $P_{\rm rad,c}/P_{\rm gas,c} = \tilde{m}  \, a_{\rm rad}\, T_c^3/ (3k \, \rho_c) $,
    %As a result of substitution of \eqref{eq.entropy},
    after some transformations, we obtain
    \begin{equation}
     Q_{\rm adv} = \frac{\Pi_3^{\rm adv}}{2} \, P_c \, v_r \,\frac{ z_0}{r} \,  \left\{\Big(12-\frac{21}{2}\,\beta\Big)\,\frac{{\rm d} \ln T_c}{{\rm d} \ln r} - (4-3\,\beta)\,\frac{{\rm d} \ln \rho_c}{{\rm d} \ln r}\right\}\,  ,
    \label{eq.Qadv}
\end{equation}
where  $\beta = P_{\rm gas,c}/P_c$ and 
\begin{equation}
\Pi_3^{\rm adv} =  \int\limits_{-1}^{1} (1-x^2)^{n+1} \, {\rm d} x. 
% in the code Pi3adv_code = 2*   Pi3adv_paper 
\end{equation}
The last dimensionless parameter appears because we assume here a fixed vertical effective polytropic index $n$, defined through the equation for density $\rho =\rho_c \, (1- (z/z_0)^2)^n$ and pressure $P = P_c\, (1-(z/z_0)^2)^{n+1}$~\citep{Paczynski-Jaroszynski1978}.

The semi-thickness $z_0$ is found at each radius by a minimisation procedure. Without an outflow,  a solution is found by a single integration starting from an external geometrically thin disc,  \revtri{where we set $Q_\mathrm{adv}=0$. The higher $\dot M$, the larger the starting radius of the calculation.}

\revtri{The matching of the S-curves involves the following steps: 
(1) for specific radius $r$, calculate the S-curve in the geometrically thin regime in the \vertcode code to determine dependences of all disc parameters, including $\Pi$-values, on $\Sigma_0$ (see Appendices \ref{s.thin_disc_code} and \ref{s.selfsim});  (2) identify $\dot M_{\times}$ where $P_\mathrm{rad}/P_\mathrm{gas} = 0.5$ at $r$ to obtain $\Pi$ values in the regime when the disc is still thin but quite hot at the same time; (3) for these fixed $\Pi$ values, calculate radial models ($\Sigma_0(r)$, etc.) of advective discs for different accretion rates $\dot M\geq\dot M_{\times}$; (4) interpolate the results to derive the dependence $\dot M (\Sigma_0)$ at $r$. Thanks to the choice of $\Pi$ and the fact that $\Pi$s vary very slowly in the ionised disc, the sequence of equilibrium points for the slim disc (i.e. its \hbox{S-curve}) smoothly matches that for the geometrically thin disc.
} 

 \section{Analytic relations for $e$-folding time and peak accretion rate}\label{s.analyt}

\begin{figure}
\includegraphics[width=1.0\columnwidth]{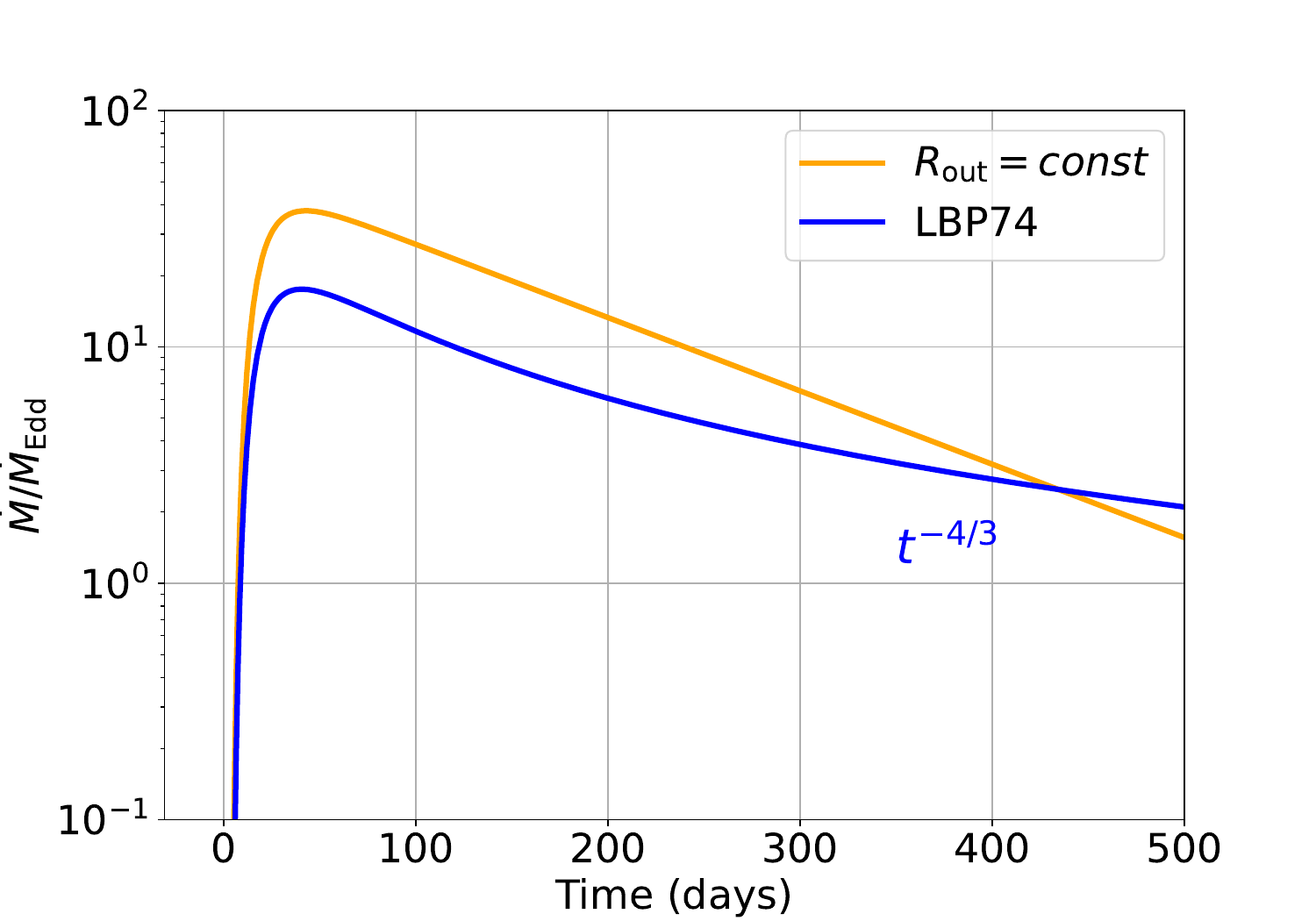}
	\caption{Analytic solutions for accretion rate evolution in disc without outflow with fixed  $z_0/r$  ($l=1/3$). At late times, the evolution is exponential in the disc with the constant outer radius \citep[orange;][]{lipunova2015} and power-law, when the outer radius is expanding~\citep[blue;][]{lyn-pri1974} For the both solutions, 
    %the viscous time is 390~d and 
    the mass is $\Mdisc = 4 M_\odot$ and initial ring radius $R_{\rm out} = 90\,\rg$, which roughly correspond to the minimum flare  in a disc around $10^7\, M_\odot$ SMBH in  Table~\ref{tab:props}.
 }
	\label{fig:Mdot_t_analyt}
\end{figure}

 If the mass, radius, and viscosity of a spreading ring are known, the time-dependent accretion rate can be found either analytically or numerically.  It also depends on the radial boundary conditions in the disc, as illustrated by Fig.~\ref{fig:Mdot_t_analyt}, which compares two analytic solutions with different outer boundary conditions.
 
 To estimate the upper limit of the observable time $\texp$, we assume that the hot advective disc during a giant flare can be treated as a Keplerian disc with a fixed outer radius and constant aspect ratio, so that both the viscosity $\nut(R)$ and the viscous time $t _ {\rm vis}=R^2/\nut$ are defined by \eqref{eq.tvis}.  For example, $t_{\rm vis} \approx 1$~yr if 
 %$\Rtrunc=60\rg$  
 $\Rout= 90\,\rg$ and $M= 10^7 M_\odot$.
 The $e$-folding decay time of such a disc
 is related to the viscous time as follows:
 \begin{equation}
   \texp = \frac{16 \,l^2}{3\,k_1^2} \,\frac{\Rout^2}{\nut (\Rout) }\, ,
   \label{eq.texp_an}
 \end{equation}
 where $l$ is the parameter defined by the viscosity type, introduced by \citet{lyn-pri1974}; $l=1/3$ is appropriate for discs with constant aspect ratio. The parameter $k_1=1.24$~\citep{lipunova2015}, and the dimensionless factor in \eqref{eq.texp_an} is about 0.38.
 
 The peak accretion rate in the solution of \citet{lyn-pri1974} for the infinite disc is 
 \begin{equation}
     \dot M_\mathrm{peak} = \frac{\Mdisc}{\tpeak} \, \frac{(1+l)^l}{e^{1+l}\,\Gamma(l)\, }
 \end{equation}
with 
\begin{equation}
\tpeak = \frac{4}{3}\, \frac{l^2}{(1+l)} \, \frac{\Rout^2}{\nut  }\, \qquad\mathrm{or}\qquad \tpeak=\frac{\tvis}{9}
\label{eq.tpeak}
\end{equation}
~\citep[see Appendix of ][]{lipunova2015}.
\begin{figure*}
\includegraphics[width=0.5\linewidth]{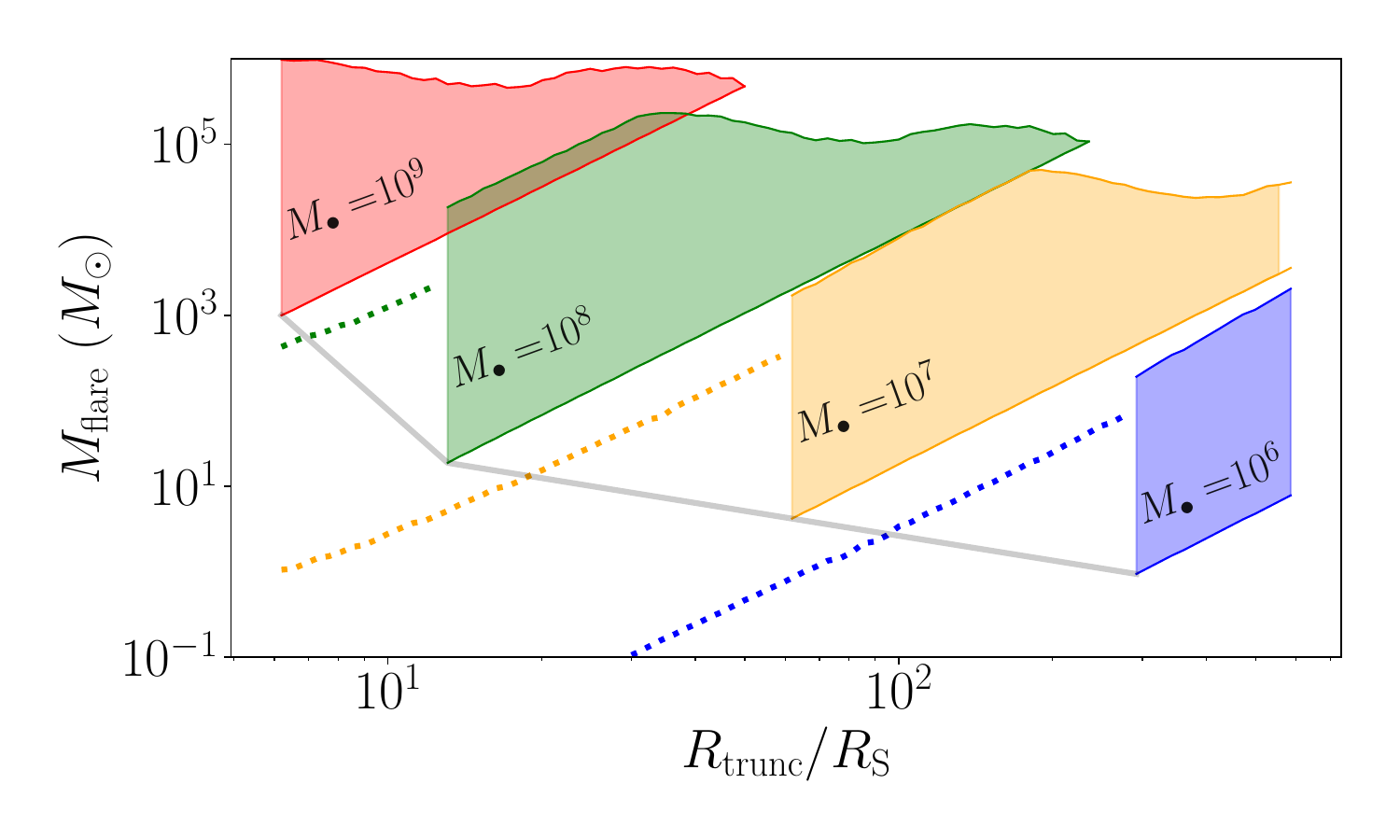}\hfill
\includegraphics[width=0.5\linewidth]{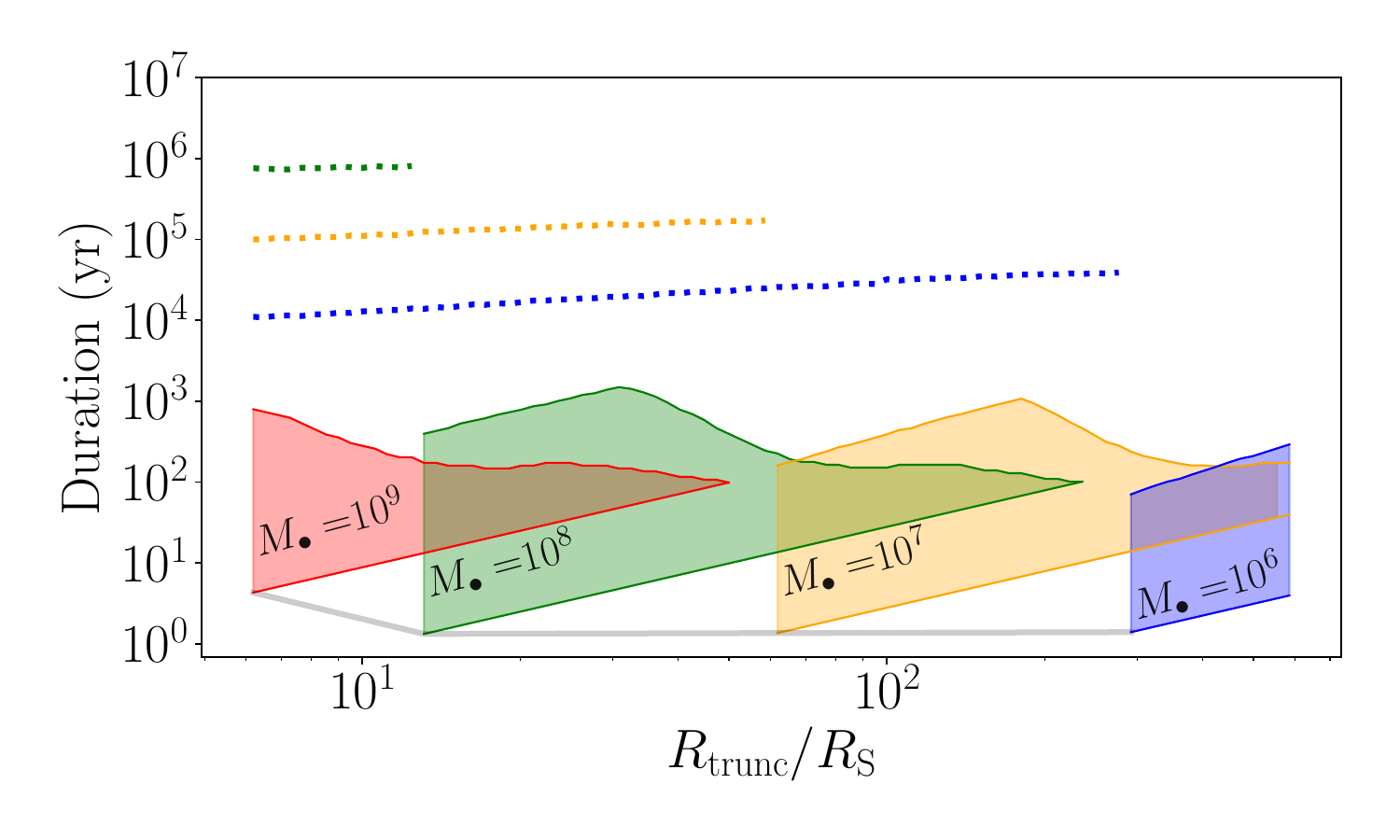}
\includegraphics[width=0.5\linewidth]{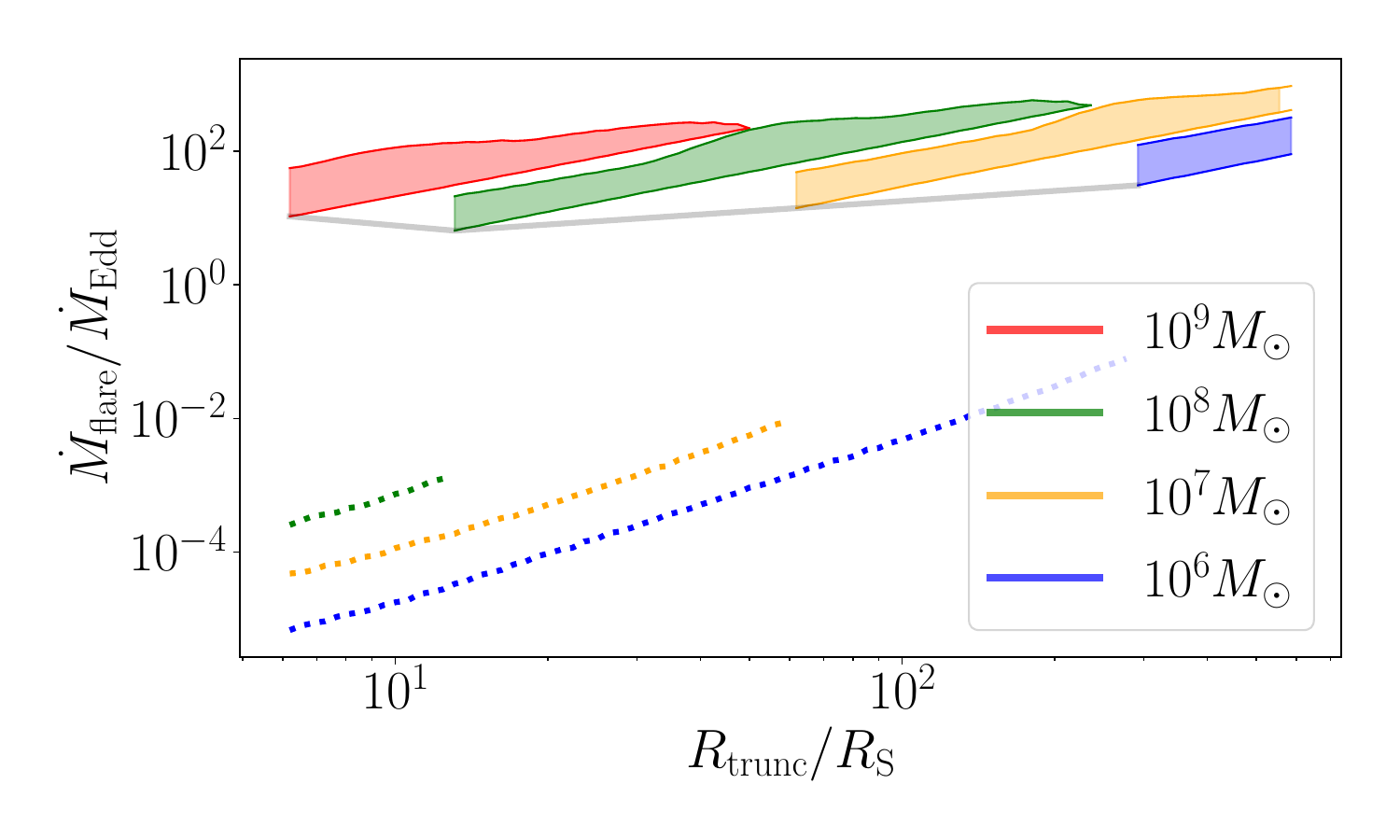}\hfill
\includegraphics[width=0.5\linewidth]{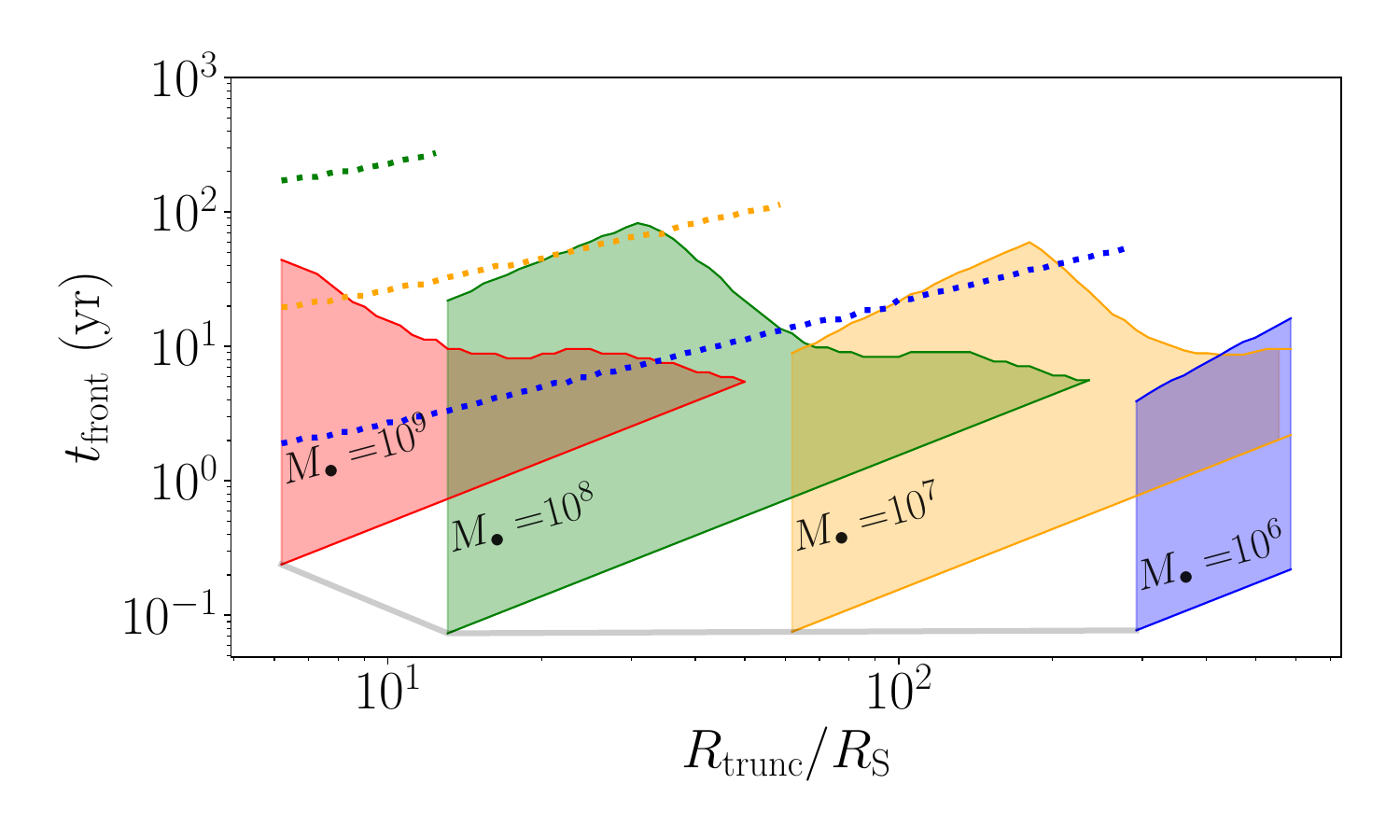}
	\caption{Outburst parameters for  giant flares (coloured areas) and normal flares (dotted lines) versus inner truncation radius of  quiescent geometrically thin disc. From top to bottom, from left to right: mass involved in a flare \eqref{eq.flare_estimates}, estimated peak accretion rate \eqref{eq.dotMpeak}, characteristic flare duration \eqref{eq.tvis},  and rise time \eqref{eq.t_front}.
 The upper boundaries of the coloured areas correspond to the case, when the disc is heated to the radius $\Rmax$; the lower boundaries, to the conservative assumption $\Rout = \Rtrunc+z_0$ (see Sect.~\ref{s.properties}). The lower grey  line illustrates the limits for intermediate masses.  For $10^9 M_\odot$, ionisation instability always ignites a giant flare.} 
	\label{fig:Flares_all_masses}
\end{figure*}
    Combining above formulae, we arrive at
    \begin{equation}
  \dot M_\mathrm{peak} = \frac{\Mdisc}{\Rout^2/\nut} \, \frac{3\,(1+l)^{1+l}}{4\,l^2\,e^{1+l}\,\Gamma(l)\,} = C \, \frac{\Mdisc}{\Rout^2/\nut}
  \label{eq.Mdotpeak}
  \end{equation}
  with $C\approx 0.97$ for $l=1/3$.
The peak accretion rate of the disc with  constant outer radius is about twice as high as \eqref{eq.Mdotpeak}~\citep{lipunova2015}.
%Most probably, it is the outflow parameters in the super-Eddington disc that determine the rate at which matter approaches the black hole.

\begin{figure*}
    \center
	\begin{minipage}{0.45\linewidth}
\center{\includegraphics[width=1.0\linewidth]{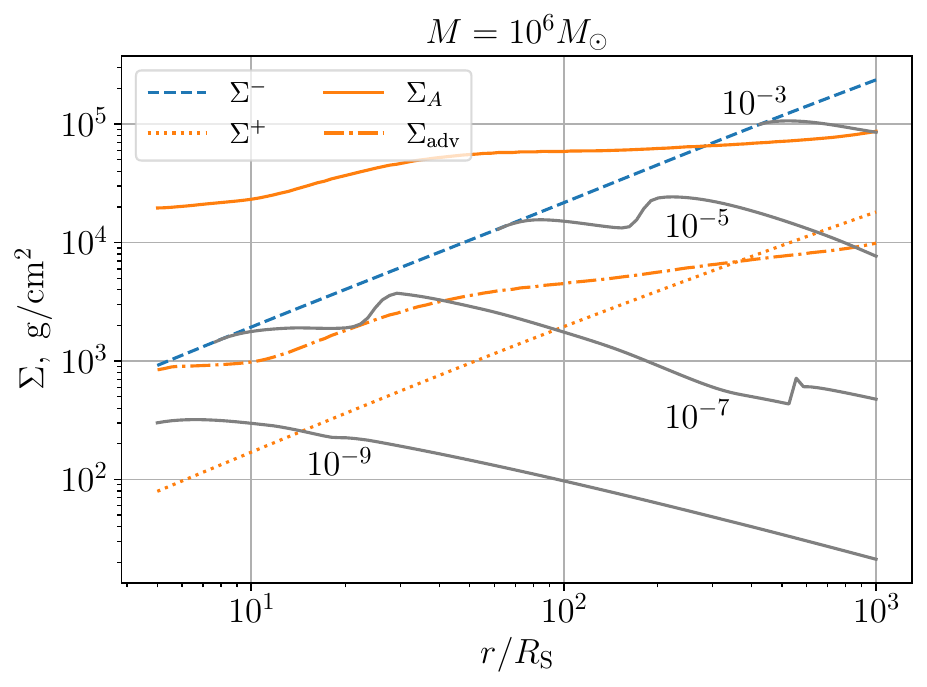}}
	\end{minipage}
	\begin{minipage}{0.45\linewidth}
\center{\includegraphics[width=1.0\linewidth]{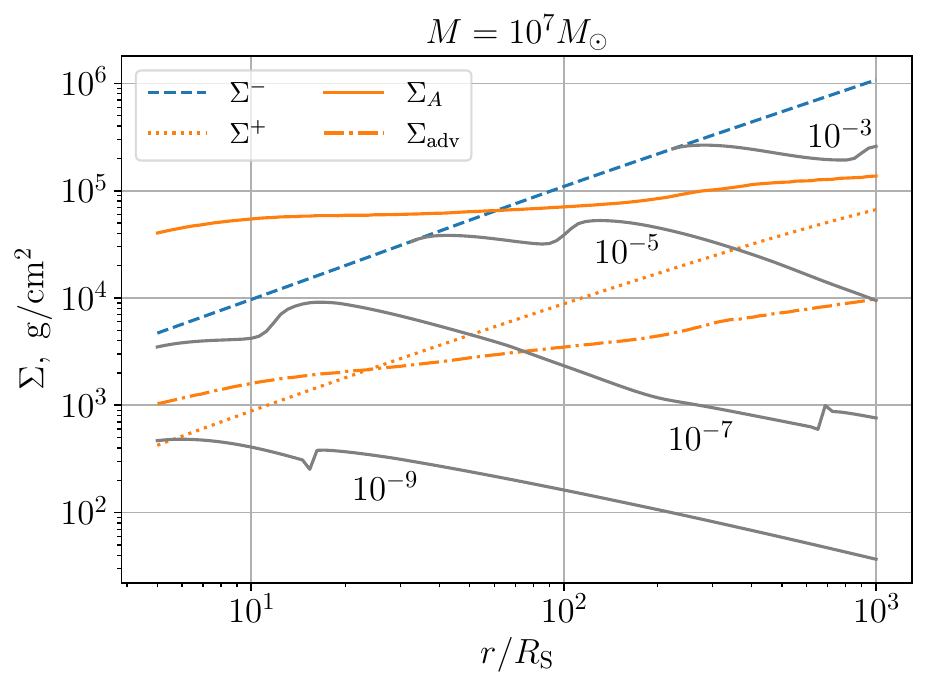}}
	\end{minipage}
	\begin{minipage}{0.45\linewidth}
\center{\includegraphics[width=1.0\linewidth]{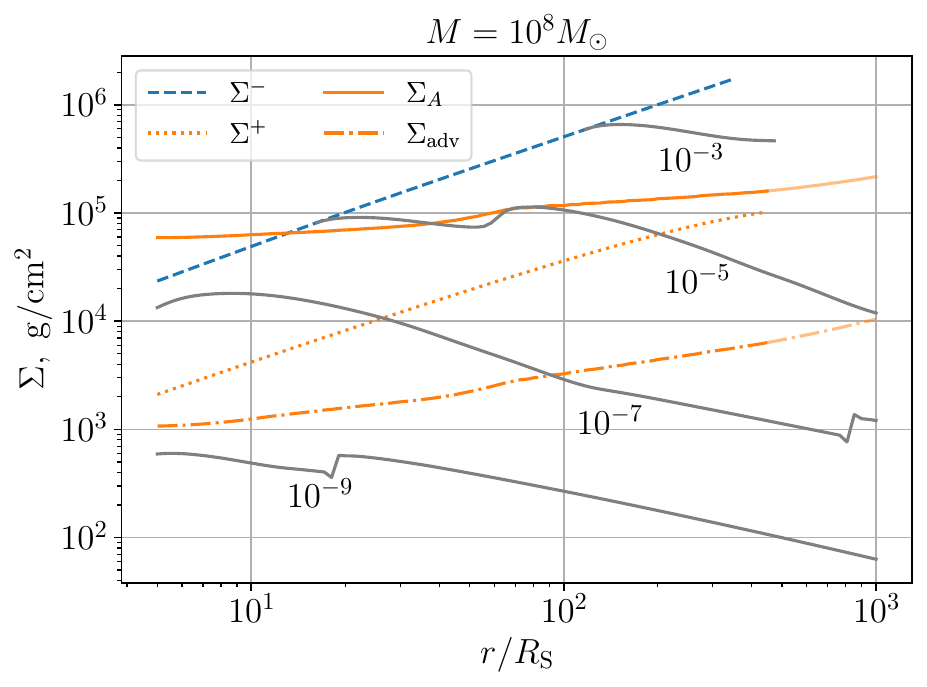}}
	\end{minipage}
	\begin{minipage}{0.45\linewidth}	\center{\includegraphics[width=1.0\linewidth]{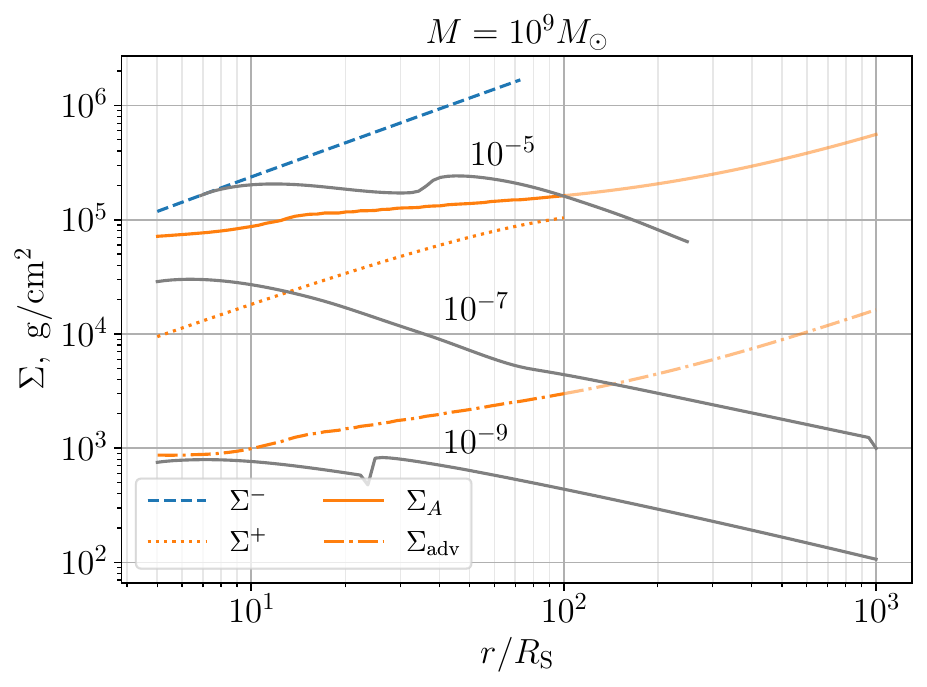}}
	\end{minipage}
    \caption{ 
    Radial distributions of $\Sigma^{-}$, $\Sigma_{\rm A}$, $\Sigma^{+}$, and $\Sigma_{\rm adv}$ for different masses of  central SMBH, see \eqref{eq.Sigma_R}. Also shown is the stationary distribution $\Sigma_{\rm stat}$ (grey lines) for different $\dot{M}$ (in $\dotMEdd$ units) for $\alpha_{\rm cold} = 0.01$ (see description in Sect.~\ref{s.points} and Appendix~\ref{s.analyt}). {Minimum} critical $\Sigma^- = \SigmaA \sim 7\times 10^{4}$~g cm$^{-2}$ for $10^6-10^8~M_\odot$ SMBH.
    }
    \label{fig:SigmaAllDots}
\end{figure*}

Figure~\ref{fig:Flares_all_masses} \revtri{summarises the mass $\Mdisc$, accretion rate $\Mdisc/t_{\rm vis}$, estimated duration $t_{\rm vis}$ and  rise time $t_{\rm front}$ for different masses of SMBH.}
The masses $\Mdisc$ involved in flares are calculated using the critical surface densities; see Sect.~\ref{s.properties}. 
The radial distributions of the critical densities $\Sigma^{-}$, $\Sigma_{\rm A}$, $\Sigma^{+}$, and $\Sigma_{\rm adv}$ are presented for different masses in Fig.~\ref{fig:SigmaAllDots}.
For $10^8$ and $10^9~M_\odot$, the stationary distributions $\Sigmastat(r)$ are plotted only for $r<\Rsg$ defined by condition \eqref{eq.Rself-grav}.
For these SMBH masses, critical $\Sigma^+$ could not be found for large $r$ because the positive hot geometrically thin branch of the S-curve disappears. In this case, the values $\SigmaA$ and $\Sigmadv$, which require $\Pi_{1..4}$ at a point on this branch, are extrapolated from smaller radii. These intervals are shown as transparent in the lower panels of Fig.~\ref{fig:SigmaAllDots}. The extrapolated values affect only the flare characteristics calculated if $\Rout=\Rmax$.

 \section{Magnetic Prandtl number}\label{s.Prandtl}

% see also Appendix B in https://arxiv.org/pdf/astro-ph/9411059
%about radiative viscosity alpha

The magnetic Prandtl number is the ratio of momentum diffusivity (i.e. microscopic kinematic coefficient of viscosity $\nu$) to the magnetic diffusivity $\magdif$:
\begin{equation}
    \prm = \frac{\nu}{\magdif}\, .
\end{equation}
In terms of the Reynolds number Re and magnetic Reynolds number $\Rem$, defined as  Re  $=  V L/\nu$  and $\Rem =  V L/\magdif$, where $V$ and $L$ represent some reference velocity and length, respectively, 
$\prm = \Rem/$Re.

The processes of momentum exchange between the  particles define the viscosity in the medium. Collisions of neutrals-neutrals,  ion-neutrals, electron-neutrals are considered below, as well as collisions between photons and electrons (scattering). For simplicity, in the formulas below, we assume the pure hydrogen plasma and substitute the charge $Z=1$ and $n_i = n_e$, but to calculate the electron volume density $n_e$ for Fig.~\ref{fig:Prandtl} we use the mean number of free electrons per nucleon $ 1/\mu_e = n_e\, m_p/\rho $, 
%ionisation_degree = n_i/ (n_i+n_a)  ,
% where n_i - only ions volume density!
%where $1/\mu = 1/\mu_i + 1/\mu_e $
 as provided by the \textsc{EOS-MESA} module for solar abundances, called from \vertcode, see Appendix~\ref{s.thin_disc_code}.  \revtri{Hydrogen atom density is   $n_{\rm H} \approx ((1+X)/2 - 1/\mu_e ) \rho/m_p$, where $X=0.7$ and the maximum number of free electrons per nucleon is $(1+X)/2$.}
 
 %Collisions of ions and atoms, along with scattering of photons off the electrons, provide microscopic viscosity in AGN disc, while the electrons determine the disc resistivity.

%The following types of collisions  contribute to microscopic viscosity: between neutral atoms (HH),  ions (ii or pp), ions with neutrals (pH), and electron-photon collisions (e$\gamma$).  

 We assume the same temperature for all species and ignore the magnetic field.  
 For such plasma, the
%the total viscosity is the sum of viscosity from   the massive particles, 
viscosity of electrons can be ignored~\citep{zhdanov2002transport}, $\eta =  \eta_{i} +\eta_{H}$.
% page 119, $ 5.1, $ 7.1
%https://scipub.euro-fusion.org/wp-content/uploads/2014/11/EFDR07001.pdf
Approximately, we calculated the total kinematic viscosity as the sum
\begin{equation}
\label{eq.total_visc}
    \nu = \eta/\rho + \nu_{e\gamma}\, ,
\end{equation} 
%Following \citet{Potter-Balbus2014}, we take 
where the radiation viscosity is~\citep{mihalas-mihalas1984}:
% eq 97.40
\begin{equation}
\label{eq.phot_visc}
\nu_{e\gamma} = {\frac {4}{15}}\,{\frac {{a_\mathrm{rad}}\,{T}^{4}}{c\,\kappa\,{\rho}^{2}}} \, ,
%6.73 \times 10^{-26}  \, \frac{T^4}{\kappa \, \rho^2} \text{cm}^2/\text{s}\, ,
\end{equation}
where $a_\mathrm{rad}\approx 7.566 \times 10^{-15}$~g\,cm$^{-1}$\,s$^{-2}$\,T$^{-4}$ is the radiation density constant  and $\kappa$ is the Rosseland opacity [cm$^2$g$^{-1}$], which we calculate using the \textsc{EOS-MESA} module of \vertcode.  

{We note that the resulting increase of $\prm(\rho,T)$ at high temperature, see Fig.~\ref{fig:Prandtl}, is largely determined by the radiation viscosity~\citep[see also][]{Potter-Balbus2017}.  
At the same time, it is important to recognise that Eq. ~\eqref{eq.phot_visc} requires more consideration.
On the microscopic scale, where the material viscosity works, the radiation cannot be treated in the diffusion approximation, under which the shear viscosity is derived \eqref{eq.phot_visc}~\citep[][]{Agol-Krolik1998, Clouet-Soulard2021}. Nevertheless, the radiation reaction forces or bulk viscosity still work to dampen the velocity fluctuations and lower the Reynolds number Re. In the end, this leads to an increase in the Prandtl number~\citep{Jiang+2013,Blaes2014}.}

\revtri{Formulae below are needed to describe not-fully ionised regimes when the Spitzer formulae do not apply. Including only the ionic viscosity in \eqref{eq.total_visc} and only the $ee/ei$ resistivity in \eqref{eq.resistivity}, we get $\prm$ consistent with the expressions used in, for instance,~\citet{Balbus-Henri2008}.
}

For the viscosity of partially ionised matter,
we use expressions (4) and (7) from \citet{Vranjes2014} for the ionic and atom viscosity:
 %, following the approach of \citet{zhdanov2002transport}, presents formulas 4 and 7 :
\begin{equation}
\eta_{i} =n_i\, k\, T \, \tau_{\rm col, i} \, \xi_i/2 \, , 
\qquad
\eta_{H} = n_H\, k       
    \, T \, \tau_{{\rm col}, H}\, \xi_a /2\, , 
    \label{eq.visc_i_H}
\end{equation}
where the collision times $\tau_{\rm col}$, with the subscript $i$ or $H$, are combinations of  specific collision times between ions  and H atoms~\citep[in the approximation of collisions of hard spheres or Coulomb interactions, see][\S5.1]{zhdanov2002transport}:
\begin{equation}
\tau^{-1}_{{\rm col}, i} = 0.3 \tau^{-1}_{pp} + 0.4\tau^{-1}_{pH}\, ,\qquad
 \tau^{-1}_{{\rm col}, H} = 0.3 \tau^{-1}_{HH} + 0.4 \tau^{-1}_{Hp} \, ,
 \label{eq.tau_col_p_H}
\end{equation}
% \begin{equation}
%  \nu_{\rm col, H} = 0.3 \nu_{HH} + 0.4 \nu_{Hp} \, , \qquad
%  \nu_{\rm col, i} = 0.3 \nu_{pp} + 0.4\nu_{pH}\, ,
% \end{equation}
and dimensionless coefficients $\xi$ lie in the interval 1$-$2, see expressions~(\ref{eq:xi_coeffs}) below.  

The magnetic diffusivity $\magdif$ is 
% "омическая диффузия" по Дудорову, Хабрахманову
related to the electrical conductivity $\sigma$ as $\magdif \equiv c^2/(4\pi \sigma)$ (in CGS units, used by us, while $\magdif = 1/(\mu_0 \sigma)$ in SI), where the electrical conductivity $\sigma$ (inverse resistivity)
%!! electrical resistivity is reverse to the electrical conductivity 
in the ionised plasma is~\citep{Spitzer1962book, NRLPF2013}:
\begin{equation}
\sigma  = n_e\,Z^2\, e^2\, \tau_{\rm col}/ m_e\,   ,
\label{eq.conductivity}
\end{equation}
%\begin{equation}
%\sigma  = \frac{1.96 n_e q_e^2 \tau_e}{m_e} 
%\end{equation}
where $\tau_{\rm col}^{-1}$ is the collision frequency of electrons.
%Notice that magnetic diffusivity $\magdif \propto \tau_{\rm col}^{-1} $.

We consider that the total magnetic diffusivity results from three types of collisions:  electron-ion ($ei$), electron-electron ($ee$), and electron-neutral ($eH$):
\begin{equation}
\label{eq.resistivity}
    \magdif = \frac{c^2\, m_e}{4\pi \, n_e \,e^2} \, (\tau^{-1}_{ei,ee} + \tau^{-1}_{eH})\, ,
\end{equation}
where the specific collision times are presented below.
%[ electron-photon resistivity is very small ]

\paragraph{Collision coefficients}

For the Lorentzian gas (collisions between electrons and protons only), the collision frequency is~\citep{NRLPF2013}:
\begin{equation}
    \tau^{-1}_{ei} = \frac 43\, \frac{\sqrt{2\pi/m_e}\, e^4 \,n_i\,\lambda_{ei}}{(kT_e)^{3/2}} \, .
    \label{eq.nuei}
\end{equation}
A modification taking into consideration electron-electron collisions yields the value $\tau^{-1}_{ei,ee} = \tau^{-1}_{ei}/0.582$~\citep{Spitzer1962book}. 

The collision frequency between ions is \citep{NRLPF2013}:
\begin{equation}
    \tau^{-1}_{ii} = \frac 43\, \frac{\sqrt{\pi/m_p}\, (Ze)^4 \,n_i\,\lambda_{ii}}{(kT_i)^{3/2}} \, .
    \label{eq.nuii}
\end{equation}

%\citep{Chae-Litvinenko2021}. 
The Coulomb logarithms $\lambda_{ii} = \lambda_{ei}
%\approx \lambda_{ee}
\sim  5$ for $n_e \sim  10^{16} \mathrm{cm}^{-3}$ and $T\sim 10^4 $~K~\citep[][]{NRLPF2013}.
%for $T <10^5$~K;
%The Coulomb logarithm $\ln \Lambda $(introduced by Spitzer 1962) describes the cumulative effect of  small angle deflections related to Coulomb-type collisions.  

General expression for the collision frequency of particles `x' (electrons $e$, ions $i$, or atoms $H$) by the neutral hydrogen atoms is as follows:
\begin{equation}
  \tau^{-1}_{\mathrm{x}H}  =  n_H \, \sigma^{{\mathrm{x}|H}}_s\, \sqrt{k\,T_{\rm x}/m_{\rm x}} \, ,
  \label{eq.nueh}   
\end{equation}
where we took as a rough upper estimate $\sigma_s = 5\times 10^{-15}$ cm$^2$ for each of the cross-sections $\sigma^{e|H}_s$,
$\sigma^{p|H}_s$, and $\sigma^{H|H}_s$, see \citet{NRLPF2013}, the `Weakly ionised Plasma' section, and \citet{Vranjes-Krstic2013}. Further, for (\ref{eq.tau_col_p_H}) we need $\tau_{Hp}$ which is found from the momentum conservation:
\begin{equation}
    \tau_{Hp} = \tau_{pH}\, n_H/n_i \, .
\end{equation} 
%τ ai = τ ia ma na/(mi ni ) 

%?It can be noted that the frequency of H-e collisions is greater than that of H-p, and thus the latter can be ignored in the result, since the magnetic diffusivity $\propto$ the collision frequency.

 In the scope of the mentioned assumptions (no magnetic field, single temperature, collisions of hard spheres or Coulomb interactions), the dimensionless coefficients $\xi$ entering the expressions for the viscosity (\ref{eq.visc_i_H}) can be calculated as follows~\citep[\S7.1 of][]{zhdanov2002transport,Vranjes2014}:
\begin{equation}
\xi_a = 1 + \frac{ n_p/n_{H}}{3 \, \tau_{pH}/\tau_{pp}+4};
\, , \qquad
\xi_i = 1 + \frac{1}{3\,\tau_{pH}/\tau_{HH} + 4 \, n_p/n_{H}}\, . 
\label{eq:xi_coeffs}
\end{equation}

%where the bulk density $\rho = m_p \, (n_i + n_{\mathrm H})$,
%\citep{Vranjes2014}, Braginskii (1965).

% The first of these gross inequalities arises because thermal ion speeds are less than thermal electron speeds, by (me/mi)1/2 if Te ≈ Ti , and so ions take longer to meet each other. The second reflects the fact that the electrons are not very effective in deflecting the much heavier ions.

\end{appendix}

\label{lastpage}
\end{document}